\documentclass[tightenlines,floatfix,11pt]{article}

\usepackage{epsfig}
\usepackage{amsfonts}
\usepackage{amsmath}
\usepackage{amssymb}
\usepackage{dsfont}
\usepackage{hyperref}
\usepackage{color}
\usepackage{cite}

\newcommand{\MS}{$\overline{\text{MS}}$ }

\newcommand{\A}{\mathcal{A}}
\newcommand{\kk}{\mathbf{k}}
\newcommand{\pp}{\mathbf{p}}
\newcommand{\Order}{\mathcal{O}}

\newcommand{\GeV}{\,\text{GeV}}
\newcommand{\MeV}{\,\text{MeV}}
\newcommand{\fm}{\,\text{fm}}

\newcommand{\dt}{\delta}

\newcommand{\slashpi}{\protect{\slash\hspace{-0.5em}\pi}}
\newcommand{\spacevec}[1]{{\mathbf #1}} 

\newcommand{\cq}{\mathcal Q}

\newcommand{\mpi}{M_{\pi}}

\newcommand{\dslash}[1]{#1 \llap{/\kern-0.5pt}}
\newcommand{\Dslash}[1]{#1 \llap{/\kern+1.5pt}}
\newcommand{\DDslash}[1]{#1 \llap{/\kern+2.3pt}}
\newcommand{\dslashh}[1]{#1 \llap{/\kern+1pt}}

\newcommand{\boldtau}{\mbox{\boldmath $\tau$}}

\newcommand{\bea}{\begin{eqnarray}}
\newcommand{\eea}{\end{eqnarray}}
\newcommand{\be}{\begin{equation}}
\newcommand{\ee}{\end{equation}}
\newcommand{\bma}{\begin{pmatrix}}
\newcommand{\ema}{\end{pmatrix}}
\newcommand{\nn}{\nonumber}

\numberwithin{equation}{section}

\allowdisplaybreaks[1]

\begin{document}
\begin{titlepage}

\begin{flushright}
LA-UR-21-20994
\end{flushright}

\vspace{1cm}

\begin{center}
{\LARGE  \bf  
Determining the 
leading-order contact term in\\[2mm] 
neutrinoless double $\boldsymbol{\beta}$ decay
}
\vspace{1cm}

{\large \bf  Vincenzo Cirigliano,$^a$ Wouter Dekens,$^{b}$
  Jordy de  Vries,$^{c,d,e,f}$\\[1mm] 
  Martin Hoferichter,$^g$ 
  Emanuele Mereghetti$^a$}
\vspace{0.5cm}

\vspace{0.25cm}

{\large 
$^a$ 
{\it Theoretical Division, Los Alamos National Laboratory, Los Alamos, NM 87545, USA}}

{\large 
$^b$ 
{\it  Department of Physics, University of California at San Diego, La Jolla, CA 92093, USA}}

{\large 
$^c$ 
{\it Institute for Theoretical Physics Amsterdam and Delta Institute for Theoretical Physics, University of Amsterdam, Science Park 904, 1098 XH Amsterdam, The Netherlands}}

{\large 
$^d$ 
{\it Nikhef, Theory Group, Science Park 105, 1098 XG, Amsterdam, The Netherlands}}

{\large 
$^e$ 
{\it Amherst Center for Fundamental Interactions, Department of Physics, University of Massachusetts, Amherst, MA 01003, USA}}

{\large 
$^f$ 
{\it RIKEN BNL Research Center, Brookhaven National Laboratory,
Upton, New York 11973-5000, USA}}

{\large 
$^g$ 
{\it Albert Einstein Center for Fundamental Physics, Institute for Theoretical Physics, University of Bern, Sidlerstrasse 5, 3012 Bern, Switzerland}}

\end{center}

\begin{abstract}
We present a method to determine the  leading-order (LO) contact term contributing 
to the  $nn \to pp e^-e^-$  amplitude through the exchange of light Majorana neutrinos. 
Our approach  is based on the representation of the 
 amplitude as the momentum integral of 
a known kernel (proportional to the neutrino propagator)  times the generalized forward 
Compton scattering amplitude  
$n(p_1)   n(p_2)   W^+ (k)  \to   p(p_1^\prime)  p(p_2^\prime)  W^- (k)$, 
in analogy to the Cottingham formula for the electromagnetic contribution to hadron masses. 
We construct  model-independent representations of the integrand   in the low-  and high-momentum  
regions, through  chiral EFT and the operator product expansion, respectively. 
We then construct a  model for the full amplitude by interpolating between these two regions, 
using  appropriate nucleon factors for the weak currents and information on nucleon--nucleon ($N\! N$) scattering 
in the $^1S_0$ channel away from threshold.  
By matching the amplitude obtained in this way to the LO  chiral EFT amplitude 
we obtain the relevant LO contact term and  discuss various sources of uncertainty. 
We validate the approach by computing the analog  $I = 2$  $N\! N$ contact term  
and by reproducing, within uncertainties,  the charge-independence-breaking contribution to the $^1S_0$  $N\! N$ scattering lengths. 
While our analysis is performed in the \MS scheme, we express our final result in terms of  the scheme-independent 
renormalized amplitude $\A_\nu(|\pp|,|\pp^\prime|)$ 
at a set of kinematic points near threshold. 
We illustrate for two cutoff schemes how, 
using our  synthetic data for $\A_\nu$, one can determine the  contact-term contribution in any 
regularization scheme, in particular the ones  employed in nuclear-structure calculations for isotopes of experimental interest.

\end{abstract}

\vfill
\end{titlepage}

\tableofcontents

\section{Introduction}

Neutrinoless double $\beta$ decay ($0\nu\beta\beta$) is the process in which two neutrons in a nucleus convert into two protons by emitting two electrons and
 no  neutrinos~\cite{Furry:1939qr}. 
This process  is by far the most sensitive laboratory probe of lepton number violation (LNV) and 
its observation would prove that neutrinos are Majorana fermions~\cite{Schechter:1981bd},  
constrain neutrino mass parameters, and provide experimental validation for  leptogenesis scenarios~\cite{Davidson:2008bu,Rodejohann:2011mu}.  
If $0\nu\beta\beta$ decay is caused by the exchange of light Majorana neutrinos, as we assume throughout this paper, the amplitude is proportional to the effective neutrino mass $m_{\beta\beta}=\sum_i U_{ei}^2 m_i$, where the sum runs over light neutrino masses $m_i$ and $U_{ei}$ are elements of the neutrino mixing matrix. $0\nu\beta\beta$ decay is a complicated process encompassing aspects from particle, nuclear, and atomic physics, with the interpretation of current experimental limits~\cite{KamLAND-Zen:2016pfg,Arnold:2016qyg,Albert:2017owj,Aalseth:2017btx,Adams:2019jhp,Agostini:2020xta}  and of  potential future discoveries limited by substantial uncertainties in the calculation of hadronic and nuclear matrix elements~\cite{Avignone:2007fu,Menendez:2008jp,Vergados:2012xy,Simkovic:2013qiy,Vaquero:2014dna,Barea:2015kwa,Senkov:2015juo,Engel:2016xgb,Dolinski:2019nrj}.       

It has been realized in recent years that chiral effective field theory (EFT)~\cite{Weinberg:1978kz,Weinberg:1990rz,Weinberg:1991um,Kaplan:1996xu,Kaplan:1998tg,Kaplan:1998we}  
can play a central role in addressing these uncertainties.  Nuclear structure, \textit{ab-initio} calculations based on chiral-EFT interactions~\cite{Epelbaum:2008ga,Machleidt:2011zz,Hammer:2019poc} 
have recently become available for some  phenomenologically relevant nuclei~\cite{Yao:2019rck,Belley:2020ejd,Novario:2020dmr}.  In addition,
the issue of $g_A$ quenching in single $\beta$ decays has been demonstrated to arise from the combination of two-nucleon 
weak currents  and strong correlations in the nucleus~\cite{Towner:1987zz,Pastore:2017uwc,Gysbers:2019uyb}, and
the few-nucleon amplitudes used as input in nuclear structure calculations have been scrutinized in chiral EFT
for various sources of LNV~\cite{Prezeau:2003xn,Menendez:2011qq,Cirigliano:2017djv,Cirigliano:2017tvr,Pastore:2017ofx,Cirigliano:2018hja,Wang:2018htk,Cirigliano:2018yza,Cirigliano:2019vdj,Dekens:2020ttz}. 
In the context of light-Majorana-neutrino exchange, using 
naive dimensional counting, the leading contribution in the chiral-EFT expansion arises from a 
neutrino-exchange diagram, in which the LNV arises from insertion of the $\Delta L=2$ effective neutrino mass $m_{\beta \beta}$. 
When considering the $^1S_0$ channel, 
in analogy to the nucleon--nucleon ($N\!N$) potential itself~\cite{Kaplan:1996xu,Kaplan:1998tg,Kaplan:1998we}
and  external currents~\cite{Valderrama:2014vra}, this conclusion no longer holds when demanding manifest renormalizability of the amplitude.  
In fact, it has been shown that renormalization  requires the promotion of an $nn\to ppe^-e^-$ contact operator to LO~\cite{Cirigliano:2018hja,Cirigliano:2019vdj}, which encodes  the exchange of neutrinos with energy/momentum  greater than the nuclear scale and thus cannot be resolved in chiral EFT. 
As discussed in greater detail in Refs.~\cite{Cirigliano:2019vdj,Cirigliano:2020yhp}, 
the new coupling encodes  a non-factorizable   two-nucleon effect,  beyond the factorizable one-nucleon corrections 
captured by  the radii of  weak form factors, which also give a short-range neutrino
potential.   Moreover, the new short-range coupling is not captured by the  so-called
short-range correlations~\cite{Miller:1975hu,Simkovic:2009pp,Engel:2011ss,Benhar:2014cka},
as it is  needed even when  one works with fully correlated wave functions, i.e.,  exact solutions of 
the Schr\"odinger equation with the appropriate strong potential.  
The situation is  analogous to single  $\beta$ decay,  where two-nucleon weak currents and short-range correlations
are both present, and the combination of both 
leads to the apparent quenching of $g_A$~\cite{Pastore:2017uwc,Gysbers:2019uyb}.

To leading order in chiral EFT,  a contact interaction is needed only in the $^1S_0$ channel and not in higher partial waves~\cite{Cirigliano:2019vdj}. 
However, it  is worth emphasizing that the effect of the contact term in the $^1S_0$ channel   
is  {\it amplified}  in nuclear matrix elements  by the cancellation 
 between the contribution of  $N\!N$ pairs  in the  $^1S_0$ channel and in states with  higher  total angular momentum.  
 This is seen quite dramatically in the matrix element densities for light nuclear transitions studied 
 in Refs.~\cite{Cirigliano:2018hja,Cirigliano:2019vdj}. 
 These {\it ab-initio} results in light nuclei are in  qualitative agreement with the behavior observed in heavy nuclei,   first  discussed   
 in Ref.~\cite{Simkovic:2007vu}. 
 A complete discussion of the $nn \to pp$ transition operator in chiral EFT 
can be found in Ref.~\cite{Cirigliano:2019vdj} (leading order) and 
Refs.~\cite{Cirigliano:2017tvr,Wang:2018htk}  (higher orders).

The value of the  short-range coupling in the  $^1S_0$ channel then has to be
 either  extracted from other processes related by chiral symmetry or calculated from first principles in lattice QCD~\cite{Feng:2018pdq,Tuo:2019bue,Cirigliano:2020yhp,Detmold:2020jqv,Feng:2020nqj,Davoudi:2020ngi,Davoudi:2020gxs} (see Ref.~\cite{Richardson:2021xiu} for a large-$N_c$ analysis). Currently, however, 
 the size of this contact operator is unknown,
 leading to substantial uncertainties in the interpretation of $0\nu\beta\beta$ decays besides the nuclear-structure ones, especially given that its impact is enhanced in $\Delta I=2$ nuclear transitions due to a node in the matrix element density~\cite{Cirigliano:2018hja,Cirigliano:2019vdj}.  
In this work  we present in some detail  the method used to obtain a first  estimate of the complete $nn\to pp e^-e^-$ amplitude including this contact-term contribution~\cite{Cirigliano:2020dmx}.

The hadronic component  of the 
light-Majorana-neutrino-exchange amplitude has the structure 
\be
{\cal A}_\nu  \propto  
\int \frac{d^4k}{(2\pi)^4} \frac{g_{\alpha \beta}}{k^2 + i \epsilon}  
\int d^4x\, e^{i k \cdot x} 
 \langle pp | 
T\{j_\text{w}^\alpha(x) j_\text{w}^\beta(0)\}
|nn \rangle\,,
\label{eq:M1}
\ee
and is controlled by the 
two-nucleon 
matrix element of the time-ordered product 
$T\{j_\text{w}^\alpha(x) j_\text{w}^\beta(0)\}$
of two weak currents.  
Such matrix elements with the weak current replaced by the electromagnetic current $j_\text{em}^\alpha(x)$ 
appear in 
the electromagnetic contributions to hadron masses or scattering processes, in which case a relation exists between the forward Compton scattering amplitude and its contraction with a  massless propagator, 
as given in Eq.~\eqref{eq:M1}.
The relation, known as the Cottingham formula~\cite{Cottingham:1963zz,Harari:1966mu}, has been used to estimate the electromagnetic contributions to the masses of pions~\cite{Ecker:1988te,Bardeen:1988zw,Donoghue:1993hj,Baur:1995ig,Donoghue:1996zn} and nucleons~\cite{Gasser:1974wd,Gasser:1982ap,WalkerLoud:2012bg,Thomas:2014dxa,Erben:2014hza,Gasser:2015dwa,Gasser:2020mzy,Gasser:2020hzn}. 
Since the matrix elements in these cases have
precisely the same structure as required for the light-Majorana-neutrino-exchange contribution to the $0\nu\beta\beta$ decay $nn\to pp e^-e^-$, our method aims to constrain the corresponding amplitude by generalizing the Cottingham approach to the two-nucleon system, and then determine the contact-term contribution by matching to chiral EFT. 

The application of the Cottingham approach to the pion and nucleon mass difference has a long history 
and in both cases the  by far dominant contribution arises from elastic intermediate states.  
The pion-pole contribution gives more than $80\%$ of the pion mass difference~\cite{Ecker:1988te,Bardeen:1988zw,Donoghue:1993hj,Baur:1995ig,Donoghue:1996zn} and, similarly, the nucleon pole provides the bulk of the electromagnetic part of the proton--neutron mass difference $m_{p-n}^\text{el}=0.75(2)\MeV$. Despite the tension
between the estimate of the inelastic contributions in lattice QCD, $m_{p-n}^\text{inel}=0.28(11)\MeV$~\cite{Borsanyi:2014jba,Brantley:2016our,Horsley:2019wha}, and from nucleon structure functions, $m_{p-n}^\text{inel}=-0.17(16)\MeV$~\cite{Gasser:2020mzy,Gasser:2020hzn,Gasser:2015dwa,Gasser:1974wd}, also in this case 
the elastic estimate is accurate at the $30\%$ level. 
The main complication in the generalization to $0\nu\beta\beta$ decay arises from the two-particle nature of initial and final states. First, due to the 
proliferation of kinematic variables and scalar functions in a Lorentz decomposition of the general amplitude, it becomes extremely cumbersome to try and set up a strict derivation 
of the elastic contribution via dispersion relations. Second, the $N\!N$ scattering amplitude itself gives rise to an additional source of momentum dependence that adds to the physics included in terms of pion and nucleon form factors in the standard Cottingham approach. Accordingly, we do not attempt a comprehensive analysis of all scalar functions describing the full two-particle problem, but instead
include the most important intermediate states in terms of the respective form factors---in close analogy to the elastic results for the pion and nucleon Cottingham formula---as well as the momentum dependence of the $N\!N$ scattering amplitude.   
To validate this approach, 
we also consider the two-nucleon matrix element  with two electromagnetic currents, 
which can be accessed experimentally in terms of
charge independence breaking (CIB) in the $N\!N$ scattering lengths. 
Comparison with data then allows us 
to confirm the expectation  
of an accuracy around $30\%$ if only elastic contributions are kept, as suggested by the Cottingham result for the proton--neutron mass difference. 
A determination at this level already has a valuable  impact in bounding the size of the 
contact-term contribution to $0\nu\beta\beta$ decay.

The derivation is organized as follows: 
in Section~\ref{sect:integral} we present the general integral representation of the amplitude and our matching strategy. 
In Section~\ref{sect:chiEFT} we recast the LO chiral EFT amplitude in a form  suitable for matching purposes. 
In Section~\ref{sect:full} we present our construction of our full $nn \to pp$ amplitude, followed by 
matching to the EFT result and  extraction of the contact term in Section~\ref{sect:matchLL}. 
In Section~\ref{sect:VV} we present the analysis for the  two-nucleon $I=2$ electromagnetic amplitude 
and the validation of the method through comparison with the experimental data on CIB in the  $N\!N$ scattering lengths. 
This section is fairly technical and can be omitted by readers primarily interested in $0\nu\beta\beta$ decay.
In Section~\ref{sect:synthetic} we return to the LNV amplitude $nn \to pp$ and present synthetic data at kinematic points near threshold, 
illustrating how these can be used to extract the contact term in any regularization and renormalization scheme. 
We present our concluding remarks in Section~\ref{sect:conclusion}.
Details on the half-off-shell behavior of the $N\!N$ scattering amplitude (Appendix~\ref{sect:HOST}), 
the operator product expansion (OPE) (Appendix~\ref{sec:LROPE}), the size of typical inelastic contributions (Appendix~\ref{sect:inel}), 
the electromagnetic pion mass splitting (Appendix~\ref{app:Zm}), and the CIB $N\!N$ scattering lengths (Appendix~\ref{sect:AppCIB}) are provided in the appendices.

\section{Integral representation  and matching strategy}
\label{sect:integral}

\subsection{Generalities}

Including the effect of LNV from the dimension-five Weinberg operator~\cite{Weinberg:1979sa}, 
the  low-energy  effective Lagrangian at scale  $\mu   \gtrsim \Lambda_\chi \sim 1\GeV$  is given by
\be
{\cal L}_{\rm eff} =    {\cal L}_{\rm QCD}  \ -\left\{ \   2 \sqrt{2} G_F V_{ud} \    \bar{u}_L \gamma^\mu d_L \, \bar{e}_L \gamma_\mu \nu_{eL} 
\ + \   \frac{1}{2} m_{\beta \beta}^*  \  {\nu}_{eL}^T  C  \nu_{eL}   
\ - \     C_L    \,  O_{L} +{\rm h.c.}\right\}\,. 
\label{eq:Seff0}
\ee
The second term in Eq.~\eqref{eq:Seff0} 
represents the Fermi charged-current weak interaction. 
The last two terms encode LNV through the neutrino  Majorana  mass, 
given by $m_{\beta \beta} = \sum_i  U_{ei}^2  m_i$ in terms of mass eigenstates and elements of the neutrino mixing matrix,\footnote{The effective mass probed in $0\nu\beta\beta$ decay is often defined as 
$m_{\beta \beta} = \big|\sum_i  U_{ei}^2  m_i\big|$, but to simplify the notation at the Lagrangian level we formally keep its phase.} 
and  a dimension-nine $\Delta L = 2$ operator  
generated at the electroweak threshold: 
 \be 
 O_{L} =  \bar{e}_L     e_L^c \  \bar{u}_L \gamma_\mu d_L \  \bar{u}_L  \gamma^\mu d_L 
 \equiv \bar{e}_L     e_L^c \ O_1 \,,
 \label{eq:O1def}
 \ee
  with $e_L^c = C \bar{e}_L^T$. 
 Since 
 $C_L =    (8 V_{ud}^2  G_F^2 m_{\beta \beta})/M_W^2   \times  (1 + \mathcal O(\alpha_s/\pi)) $, 
the effect of the latter term on the $0\nu\beta\beta$ amplitude is suppressed by $(k_F/M_W)^2$  (where $k_F \sim\Order(100)\MeV$ is the typical Fermi momentum of nucleons in a nucleus)  compared to light-neutrino exchange and can be safely neglected at this stage.  
However, the  isotensor four-quark local operator $O_1$ itself will play an important role in the following analysis. 

The interactions of Eq.~\eqref{eq:Seff0} induce  $\Delta L= 2$  transitions (such as  $\pi^- \pi^- \to e^- e^-$, $n n  \to  p p e^- e^-$, 
$^{76}{\rm Ge} \to  \,  ^{76}{\rm Se} \,  e^- e^-$,  $^{136}{\rm Xe} \to \,    ^{136}{\rm Ba} \, e^- e^-$, \ldots) through  the non-local effective action  
obtained by contracting the neutrino fields in the two weak vertices, 
\be
S_{\rm eff}^{\Delta L = 2} 
=  \frac{ 8  G_F^2 V_{ud}^2  m_{\beta \beta} }{2!}    \int d^4 x  d^4y    \ S(x-y) \times   \bar{e}_L   (x)  \gamma^\mu  \gamma^\nu e_L^c (y)  
\times  T \Big( \bar{u}_L \gamma_\mu d_L(x)   \ \bar{u}_L \gamma_\nu d_L(y)  \Big)\,\,,
\label{eq:Seff1-v0}
\ee
where
\be  \label{eq:EffProp}
S(r) = \int \frac{d^4k}{(2 \pi)^4}   \frac{e^{-i k \cdot r}}{k^2  + i \epsilon}  
\ee
is the scalar massless propagator, a remnant of the neutrino propagator. 
Computing matrix elements of $S_{\rm eff}^{\Delta L=2}$ in hadronic and nuclear states
is a notoriously difficult task.  
The multi-scale nature of the problem can be seen 
more explicitly by going to the Fourier representation 
\footnote{In deriving  Eq.~\eqref{eq:Sfourier1}  from 
Eq.~\eqref{eq:Seff1-v0}
we approximate 
$\bar{e}_L   (x)  \gamma^\mu  \gamma^\nu e_L^c (y) \simeq  \bar{e}_L   (x)  \gamma^\mu  \gamma^\nu e_L^c (x)  = g^{\mu \nu}   \bar{e}_L   (x)  e_L^c (x)$, 
which amounts to neglecting  the difference in electron momenta,  a safe assumption given that $|p_{e1} - p_{e2}|/k_F \ll 1$.} 
\be
\langle e_1 e_2 h_f |  S_{\rm eff}^{\Delta L = 2}  |h_i \rangle 
=   \frac{ 8  G_F^2 V_{ud}^2  m_{\beta \beta} }{2!}    \int d^4 x  \  
\langle e_1 e_2 |    \bar{e}_L   (x)  e_L^c (x)  | 0 \rangle 
\int \frac{d^4k}{(2 \pi)^4} 
\frac{  g^{\mu \nu} 
\overline{T}_{\mu \nu} (k, p_{\rm ext},x) 
 }{ k^2  + i \epsilon}\,\,, \ \ 
\label{eq:Sfourier1}
\ee
where 
\begin{align}
\overline{T}_{\mu \nu} (k, p_{\rm ext},x)   
&=   \langle h_f  (p_f) | \   \hat{\Pi}_{\mu \nu}^{LL} (k,x)  \ | h_i  (p_i) \rangle\,,\notag
\\
\hat{\Pi}_{\mu \nu}^{LL} (k,x) &=  \int d^4 r  \, e^{i k \cdot r} \     T  \Big\{  J_\mu^L (x +r/2)   \ J_\nu^L(x - r/2)  \Big\}\,,\qquad 
 J^L_\mu  =  \bar u_L \gamma_\mu d_L\,,
\label{eq:correlator}
\end{align}
and $p_{\rm ext}$ denotes generically the hadronic external momenta $p_f$ and $p_i$.
Using translational invariance one has   
\begin{align}
\overline{T}_{\mu \nu} (k, p_{\rm ext},x)   
&=   \langle h_f  (p_f) | \ e^{i x \cdot P} \    \hat{\Pi}_{\mu \nu}^{LL} (k,0)  \ e^{ - i x \cdot P} \ | h_i  (p_i) \rangle \notag
\\
& =  e^{i x \cdot (p_f - p_i)}  \langle h_f  (p_f) | \      \hat{\Pi}_{\mu \nu}^{LL} (k,0)    \ | h_i  (p_i) \rangle  \notag
\\
& =  e^{i x \cdot (p_f - p_i)}  \ \overline{T}_{\mu \nu} (k, p_{\rm ext},0)\,.
\label{eq:compton}
\end{align}
Therefore, defining\footnote{In what follows we will  suppress the space-time label in the  correlator:  
 $\hat{\Pi}^{LL}_{\mu \nu} (k,0) \to \hat{\Pi}^{LL}_{\mu \nu} (k)$.}
\be
T_{\mu \nu} (k, p_{\rm ext}) 
\equiv \overline{T}_{\mu \nu} (k, p_{\rm ext},0) 
=  \langle h_f  (p_f) | \   \hat{\Pi}_{\mu \nu}^{LL} (k,0)  \ | h_i  (p_i) \rangle \,,
\ee
one arrives at 
\begin{align}
\langle e_1 e_2 h_f |  S_{\rm eff}^{\Delta L = 2}  |h_i \rangle 
&=  (2 \pi)^4 \, \delta^{(4)}  (p_{e1} + p_{e2} + p_f - p_i)  
\Big( 4  G_F^2 V_{ud}^2  m_{\beta \beta} \   \bar{u}_L   (p_1)  u_L^c (p_2)  \Big)  
\times 
 {\cal A}_\nu\,,\notag   \\
{\cal A}_\nu  & =
2\, \int \frac{d^4k}{(2 \pi)^4} 
\frac{    {\cal T} (k, p_{\rm ext}) 
 }{ k^2  + i \epsilon}\,, 
 \qquad
 {\cal T} (k, p_{\rm ext})  \equiv
g^{\mu \nu}  T_{\mu \nu} (k, p_{\rm ext}) \,. 
\label{eq:M}
\end{align}
The  hadronic amplitude   ${\cal A}_\nu$ in Eq.~\eqref{eq:M}  
receives contributions from neutrino virtualities $k^2$  ranging from the weak scale all the way down to the 
infrared (IR) scale of nuclear bound states.  

To estimate the LO contact term arising in  chiral EFT, 
we will employ  the  representation~\eqref{eq:M} to obtain the amplitude in the ``full theory,'' 
and then match to the appropriate EFT expression. 
Since the contact term arises in the $^1S_0$ channel, 
we will take as external states  $nn$ and $pp$  in the $^1S_0$ state and 
$T_{\mu \nu} (k, p_{\rm ext})$ will  be thought of as 
the generalized forward Compton amplitude 
\be
n(p_1) \   n(p_2)  \ W^+ (k)  \ \to  \ p(p_1^\prime) \ p(p_2^\prime) \ W^- (k)  \,.
\ee
Since the low-energy constants (LECs) do not depend on the IR details,  we will perform the matching calculation  at the simplest   kinematic point, in which  the two electrons are emitted with zero three-momentum in the center-of-mass frame of the incoming neutron pair~\cite{Cirigliano:2018hja,Cirigliano:2019vdj}. 
Explicitly we have  
\begin{align}
 p_{1 \mu} &= (E, \spacevec{p})\,, & p_{2 \mu} &= (E, - \spacevec{p})\,, & E &= \sqrt{\spacevec{p}^2 + m_n^2}\,,\notag\\
 p^\prime_{1 \mu} &= (E^\prime, \spacevec{p}^\prime)\,,& 
 p^\prime_{2 \mu} &= (E^\prime, - \spacevec{p}^\prime)\,,& 
 E^\prime &= \sqrt{\spacevec{p}^{\prime 2} + m_p^2}\,,
\end{align}
where $2 E = 2 E^\prime + 2 m_e$. 
Free two-nucleon states with vanishing total three-momentum and individual three-momenta given by $\pm \spacevec{q}$ will be denoted by $| \spacevec{q} \rangle$~\cite{Kaplan:1996xu}, so for example  for the initial and final state we will have $| i_0 \rangle = | \spacevec{p} \rangle$ and 
$| f_0 \rangle = | \spacevec{p}^\prime \rangle$, respectively.

\subsection{Matching strategy}

\begin{figure}[!t]
\begin{center}
\includegraphics[width=0.6\textwidth]{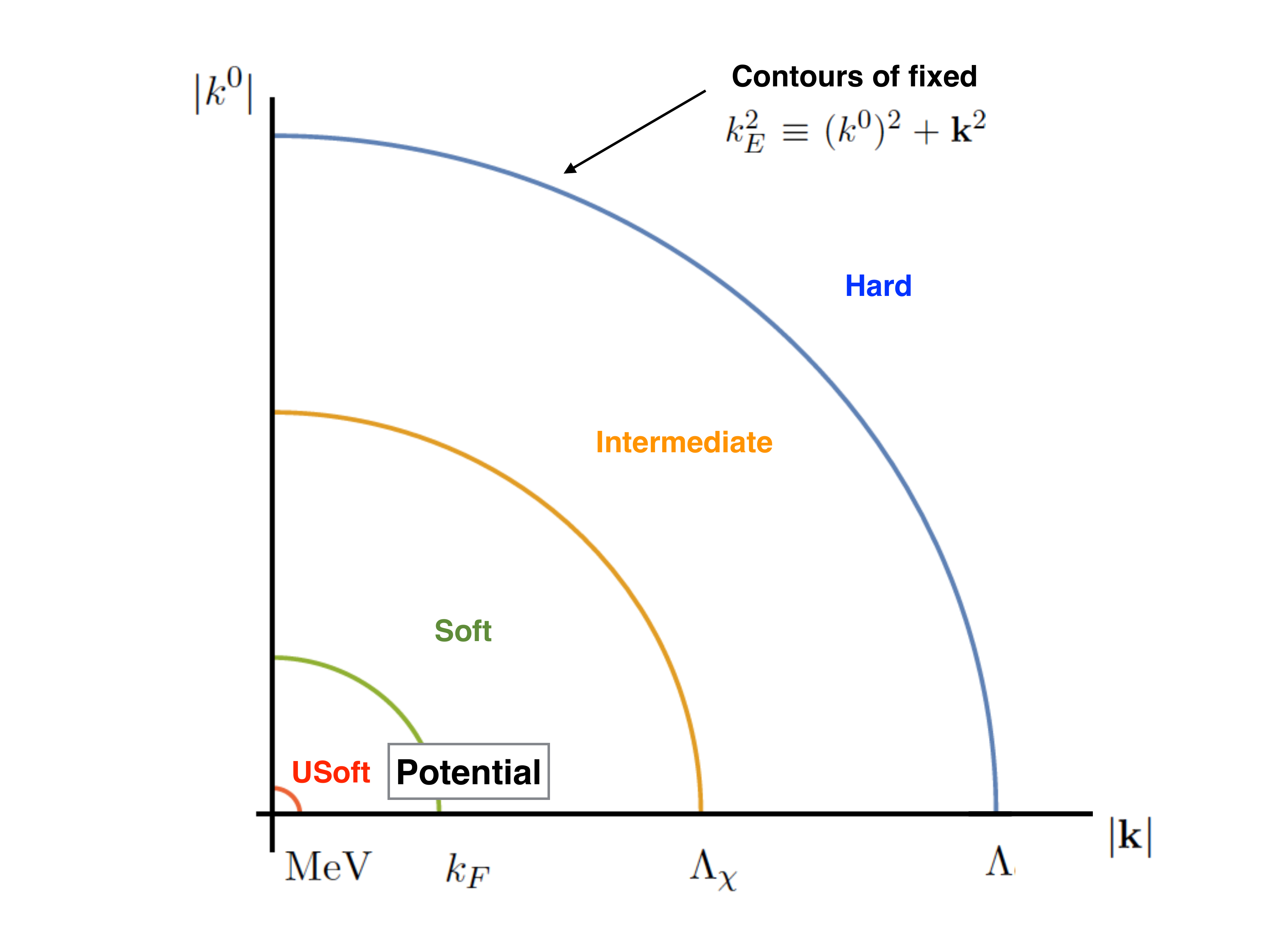}
\end{center}
\vspace{-0.6cm}
\caption{Schematic representation of the  regions of neutrino virtuality contributing to the amplitude in Eq.~\eqref{eq:M}.
The boundaries between various regions are given by  $k_F \sim 100\MeV$,  $\Lambda_\chi  \lesssim 1\GeV$,  and $\Lambda \gtrsim 1.5\GeV$.
 }\label{fig:regions}
\end{figure}

The   amplitude for the process $nn \to pp$ is given in Eq.~\eqref{eq:M} as the integral of the product  of a massless propagator 
(remnant of the Majorana neutrino  propagator)  
with the contracted hadronic tensor ${\cal T} (k,p_{\rm ext})= g_{\mu \nu} T^{\mu \nu}$. 
The neutrino  four-momentum regions relevant for the integration over $d^4 k$ are 
schematically depicted in Fig.~\ref{fig:regions}. 
Denoting the Euclidean four-momentum squared  by $k_E^2 \equiv (k^0)^2 + \spacevec{k}^2$, 
one can introduce   hard  ($k_E^2 > \Lambda^2$), intermediate  ($\Lambda_\chi^2 < k_E^2 < \Lambda^2$), and low-energy 
($k_E^2 < \Lambda_\chi^2$) regions,  separated by  $\Lambda_\chi$ (the breakdown scale of the low-energy hadronic EFT)  and 
$\Lambda$ (scale  at which the  OPE becomes reliable).
The low-energy region further includes  the soft   ($| k^0| \sim |\spacevec{k}| \sim  k_F$),  potential   ($|k^0| \sim  k_F^2/m_N$,  $|\spacevec{k}| \sim k_F$), 
 and ultrasoft  ($|k^0| \sim |\spacevec{k}|  \ll   k_F$)     regions, essential to reproduce the  IR behavior of the amplitude.

The basic idea behind our approach is that model-independent representations of the integrand in Eq.~\eqref{eq:M} 
can be constructed  in the low-energy  region (via pionless and chiral EFT) and  in the hard region (via the OPE). 
 Given this,  a  model for the full amplitude can be constructed by  interpolating between these two regions. 
This approach uses model-dependent input for the intermediate momentum region, which we anchor to known constraints from QCD at low and high momenta. 

In practice, given the non-relativistic nature of the process of interest, 
we will not use $k_E^2$ as matching  variable. Instead, we decompose 
$d^4 k = d k^0 d^3\spacevec{k}$, first perform the $k^0$ integral  in the appropriate regions via Cauchy's theorem, 
and then carry out the angular integrations in $d^3\spacevec{k}$  to reduce the amplitude to an integral over $d |\spacevec{k}|$. 
To LO in the expansion in {\it external} momenta $|\pp|, |\pp^\prime| \sim Q$ (we denote by $\mu_\chi \sim Q$ the soft scale in the EFT), 
we  write the full amplitude as  an integral over the  {\it internal} neutrino three-momentum $\spacevec{k}$,  
which we split into a low- plus  intermediate-momentum region and  a high-momentum region 
\begin{align}
{\cal A}_\nu^{\rm full}  &= 
 \int_0^\infty \,  d |\spacevec{k}|  \ a^{\rm full}  (|\spacevec{k}|) = 
 {\cal A}^< + { \cal A}^>\,,  
\notag
\\
{\cal A}^<  &=    \int_0^{\Lambda} \,  d |\spacevec{k}|  \ a_< (|\spacevec{k}|)\,,\notag
\\
{ \cal A}^>  &=  \int_{\Lambda}^\infty \,  d |\spacevec{k}|  \ a_> (|\spacevec{k}|)\,, 
\label{eq:Msplit}
\end{align}
separated by  the    scale $\Lambda$ that represents the onset of the 
asymptotic  behavior for the current--current correlator, controlled by the OPE. 
 This representation introduces model dependence through:   
(i)  The choices made to extend the model-independent  integrand $a_\chi (|\spacevec{k}|)$ 
 dictated by chiral EFT   in the region $|\spacevec{k}|  <  \Lambda_\chi$ to the 
 function  $a_< (|\spacevec{k}|)$   valid  up to     $|\spacevec{k}| \sim  \Lambda$. 
We will  provide  a simple parameterization of   $a_< (|\spacevec{k}|)$  that reduces to $a_\chi (|\spacevec{k}|)$  
for  $|\spacevec{k}| < M_\pi$   and incorporates  phenomenological input  such as nucleon form factors of the weak current and resonance contributions to the strong-interaction potential. 
(ii) The choice of $\Lambda$  that determines the boundary  of integration regions in the variable $|\spacevec{k}|$. 
Once a representation for ${\cal A}_\nu^{\rm full}$ is obtained, along with an estimate of the associated uncertainties, 
we will estimate the LEC appearing in ${\cal A}_\nu^{\chi {\rm EFT}}$ by  
enforcing the matching condition 
\be
{\cal A}_\nu^{\chi {\rm EFT}} =  {\cal A}^< + { \cal A}^> \,.
\label{eq:match0}
\ee
In the following sections, we will describe the construction of ${\cal A}^{<,>}$ and the matching to ${\cal A}_\nu^{\rm \chi {\rm EFT}}$, 
starting with the spectral representation in Section~\ref{sec:spectral}.

\subsection{Spectral representation}
\label{sec:spectral}

In this section we provide a spectral representation for the $nn \to pp$ amplitude, 
which will prove very useful in identifying and organizing  
the various intermediate-state  contributions to ${\cal A}^{<}$ and carrying out the analysis in analogy to the Cottingham formula~\cite{Cottingham:1963zz,Harari:1966mu}.

We begin by recalling some elements of the formal  theory of scattering  that we will use in various parts of the discussion. 
We denote by $\hat H = \hat H_0 + \hat V$ the total Hamiltonian, split into a free and interaction term ($\hat V$, not to be confused with the potential).
Retarded and advanced Green's functions in the interacting theory are given by
\be
\hat G_\pm  (E) =  \frac{1}{E - \hat H \pm i \epsilon}  = \int dt \, e^{i E t} \hat{G}_\pm (t)~, \qquad 
\hat G_\pm  (t) = \mp i \theta (\pm t) \, e^{- i \hat H t}\,, 
\ee
with analogous definitions for the  free-theory   ones,  denoted by $\hat{G}_\pm^{(0)} (E)$, with the replacement $\hat{H} \to \hat H_0$. 
The scattering operator $\hat T(E)$ is   formally given by
\be
\hat T (E) = \hat V \ \left(I - \hat G^{(0)}_+ (E) \,  \hat V \right)^{-1} 
\ee 
and satisfies $\hat G_+ (E) \hat V = \hat G^{(0)}_+ (E) \hat T (E)$. 
The scattering states $ \langle f_{-} |$ and $ | i_+ \rangle$ are related to the free states $\langle f_0|$ and $|i_0 \rangle$ by 
\begin{align}
| i_+ \rangle  &=  \left( I + \hat G_+^{(0)} (E) \, \hat T (E) \right)  |i_0 \rangle\,,\notag 
\\
\langle f_- |  & =  \langle f_0 | \left( \hat T (E^\prime) \hat G_+^{(0)} (E^\prime) +  I \right)\,.
\label{eq:scatteringstates}
\end{align}

In terms of the scattering states, 
the amplitude for $n n \to p p$ can be written as 
\be
{\cal A}_\nu = \int \frac{d^3 \spacevec{k}}{(2 \pi)^3} \ \langle f_- | \,  \hat{O}^{LL} (\spacevec{k})  \, | i_+ \rangle\,,
\label{eq:Mnew}
\ee
with the weak transition operator 
\be
\hat{O}^{LL} (\kk) \equiv   2   \int \frac{dk^0}{2 \pi} \,  \frac{g^{\mu \nu}  \hat{\Pi}_{\mu \nu}^{LL} (k)}{k^2 + i \epsilon} \,.
\label{eq:OLLdef}
\ee
From the definition of the correlator in Eq.~\eqref{eq:correlator} one obtains the following representation 
for   $\hat{\Pi}_{\mu \nu}^{LL} (k) $  in terms of Green's functions:
\be
g^{\mu \nu} \hat{\Pi}_{\mu \nu}^{LL} (k) = i (2 \pi)^3  \, 
 J_\mu^L (0) \, \left[ 
\hat{G}_+ (k_+^0) \, \delta^{(3)} ( \hat{\spacevec{P}} - \spacevec{k}_+) 
+ 
\hat{G}_+ (k_-^0) \, \delta^{(3)} ( \hat{\spacevec{P}} - \spacevec{k}_-) 
\right]  \,  J^{L \mu} (0)\,,
\label{eq:gf1}
\ee 
where $\hat{\spacevec{P}}$ is the total three-momentum operator and we have introduced the four-vectors
\be 
k_\pm^\mu = \tilde{p}^\mu \pm k^\mu\,, \qquad  \tilde{p}^\mu = \frac{1}{2} (p_i + p_f)^\mu = (\tilde E, \tilde{\spacevec{p}})\,. 
\ee
The labels in $p_{i,f}^\mu$ refer to the initial and final states between which  $\hat  \Pi^{LL}_{\mu \nu} (k)$ is evaluated. 
Since we are considering  two-nucleon external states with vanishing total three-momentum (and total momentum 
is conserved at each vertex) we have $\tilde{\spacevec{p}}=0$ and hence $\spacevec{k}_\pm = \pm \spacevec{k}$. 
In Eq.~\eqref{eq:gf1}  the dependence on $k^0$ is very simple, 
as $k^0$ appears only through the energy denominators of $\hat{G}_+( k_\pm^0)$. 
Performing the integration over $k^0$ in Eq.~\eqref{eq:OLLdef} with Cauchy's theorem,\footnote{
For each term in Eq.~\eqref{eq:gf1}, one can close the contour in the upper or lower $k^0$ plane so that the integral is given by the residue 
at the $k^0$ pole from the neutrino  propagator in Eq.~\eqref{eq:OLLdef}. }
one arrives at 
\be
\hat{O}^{LL} (\spacevec{k})  
= \frac{1}{|\spacevec{k}|} \ 
 J_\mu^L (0) \, \hat{G}_+ (\tilde{E} - |\spacevec{k}|) \    (2 \pi)^3  \left[  \delta^{(3)} ( \hat{\spacevec{P}} - \spacevec{k}_+) 
+   \delta^{(3)} ( \hat{\spacevec{P}} - \spacevec{k}_-) 
\right]  \,  J^{L \mu} (0)\,. 
\label{eq:gf2}
\ee 
Further inserting a complete set of states  between the current operators in Eq.~\eqref{eq:gf2}  leads to the spectral representation for the 
amplitude\footnote{The summation is over  intermediate states $| n (\kk_\pm) \rangle$ of total three-momentum $\kk_\pm$, 
enforced by the $\delta$-functions in Eq.~\eqref{eq:gf2}. 
Therefore, for an $N$-particle intermediate state  $\scriptsize{\sum_n}$  involves phase space integrals over the $N-1$ internal momenta 
(the total momentum being fixed to  $\kk_\pm$) and  carries non-zero mass dimension. 
For example, for  two-nucleon intermediate states, using non-relativistic normalizations for the states 
$\langle \pp_n | \pp_n^\prime \rangle = (2 \pi)^3 \delta^{(3)} (\pp_n - \pp_n^\prime)$ one has 
$\sum_n \to  \int d^3 \pp_n/(2\pi)^3$,  where $\pp_n$ is the relative momentum of the two-nucleon pair.
In general the summation $\sum_n \  | n (\kk_\pm) \rangle \langle n (\kk_\pm)|$ carries mass dimension $-3$.}    
\be
{\cal A}_\nu = 
-  \sum_n 
 \int \frac{d^3 \bf{k}}{(2 \pi)^3} \, \frac{1}{|\spacevec{k}|} \, 
 \left[
\frac{\langle f_- | J^L_\mu | n (  \spacevec{k}_+) \rangle \langle n (\spacevec{k}_+) | J^{L \mu} | i_+ \rangle}{|\spacevec{k}| + (E_n (\spacevec{k}_+) - \tilde{E}) -  i \epsilon}   
\ +  \
\frac{\langle f_- | J^L_\mu | n (  \spacevec{k}_-) \rangle \langle n (\spacevec{k}_-) | J^{L \mu} | i_+ \rangle}{ |\spacevec{k}| +  (E_n (\spacevec{k}_-) -  \tilde{E}) -  i \epsilon}   
\right]\,. 
\label{eq:spectral2}
\ee 

The representations~\eqref{eq:Mnew} and~\eqref{eq:spectral2} are  quite general.   
The asymptotic behavior of the integrand in Eq.~\eqref{eq:spectral2} at large $|\spacevec{k}|$ is dictated by the OPE 
for $\hat \Pi^{LL}_{\mu \nu} (k)$ or, equivalently, $\hat O^{LL} (\spacevec{k})$.  
An  explicit calculation to be described below shows  the behavior 
$d^3\spacevec{k}  / |\spacevec{k}|^5$, 
so the amplitude in the full theory is finite. 
Moreover,   Eq.~\eqref{eq:spectral2}  shows that  
once $|\spacevec{k}| > k_F $,   so that $\spacevec{k}^2 / m_N$   is above the typical nuclear binding energies, one expects   
$(E_n (\spacevec{k}_\pm) -  \tilde{E}) > 0$  even for bound intermediate states (such as the deuteron),  and therefore the energy denominators in Eq.~\eqref{eq:spectral2} will not lead to any singular behavior in the variable $|\spacevec{k}|$. 
The matrix elements in the numerator are also expected to have a smooth behavior 
in $|\spacevec{k}|$, dictated by single- and multi-hadron form factors, 
as shown by  explicit EFT calculations.
Based on these considerations,  we conclude that  a smooth interpolation between the 
calculable regimes of  $|\spacevec{k}| \lesssim \Lambda_\chi$ and $|\spacevec{k}|  \gtrsim \Lambda$  is adequate. 

In order to make the integrand in Eqs.~\eqref{eq:Mnew} and   \eqref{eq:spectral2} more explicit,     
we use the  expression for the scattering states~\eqref{eq:scatteringstates} in Eq.~\eqref{eq:Mnew} and  arrive at 
\begin{align}
{\cal A}_\nu &=
 \int \frac{d^3 \spacevec{k}}{(2 \pi)^3} \ 
\langle f_0 | \left( \hat T (E^\prime) \hat G_+^{(0)} (E^\prime) +  I \right)   
\,  \hat{O}^{LL} (\spacevec{k})  \,
  \left( I + \hat G_+^{(0)} (E) \, \hat T (E) \right)  |i_0 \rangle 
\label{eq:Mnew2}
\\
&= 
 \int \frac{d^3 \spacevec{k}}{(2 \pi)^3} \ 
 \Bigg\{
  \langle f_0 |  \,  \hat{O}^{LL} (\spacevec{k}) \, |i_0 \rangle \notag\\
&\qquad+ \sum_m  \     
\langle f_0 | \hat T (E^\prime)   | m  \rangle 
\left[ G^{(0)}_+ (E^\prime) \right]_{m}   
\langle m |  \,  \hat{O}^{LL} (\spacevec{k}) \, | i_0  \rangle  
\nn \\
&\qquad+ \sum_m  \ 
\langle f_0 |  \,    \hat{O}^{LL} (\spacevec{k}) \, | m  \rangle  
 \left[ G^{(0)}_+ (E) \right]_{m}  \langle m | \hat T (E)   |i_0 \rangle
\nn \\
&\qquad+ \sum_{m, m^\prime} \  
 \langle f_0 | \hat T (E^\prime)   | m^\prime  \rangle      
 \left[ G^{(0)}_+ (E^\prime) \right]_{m^\prime}      
  \langle m^\prime |  \,    \hat{O}^{LL} (\spacevec{k}) \, | m  \rangle     
  \left[ G^{(0)}_+ (E) \right]_{m}   
\langle m |  \  \hat T (E) \    | i_0  \rangle  
\Bigg\}\,,
\notag
\end{align}
where 
\be
\left[ G^{(0)}_+ (E) \right]_{m}   =  \frac{1}{E - E^{(0)}_m + i \epsilon}\,,
\ee
$E^{(0)}_m$ denotes the energy associated with the free Hamiltonian $\hat H_0$,
and
\be
  \langle a  | \,    \hat{O}^{LL} (\spacevec{k}) \, | b  \rangle  = 
-  \sum_n \, \frac{1}{|\kk |} \, 
\left[
\frac{\langle a | J^L_\mu | n (  \spacevec{k}) \rangle \langle n (\spacevec{k}) | J^{L \mu} | b \rangle}{|\spacevec{k}| + (E_n (\spacevec{k}) - \tilde{E}_{ab}) -  i \epsilon}   
\ +  \
\frac{\langle a  | J^L_\mu | n ( - \spacevec{k}) \rangle \langle n (-\spacevec{k}) | J^{L \mu} | b \rangle}{ |\spacevec{k}| +  (E_n (-\spacevec{k}) -  \tilde{E}_{ab}) -  i \epsilon}   
\right]\,, 
\label{eq:spectral2v2}
\ee
with $\tilde{E}_{ab} = (E_a + E_b)/2$. 
In general the sum over hadronic intermediate states 
in Eqs.~\eqref{eq:Mnew2} and \eqref{eq:spectral2v2} 
involves  $|m\rangle$, $| m^\prime \rangle$, $|n\rangle$,    $ \in \{ N\!N,  N\!N \pi, ... \} $, 
i.e., both elastic $|N\!N\rangle$ contributions and inelastic contributions  $| m \rangle , | m^\prime \rangle,  | n \rangle  \neq |N\!N \rangle$. 
Equations~\eqref{eq:Mnew2}--\eqref{eq:spectral2v2}   make it explicit which dynamical input is needed 
for the  evaluation of the $nn \to pp$ amplitude:
\begin{enumerate}
\item One needs  the matrix elements of the  current--current operator  $\hat O^{LL} (\spacevec{k})$ among 
two-nucleon and possibly other intermediate states, namely 
$\langle m^\prime | \hat O^{LL} (\spacevec{k}) | m \rangle$, 
that can be further decomposed according to Eq.~\eqref{eq:spectral2v2}.
\item One needs the   $T$-matrix elements  $\langle m |  \hat T (E)    | i_0  \rangle$ and $ \langle f_0 | \hat T (E^\prime)   | m^\prime  \rangle$ 
involving  arbitrary intermediate states $\langle m|$, $ |m^\prime \rangle$ and on-shell two-nucleon states $ |i_0\rangle = |\pp\rangle$ (with $E =\pp^2/m_N$) and 
$\langle f_0| = \langle \pp^\prime |$ (with $E^\prime =\pp^{\prime 2}/m_N$). 
When considering the elastic contributions, these reduce to  the so-called half-off-shell (HOS)  $T$-matrix elements 
 $\langle  \pp_m |  \hat T (E)    | \pp  \rangle$ and $ \langle \pp^\prime | \hat T (E^\prime)   | \pp_{m^\prime}  \rangle$, 
 involving  loop momenta $\pp_m$ and $\pp_{m^\prime}$.  
While it is well known that the HOS $T$-matrix elements by themselves are not physical quantities 
(see for example the discussion in  Ref.~\cite{Furnstahl:2000we}),   they enter  Eq.~\eqref{eq:Mnew2}  in 
such a way that the full physical  amplitude ${\cal A}_\nu$ is free of off-shell ambiguities 
(see Appendix~\ref{sect:HOST} for an explicit check of this point). 
\end{enumerate}

To LO in  chiral EFT the amplitude $\A_\nu$ is saturated by elastic contributions, 
with all inputs in Eqs.~\eqref{eq:Mnew2}--\eqref{eq:spectral2v2} given to leading chiral order. 
The LO chiral input provides a good representation of the low-momentum part of the integrand 
but misrepresents the high-momentum component. 
In this language, the ultraviolet (UV) divergence and the need for a LO contact term arises from  
the $1/|\kk|$ behavior of the integrand, as discussed in Section~\ref{sect:chiEFT}.

On the other hand,   in our estimate of the full amplitude to be described in Section~\ref{sect:full},  
we will start from Eqs.~\eqref{eq:Mnew2}--\eqref{eq:spectral2v2} 
and use representations of  $\langle m^\prime | \hat O^{LL} (\spacevec{k}) | m \rangle$,   
$\langle  \pp_m |  \hat T (E)    | \pp  \rangle$,  and $ \langle \pp^\prime | \hat T (E^\prime)   | \pp_{m^\prime}  \rangle$ 
that go beyond leading chiral order to construct a UV convergent integrand. 
Motivated by the leading chiral EFT analysis and the analogy with the Cottingham approach 
to the pion and nucleon electromagnetic mass splitting, 
we expect the elastic two-nucleon intermediate state to provide the dominant contribution. 
While we will  mostly focus  on the elastic channel, 
we will also estimate the effect of the leading $N\!N\pi$ inelastic channel as we expect 
this to be one of the dominant sources 
of uncertainty in our final result.

\section{Chiral EFT result to leading  order}
\label{sect:chiEFT}

In this section we briefly revisit the chiral EFT result of Refs.~\cite{Cirigliano:2018hja,Cirigliano:2019vdj} 
in light of the representation given in Eq.~\eqref{eq:Mnew2}. 
This will serve two purposes: setting up the notation and pointing to a useful  way of organizing 
the integrand in the  full theory amplitude.

\begin{figure}[t]
\centering
\includegraphics[width=0.7\textwidth]{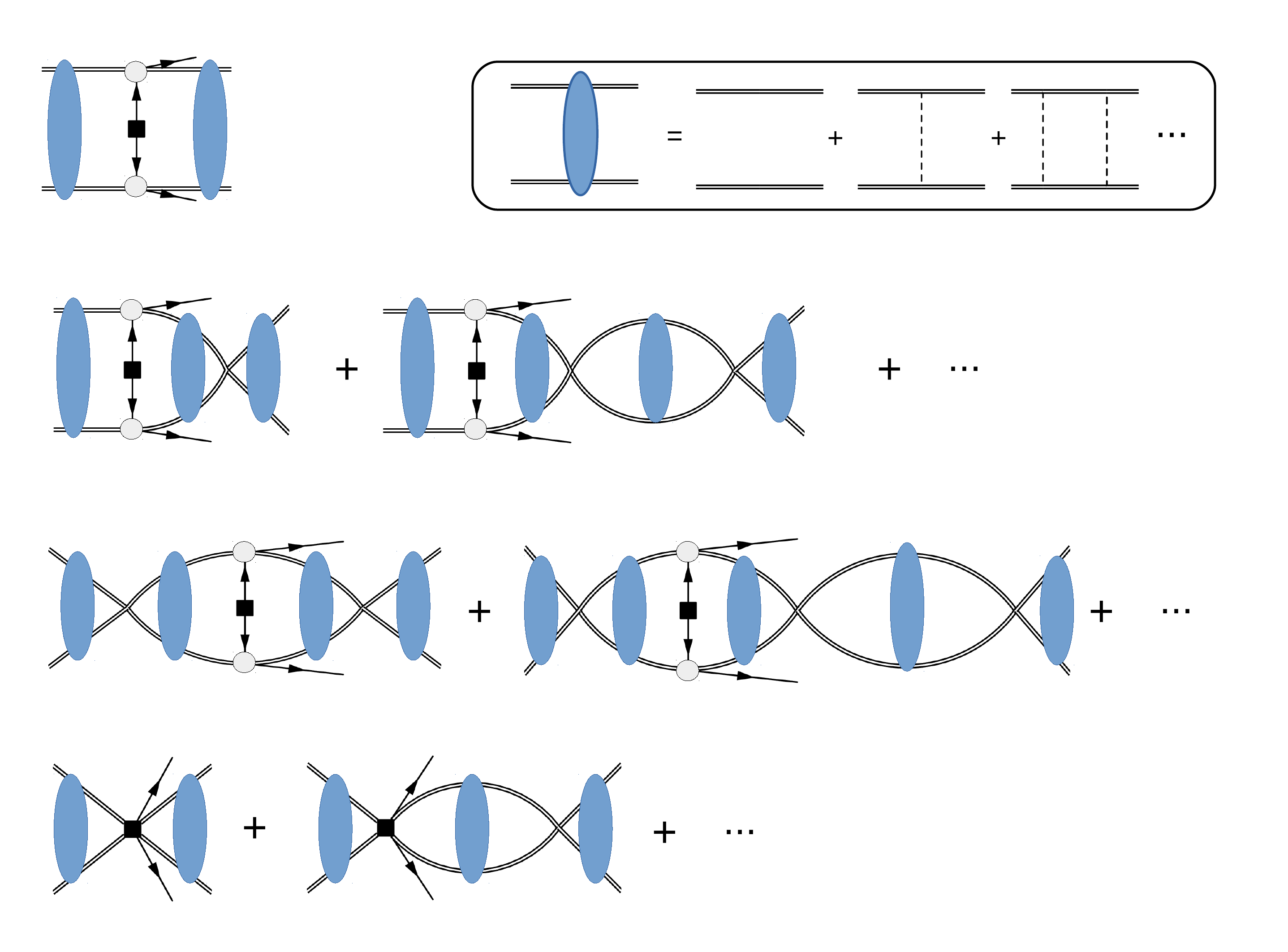}
\caption{Diagrammatic representation of LO contributions to 
$n n \rightarrow p p ee$ in chiral EFT. Double, dashed, and plain lines denote nucleons, 
pions, and leptons, respectively. 
The blue ellipse represents iteration of  $\hat V_\pi$.
Gray circles denote the nucleon axial and vector currents.  
In the first three lines the black square represents an insertion of $m_{\beta \beta}$, 
while in the fourth line it represents  an insertion of  ${\cal C}_1= g^{NN}_\nu$. 
The ellipses in the second to fourth lines denote diagrams with 
an arbitrary number of $N\!N$ bubble insertions.  
}
\label{Fig:chiEFT}
\end{figure}

The power counting for ${\cal A}_\nu$ in chiral EFT  is described in Refs.~\cite{Cirigliano:2018hja,Cirigliano:2019vdj} and we 
recall here some of its elements as needed.    Denoting by $\mu_\chi \sim Q$ the soft scale in the EFT, 
to LO in chiral counting,  i.e., $1/Q^2$,  only elastic  $N\!N$ intermediate states are relevant. 
The corresponding diagrams are reported in Fig.~\ref{Fig:chiEFT}. 
For concreteness, we regulate all the integrals dimensionally and perform 
$\overline{\rm MS}$ subtraction of the divergences  when needed.  
The LO chiral EFT results correspond to replacing in Eq.~\eqref{eq:Mnew2} the LO  form for the 
 $\hat O^{LL} (\spacevec{k})$  and $\hat T (E)$   operators, denoted by 
  $\hat O_\chi^{LL} (\spacevec{k})$  and $\hat T^\chi (E)$,   respectively, 
 and using the non-relativistic form of the free two-nucleon Green's function, 
 $[G^{(0)}_+ (E)]_{n} = 1/(E - \spacevec{p}_n^2/m_N + i \epsilon)$. 

The current--current correlator to LO  in chiral EFT is given by 
\be
_{^1 S_0}\langle \spacevec{p}^\prime | \hat O_\chi^{LL} (\spacevec{k}) | \spacevec{p} \rangle_{^1 S_0} \ = \ 
-\frac{1+ 2 g_A^2  + \frac{ g_A^2 M_\pi^4}{ (\spacevec k^2 + M_\pi^2)^2}}{\spacevec{k}^2}  
\ \frac{ (2 \pi)^3}{2}  \left[  \delta^{(3)} (\spacevec{k} + \spacevec{p} - \spacevec{p}^\prime)   + 
    \delta^{(3)} (\spacevec{k} - \spacevec{p} + \spacevec{p}^\prime)   \right]\,.
    \qquad
\label{eq:OLLchi}
\ee
The $\hat T$ operator is determined by the LO interaction Hamiltonian $\hat V = \hat V_\pi + \hat V_S$,  which 
contains  the  one-pion-exchange and a short-range contribution, parameterized to LO by the LEC $C$:
\begin{align}
\langle \spacevec{p}^\prime | \hat V | \spacevec{p} \rangle    &=    
  V_{S}  (\spacevec{p}^\prime, \spacevec{p}) 
+  V_{\pi}  (\spacevec{p}^\prime, \spacevec{p})\,, \notag
\\
  V_{S}  (\spacevec{p}^\prime, \spacevec{p})  &= C\,, \notag
  \\
 V_{\pi}   (\spacevec{p}^\prime, \spacevec{p})  & = - \frac{4 \pi \alpha_\pi}{ (\spacevec{p} - \spacevec{p}^\prime)^2 + M_\pi^2}\,,\qquad
\alpha_\pi = \frac{g_A^2 M_\pi^2}{16 \pi F_\pi^2}\,.
\end{align}
Here $F_\pi$ is the pion decay constant and $M_\pi$ denotes the pion mass. 
The LO $^1 S_0$  $N\!N$ contact coupling $C$ contains contributions from one-pion exchange as well as a contact interaction and scales as $C \sim 4 \pi/(m_N Q)$. 
The split $\hat V = \hat V_\pi + \hat V_S$ implies that the $\hat T$ operator can be similarly separated into a pion-range and short-range contribution as follows~\cite{GOL64}
\begin{align}
\hat T (E) &= \hat T_\pi (E)  + \hat T_S  (E)\,,\notag
\\
\hat T_\pi (E)  &=  \hat V_\pi  \left( I  + \hat{G}_+^{(\pi)}  (E)  \,   \hat V_\pi \right)  =   \hat V_\pi \ \left(I - \hat G^{(0)}_+ (E) \,  \hat V_\pi \right)^{-1} \,,\notag
\\
\hat T_S (E)  &= \left( I  + \hat V_\pi  \hat{G}_+^{(\pi)}  (E)  \right)   \left[   \hat V_S   
\,  \left(I - \hat G^{(\pi)}_+ (E) \,  \hat V_S \right)^{-1} 
\,  \right]
\left( I  + \hat{G}_+^{(\pi)}  (E) \,  \hat V_\pi \right)\,,
\label{eq:TS}%
\end{align}
where $\hat{G}_+^{(\pi)} (E) = 1/(E - \hat H_0 - \hat V_\pi + i \epsilon)$ is the Green's function associated with the pion-exchange interaction.

Using Eq.~\eqref{eq:Mnew2} and separating out the 
contributions with zero, one, and two insertions of $\hat T_S^\chi$,  the LO chiral EFT amplitude can be written 
as 
\be
{\cal A}_\nu^{\chi {\rm EFT}} =  {\cal M}_A^\chi  + {\cal M}_{B+ \bar B} ^\chi  +{\cal M}_C^\chi\,,
\label{eq:Mmchi}
\ee
with
\begin{align}
{\cal M}_A^\chi  &=
 \int \frac{d^3 \spacevec{k}}{(2 \pi)^3} \ 
\langle f_0 | \ \left( \hat T^\chi_\pi  (E^\prime) \hat G_+^{(0)} (E^\prime) +  I \right)   
\,  \hat{O}_\chi^{LL} (\spacevec{k})  \,
  \left( I + \hat G_+^{(0)} (E) \, \hat T^\chi_\pi (E) \right)  |i_0 \rangle 
  \,,\notag
\\
{\cal M}_{B+ \bar B} ^\chi   & = 
 \int \frac{d^3 \spacevec{k}}{(2 \pi)^3} \  
 \langle f_0 |  \left(  \hat T^\chi_S (E^\prime)  \, \hat G_+^{(0)} (E^\prime) \right)  \, \hat O_\chi^{LL} (\spacevec{k}) \, \left( 
   I + \hat{G}_+^{(0)}  (E)  \,   \hat T^\chi_\pi (E)  \right)  | i_0 \rangle
\notag\\  
  &+
 \int \frac{d^3 \spacevec{k}}{(2 \pi)^3} \  
 \langle f_0 |    \left(   I + \hat T^\chi_\pi (E^\prime)  \hat{G}_+^{(0)}  (E) \right)    
 \, \hat O_\chi^{LL} (\spacevec{k}) \, \left(  \hat G_+^{(0)} (E)     \hat T^\chi_S (E)  \right)
   | i_0 \rangle\,,  \notag
\\ 
{\cal M}_C^\chi
&=
 \int \frac{d^3 \spacevec{k}}{(2 \pi)^3} \ 
\langle f_0 | \left( \hat T^\chi_S  (E^\prime) \hat G_+^{(0)} (E^\prime) \right)   
\,  \hat{O}_\chi^{LL} (\spacevec{k})  \,
  \left(  \hat G_+^{(0)} (E) \, \hat T^\chi_S  (E) \right)  |i_0 \rangle\,.
\label{eq:ABCop}%
\end{align}
Diagrammatically,  
${\cal M}_A^\chi $, $ {\cal M}_{B+ \bar B} ^\chi$,  and ${\cal M}_C^\chi$ correspond to 
the first, second, and third row in Fig.~\ref{Fig:chiEFT}. 
Using  the definition of $\hat{T}_S$ in  Eq.~\eqref{eq:TS} and the fact that to LO $\hat V_S$ is a momentum-independent contact interaction, 
in  dimensional regularization one can show  that the 
rescattering factors $K_E, K_{E^\prime}$  
\be
K_{E} =\frac{
C}{1 - 
C G^+_{E}(\spacevec 0, \spacevec 0)} \,,  \qquad  
G^\pm_{E}(\spacevec r, \spacevec r^\prime)  = \int \frac{d^3 \spacevec q }{(2\pi)^3} 
\int \frac{d^3 \spacevec q^\prime}{(2\pi)^3}
\, e^{i \spacevec q \cdot r} e^{-i \spacevec q^\prime \cdot \spacevec r^\prime}
\langle \spacevec q | \hat{G}_\pm^{(\pi)} (E) | \spacevec q^\prime \rangle\,,
\ee
and Yukawa wave functions 
at the origin  $\chi_\spacevec{p}^+ (\spacevec 0)$,    $\chi_{\spacevec{p}^\prime}^+ (\spacevec 0)$  
\be
\chi^\pm_{\spacevec p}(\spacevec r) = \int \frac{d^3 \spacevec q}{(2\pi)^3} 
\,
 e^{i \spacevec k \cdot \spacevec r}
 \langle \spacevec k | (1 +  \hat{G}^{(\pi)}_\pm (E ) \,  \hat V_\pi) | 
\spacevec p \rangle \,, 
\\
\ee
can be factored out of the  $d^3 \spacevec k$ integrals in   $ {\cal M}_{B+ \bar B} ^\chi$,  and ${\cal M}_C^\chi$, 
 thus reproducing the chiral EFT results of 
Refs.~\cite{Cirigliano:2018hja,Cirigliano:2019vdj}
\begin{align}
{\cal M}_A^\chi &= {\cal A}_A\,,\notag
\\
{\cal M}_{B+ \bar B} ^\chi  &=  
\mathcal{\bar A}_B \, K_E \, \chi^+_{\spacevec p}(\spacevec 0) + 
\chi^+_{\spacevec p^\prime}(\spacevec 0) \, K_{E^\prime} \, \mathcal A_B\,,\notag
\\
{\cal M}_C^\chi &= \chi^+_{\spacevec p^\prime}(\spacevec 0) \, K_{E^\prime}  \, \mathcal A_C  \,  K_E \, \chi^+_{\spacevec p}(\spacevec 0)\,. 
\label{eq:mcc}
\end{align}
The divergence in $G^+_{E}(\spacevec 0, \spacevec 0)$ is absorbed by  
$C^{-1}$, so that $K_E$ 
is well defined and independent of the chosen scheme and scale~\cite{Kaplan:1996xu}.

The reduced amplitudes ${\cal A}_{A,B,C}$  correspond to the left-most  
diagram in the first, second, and third rows of  Fig.~\ref{Fig:chiEFT}
respectively (without the Yukawa iteration in the outer legs after the $\hat V_S$ insertion). 
Neglecting Yukawa interactions ($\alpha_\pi \to 0$),  i.e., in the pionless EFT ($\slashpi$EFT),  ${\cal A}_{A,B,C}$ have simple 
expressions~\cite{Cirigliano:2018hja,Cirigliano:2019vdj}. 
In general, however, they are non-perturbative objects that involve the sum of infinitely many Feynman diagrams. 
The amplitude  ${\cal A}_C$ in Eqs.~\eqref{eq:ABCop} and \eqref{eq:mcc} contains a UV-divergent term at the two-loop level, 
which we denote by ${\cal A}_C^{\rm sing}$, 
as well as UV-finite terms induced by pion exchange, which we denote by $\delta {\cal A}_C$, 
leading to the decomposition:
\be
{\cal A}_C = {\cal A}_C^{\rm sing} + \delta {\cal A}_C\,.
\label{eq:ACdec}
\ee
The UV-convergent term $\delta {\cal A}_C$ arises from 
(i)  the iteration of the pion-induced potential 
(see Fig.~\ref{Fig:chiEFT} and Eq.~\eqref{eq:ABCop}, as well as Eq.~\eqref{eq:TS} for the definition of $\hat T_S^\chi$);
(ii) 
the term proportional to $g_A^2 M_\pi^4$    in  $\hat O^{LL}_\chi (\spacevec{k})$, see Eq.~\eqref{eq:OLLchi}, 
which is one of two manifestations of the  induced  pseudoscalar form factor  of the axial current 
(the other is the change $3 g_A^2 \to 2 g_A^2$).  
Using dimensional regularization with scale  $\mu_\chi$ and  the $\overline{\rm MS}$ scheme for renormalization, 
we thus identify the singular (UV-divergent) term with 
\be
{\cal A}_C^{\rm sing}  (\mu_\chi)  =
-   m_N^2   \  \mu_\chi^{4-d} \, \int \frac{d^{d-1}k}{(2 \pi)^{d-1}} \,  \frac{1 + 2 g_A^2} {{\bf k}^2  - i \epsilon} \ {\cal I}_C ({\bf k}^2,  {\bf p}^2,  {\bf p}^{\prime 2}) \,, 
\label{eq:ACsing0}
\ee
where 
\begin{align}
{\cal I}_C ({\bf k}^2,  {\bf p}^2,  {\bf p}^{\prime 2}) 
&= \int \frac{d^3 \spacevec{q}}{(2 \pi)^3} \, 
\frac{1}{ {\bf p}^{\prime 2} - ( {\bf q}  + {\bf k})^2 + i \epsilon} 
 \  
\frac{1}{ {\bf p}^{2} - {\bf q}^2 + i \epsilon}\,,
\notag
\\
{\cal I}_C ({\bf k}^2,  {\bf p}^2,  {\bf p}^2) 
&=  \frac{1}{8 |{\bf k}|} \,  \theta(|{\bf k}| - 2 |{\bf p}|) + \frac{i}{8 \pi |{\bf k}|} \, \log \left|  \frac{1+ 2 \frac{|{\bf p}|}{{|{\bf k}|}} }{ 1-  2 \frac{|{\bf p}|}{{|{\bf k}|}}}   \right|\,.
\label{eq:IC}
\end{align}
In case of $\bf p = \bf p'$ the resulting expression becomes
\be
{\cal A}_C^{\rm sing}  (\mu_\chi) 
=
- \frac{(1+2 g_A^2)}{2}   \,  \frac{m_N^2}{(4 \pi)^2}   
\left[
\log \frac{\mu_\chi^2}{- 4 |{\bf p}|^2 - i \epsilon}    
+ 1
\right]\,.
\ee
It will prove useful to write the real part of this result in an integral representation
\be\label{intrep}
\mathrm{Re}\,{\cal A}_C^{\rm sing}  (\mu_\chi) 
=
- \frac{(1+2 g_A^2)}{2}   \,  \frac{m_N^2}{(4 \pi)^2}   
\left[1 + 2 \int_0^{\mu_\chi} \frac{d {|\bf k}|}{|{\bf k}|} \theta(|{\bf k}| - 2 |{\bf p}|) 
\right]\,.
\ee

The UV divergence is removed by introducing a new contact term 
\be
\mathcal L_{|\Delta L| =2}^{NN}  = \left( 2 \,  G_F^2 V_{ud}^2  \right)  \ m_{\beta \beta}  \, {\cal C}_1 \times
  \bar{e}_L e_L^c    \,   \bar N \mathcal \tau^+ N \, \bar N \tau^+ N\,,  
  \label{eq:C1v1}
\ee
with a LNV coupling proportional to $m_{\beta \beta}$   denoted by ${\cal C}_1$~\cite{Cirigliano:2019vdj}, 
to be identified with  $g_\nu^{NN}$ also used in the recent literature, i.e.,  ${\cal C}_1 \equiv g_\nu^{NN}$. 
A chiral-covariant form of the  contact operator in Eq.~\eqref{eq:C1v1}  will be given in Section~\ref{sect:VV}.  

Including the contact term, the  finite, renormalized amplitude is given by 
\begin{align}
{\cal A}_\nu^{\chi {\rm EFT}}
&= \mathcal A_A  + 
\chi^+_{\spacevec p^\prime}(\spacevec 0) \, K_{E^\prime} \, \mathcal A_B         
+ \mathcal{\bar A}_B \, K_E \, \chi^+_{\spacevec p}(\spacevec 0) 
\nonumber \\
&+
\chi^+_{\spacevec p^\prime}(\spacevec 0) \, K_{E^\prime}  \, \left(  \mathcal A_C^{\rm sing} (\mu_\chi)   +
  \frac{2 \,  {\cal C}_1 (\mu_\chi)  }{C^2}    + \delta {\cal A}_C \right) 
\,  K_E \, \chi^+_{\spacevec p}(\spacevec 0)\,.
\label{eq:chiEFT2}
\end{align}
The matching analysis will provide a representation for  the combination   ${\cal A}_C^{\rm sing} (\mu_\chi) + 2 \, {\cal C}_1 (\mu_\chi)/C^2$.

\section{Full theory parameterization}
\label{sect:full}

The starting point for parameterizing the full-theory amplitude is provided by Eq.~\eqref{eq:Mnew2},  in which we split the integral into a low- plus intermediate-momentum region 
and  high-momentum region, according to Eq.~\eqref{eq:Msplit}. 
In order to mimic the structure of the chiral EFT amplitude and facilitate matching,  it is also convenient to separate the 
full theory $T$ matrix into a term induced by one-pion exchange and a short-range contribution, 
i.e., $\hat T = \hat T_\pi + \hat T_S$, as done in Eq.~\eqref{eq:TS}.

\subsection[Low- and intermediate-momentum region: ${\cal A}^<$]{Low- and intermediate-momentum region: $\boldsymbol{{\cal A}^<}$}
\label{sect:MLLm}

\begin{figure}[t]
\centering
\includegraphics[width=0.5\textwidth]{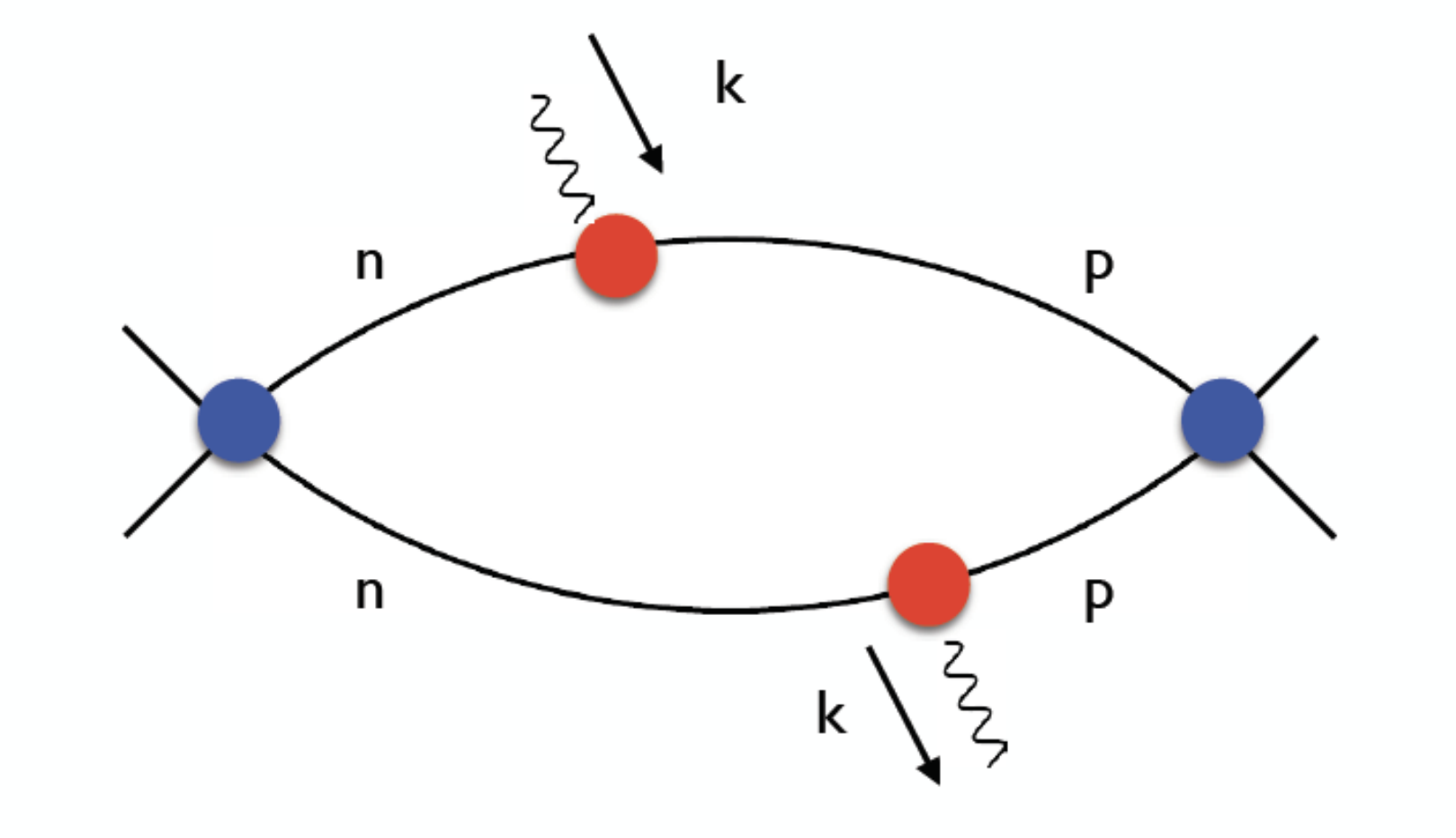}
\caption{
Diagrammatic representation of the forward Compton amplitude corresponding to the 
integrand of ${\cal A}_C^{\rm sing}$ in Eq.~\eqref{eq:ACsing0} and ${\cal A}_C^{<,\rm sing}$  
in Eq.~\eqref{eq:ACmsing}. 
Solid lines represent nucleons and  wavy lines the  currents.  
Blue circles represent short-range $N\!N$ interactions and red circles represent single-nucleon weak form factors. 
}
\label{Fig:NN}
\end{figure}

Following Eqs.~\eqref{eq:Mnew2} and  \eqref{eq:Mmchi},  we write 
\be
{\cal A}^< =  {\cal M}_A^<  + {\cal M}_{B+ \bar B} ^<  +{\cal M}_C^<\,, 
\label{eq:Mm}
\ee
with 
\begin{align}
{\cal M}^<_A  &=
 \int \frac{d^3 \spacevec{k}}{(2 \pi)^3} \ 
\langle f_0 | \ \left( \hat T^<_\pi  (E^\prime) \hat G_+^{(0)} (E^\prime) +  I \right)   
\,  \hat{O}_<^{LL} (\spacevec{k})  \,
  \left( I + \hat G_+^{(0)} (E) \, \hat T^<_\pi (E) \right)  |i_0 \rangle\,,\notag
\\
{\cal M}_{B + \bar B}^< 
  & = 
 \int \frac{d^3 \spacevec{k}}{(2 \pi)^3} \  
 \langle f_0 |  \left(  \hat T^<_S (E^\prime)  \, \hat G_+^{(0)} (E^\prime) \right)  \, \hat O_<^{LL} (\spacevec{k}) \, \left( 
   I + \hat{G}_+^{(0)}  (E)  \,   \hat T^<_\pi (E)  \right)  | i_0 \rangle
 \nonumber \\
   &+
 \int \frac{d^3 \spacevec{k}}{(2 \pi)^3} \  
 \langle f_0 |    \left(   I + \hat T^<_\pi (E^\prime)  \hat{G}_+^{(0)}  (E) \right)    
 \, \hat O_<^{LL} (\spacevec{k}) \, \left(  \hat G_+^{(0)} (E)     \hat T^<_S (E)  \right)
   | i_0 \rangle\,,\notag   
\\
{\cal M}_C^< 
&=
 \int \frac{d^3 \spacevec{k}}{(2 \pi)^3} \ 
\langle f_0 | \left( \hat T^<_S  (E^\prime) \hat G_+^{(0)} (E^\prime) \right)   
\,  \hat{O}_<^{LL} (\spacevec{k})  \,
  \left(  \hat G_+^{(0)} (E) \, \hat T^<_S  (E) \right)  |i_0 \rangle\,.
\label{eq:ABCfull}%
\end{align} 
We wish to provide a representation of the integrand 
in ${\cal M}^<_{A,B,C}$ that is valid up to  a scale $|\spacevec{k}| \sim \Lambda$. 
We build the integrand starting from the low-momentum region. 
For $|\spacevec{k}| \sim Q$ the integrand is controlled by  chiral EFT.
Using the LO chiral representation for 
the two-current operator  ($\hat O_<^{LL} =\hat O_\chi^{LL}$) and the $T$ matrix
($\hat T^<_{\pi, S} = \hat T^\chi_{\pi, S}$)  
in ${\cal M}_A$ and ${\cal M}_{B + \bar B}$  leads to convergent  integrals, 
and one recovers the chiral EFT results of Eqs.~\eqref{eq:ABCop}, up to terms of $\Order(Q/\Lambda)$ 
that are irrelevant at LO
\begin{align}
{\cal M}_A^<&=  {\cal M}_A^\chi   \times \left(1 + \Order \left (\frac{Q}{\Lambda} \right) \right)\,,\notag
\\
{\cal M}_{B + \bar B}^< &=  
{\cal M}_{B + \bar B}^\chi \times 
\left(1 + \Order \left (\frac{Q}{\Lambda} \right) \right)\,.
\end{align}
We will also show that ${\cal M}^>_{A,B + \bar B}$ does not  contribute to the amplitude at LO. 
Therefore,  ${\cal M}_A$ and ${\cal M}_{B + \bar B}$  drop out of the LO matching  
condition~\eqref{eq:match0}, which will therefore only 
involve ${\cal M}_C^<$. 
In order to have a representation of the integrand valid up to  momenta  $|\spacevec{k}| \sim \Lambda$, 
in Eq.~\eqref{eq:ABCfull} we  need to model both $\hat O^{LL} (\spacevec{k})$ and $\hat T_S$ beyond the LO chiral EFT expressions. 
We will do so by introducing appropriate single-nucleon weak form factors and HOS form factors, represented 
by the red and blue circles in Fig.~\ref{Fig:NN}, respectively.

For the  current--current operator we make the replacement $\hat O^{LL}_\chi (\spacevec{k}) \to \hat O^{LL}_< (\spacevec{k})$, 
by introducing  single-nucleon form factors in the weak current vertices, 
\be
_{^1 S_0}\langle \spacevec{p}^\prime | \hat O_<^{LL} (\spacevec{k}) | \spacevec{p} \rangle_{^1 S_0} \ = \ 
-\frac{g_V^2  (\spacevec{k}^2) + 3 h_{GT}^{LL} (\spacevec{k}^2)}{\spacevec{k}^2} \, 
\frac{(2 \pi)^3}{2}  \left[  \delta^{(3)} (\spacevec{k} + \spacevec{p} - \spacevec{p}^\prime)   + 
    \delta^{(3)} (\spacevec{k} - \spacevec{p} + \spacevec{p}^\prime)   \right]\,,
\ee
where 
\be
h^{LL}_{GT} (\spacevec{k}^2)  =  g_A^2 (\spacevec{k}^2)  + \frac{\spacevec{k}^4 g_P^2 (\spacevec{k}^2)}{12 m_N^2}  
  + \frac{\spacevec{k}^2 g_A(\spacevec{k}^2) g_P(\spacevec{k}^2)}{3 m_N} 
+  \frac{\spacevec{k}^2 g_M^2 (\spacevec{k}^2)}{6 m_N^2}\,.
\ee
In our baseline analysis, we will use the simple dipole parameterization 
for the vector, magnetic, and axial-vector form factors 
\begin{align}
g_{V} (\spacevec{k}^2) &=  \frac{1}{\left(1 + \frac{\spacevec{k}^2}{\Lambda_{V}^2}\right)^2}\,,\qquad 
g_M (\spacevec{k}^2) =  (1  + \kappa_1) \, g_V (\spacevec{k}^2)\,, \qquad \kappa_1 \simeq 3.7\,,\notag
\\
g_{A} (\spacevec{k}^2) &=  \frac{g_A}{\left(1 + \frac{\spacevec{k}^2}{\Lambda_{A}^2}\right)^2}\,,\qquad
g_{P} (\spacevec{k}^2)   =  - \frac{ 2  m_N \, g_A (\spacevec{k}^2)}{\spacevec{k}^2 + M_\pi^2}\,, 
\end{align}
with $\Lambda_V = 0.84\GeV$ and $\Lambda_A=1.0\GeV$. 
While we expect this to be a good first approximation, 
we have explored the dependence of the contact term 
on the form factor input.  
In the case of the axial-vector form factor, we have taken the   range  
$\Lambda_A\in [0.8, 1.2]\GeV$, which corresponds to an axial radius of  
$0.47(19)\fm^2$,  in good agreement with the result $0.46(16)\fm^2$ quoted in Ref.~\cite{Hill:2017wgb}.
For the vector form factor, we have explored the continued fraction expansion of Ref.~\cite{Arrington:2006hm}.  
Since input on the form factors turns out to induce  a by far subdominant uncertainty in the matching result (see discussion below), 
we will not perform a more sophisticated error analysis  based on state-of-the-art information on the nucleon form 
factors~\cite{Lorenz:2014yda,Hoferichter:2016duk,Ye:2017gyb,Hill:2017wgb}.

The  parameterization of  $\hat O^{LL}_< (\spacevec{k})$ 
in terms of the  weak   nucleon form factors then leads to 
\begin{align}
{\cal M}_C^< & = - \int \frac{d^3 \spacevec{k}}{(2 \pi)^3}  \
\left[\frac{g_V^2  (\spacevec{k}^2) + 3 h_{GT}^{LL} (\spacevec{k}^2)}{\spacevec{k}^2} \,  \right] \notag
\\
&\times 
 \int \frac{d^3 \spacevec{q}}{(2 \pi)^3}  \ 
T_S (\spacevec{p}^\prime, \spacevec{q} + \spacevec{k}) \  \frac{1}{E^\prime - (\spacevec{q} + \spacevec{k})^2/m_N + i \epsilon}   
\,   \frac{1}{E - \spacevec{q}^2/m_N + i \epsilon}   \ 
T_S (\spacevec{q}, \spacevec{p} )\,,
\label{eq:MCm}
\end{align}
where the notation for the HOS $\hat T_S$-matrix  
elements  is\footnote{These are  $S$-wave projected matrix elements 
and  depend only on $|\spacevec{p}|$ and $|\spacevec{q}|$.}
\begin{align}
T_S (\spacevec{q}, \spacevec{p} ) &=  \langle \spacevec{q} | \hat T_S (E) | \spacevec{p} \rangle\,,
& 
E&= \frac{\spacevec{p}^2}{m_N}\,,\notag
\\
T_S (\spacevec{p}^\prime, \spacevec{q} ) &=  \langle \spacevec{p}^\prime | \hat T_S (E^\prime) | \spacevec{q} \rangle\,, &
E^\prime &= \frac{\spacevec{p}^{\prime 2}}{m_N}\,.
\label{eq:host1}
\end{align}

We further write the HOS $\hat T_S$-matrix  element as the product of the on-shell $\hat T_S$-matrix  element times  the 
HOS  factors $\tilde f_S(\spacevec q, \spacevec p)$~\cite{Srivastava:1975eg}  
\be
T_S (\spacevec{q}, \spacevec{p} ) \equiv T_S (\spacevec{p}, \spacevec{p} ) \times \tilde f_S (\spacevec{q}, \spacevec{p} )\,. 
\label{eq:fS}
\ee
Using  the LO on-shell result~\cite{Kaplan:1996xu}, see Eq.~\eqref{eq:TS},
\be
T_S (\spacevec{p}, \spacevec{p} )  = \left( \chi^+_\spacevec{p} (\spacevec{0}) \right)^2 \ K_E
\ee
and defining 
\be
\bar f_S (\spacevec{q}, \spacevec{p} ) \equiv  \tilde  f_S (\spacevec{q}, \spacevec{p} )  \  \chi^+_\spacevec{p} (\spacevec{0}) 
= 
 \frac{T_S(\spacevec{q}, \spacevec{p})} {T_S (\spacevec{p}, \spacevec{p})} \  \chi^+_\spacevec{p} (\spacevec{0}) \,,  
\label{eq:fsbar}
\ee
we arrive at  
\begin{align}
{\cal M}_C^<  &= 
\chi^+_{\spacevec p^\prime}(\spacevec 0) \, K_{E^\prime}  \,  \Big[ \,  {\cal A}^<_C  \,  \Big]  \,  K_E \, \chi^+_{\spacevec p}(\spacevec 0)\,, 
\label{eq:MCm2}
\\
{\cal A}_C^< &=
-  \int \frac{d^3 \spacevec{k}}{(2 \pi)^3}  \
\left[\frac{g_V^2  (\spacevec{k}^2) + 3 h_{GT}^{LL} (\spacevec{k}^2)}{\spacevec{k}^2} \,  \right] \, 
\times \,  \tilde {\cal I}_C^< ( |\spacevec{k}|) \,, 
\label{eq:ACm}
\\
\tilde  {\cal I}_C^< ( |\spacevec{k}|)  
&=
 \int \frac{d^3 \spacevec{q}}{(2 \pi)^3}  \ 
\bar f_S (\spacevec{p}^\prime, \spacevec{q} + \spacevec{k}) \  \frac{1}{E^\prime - (\spacevec{q} + \spacevec{k})^2/m_N + i \epsilon}   
\,   \frac{1}{E - \spacevec{q}^2/m_N + i \epsilon}   \ 
\bar f_S (\spacevec{q}, \spacevec{p} )\,.
\label{eq:ICtilde}
\end{align}
Equation~\eqref{eq:MCm2} shows that ${\cal M}_C^<$ has the correct IR behavior to LO in the external momenta $\spacevec{p}$ 
and $\spacevec{p}^\prime$ and that 
the quantity ${\cal A}_C^<$  corresponds to the amplitude ${\cal A}_C$ in  chiral EFT. 
As argued in Section~\ref{sect:chiEFT},  ${\cal A}_C$ contains a singular UV-divergent term (${\cal A}_C^{\rm sing}$)
and  UV-finite contributions induced by pion exchange ($\delta {\cal A}_C$).   
In analogy to Eq.~\eqref{eq:ACdec} we can write 
\be
{\cal A}_C^< = {\cal A}_C^{<, \rm sing} + \delta {\cal A}_C^<\,, 
\label{eq:ACmdec}
\ee
where  $\delta {\cal A}_C^<$  denotes the convergent pion-exchange contributions to ${\cal A}_C^<$, satisfying 
\be
\delta  {\cal A}^<_C  - \delta  {\cal A}_C  
 \sim  {\cal A}_C \times    \Order \left( \frac{Q}{\Lambda_\chi} \right)\,.
\ee
Therefore,  these finite contributions  drop out of the matching relation~\eqref{eq:match0}  to LO and 
for the purpose of matching one only needs to identify the  singular  component  ${\cal A}_C^<$. 

${\cal A}_C^{<, \rm sing}$  is obtained from  Eq.~\eqref{eq:ACm}   
by systematically removing the convergent pion contributions. 
This requires:
\begin{enumerate}
\item  Discarding the  convergent terms due to the induced pseudoscalar form factor in $h_{GT}^{LL} (\spacevec{k}^2)$.
In practice this means replacing 
$ 3 h_{GT}^{LL} (\spacevec{k}^2) \to   2 g_A( \spacevec{k}^2 ) +   \frac{\spacevec{k}^2 g_M^2 (\spacevec{k}^2)}{2 m_N^2}$.
\item 
Setting $\alpha_\pi \to 0$ in the evaluation of $\bar{f}_S (\spacevec{q}, \spacevec{p})$ in Eq.~\eqref{eq:ICtilde}, 
which implies 
\be
\bar f_S (\spacevec{q}, \spacevec{p} )   \bigg \vert_{\alpha_\pi =0} = 
 \frac{T_S(\spacevec{q}, \spacevec{p})} {T_S (\spacevec{p}, \spacevec{p})}    \bigg \vert_{\alpha_\pi =0} 
 \equiv 
 f_S (\spacevec{q}, \spacevec{p} )\,.
\label{eq:fsbarnopi} 
\ee
 The quantity $f_S (\spacevec{q}, \spacevec{p} )$ is real-valued and can be interpreted as a form factor. 
In  light of this,  from Eqs.~\eqref{eq:MCm2}--\eqref{eq:ICtilde}  we arrive at 
\begin{align}
{\cal A}_C^{<, {\rm sing}} &=
- m_N^2 \,   \int \frac{d^3 \spacevec{k}}{(2 \pi)^3}  \
\left[\frac{g_V^2  (\spacevec{k}^2) + 2 g^2_A( \spacevec{k}^2 ) +   (\spacevec{k}^2 g_M^2 (\spacevec{k}^2))/(2 m_N^2)   
  }{\spacevec{k}^2}
 \right] \times {\cal I}_C^< ( |\spacevec{k}|)  \,,
\notag \\
{\cal I}_C^<  (|\spacevec{k}|)   
&= 
 \int \frac{d^3 \spacevec{q}}{(2 \pi)^3}  \ 
 f_S   (\spacevec{p}^\prime, \spacevec{q} + \spacevec{k}) \ 
  \frac{1}{ {\bf p}^{\prime 2} - ( {\bf q}  + {\bf k})^2 + i \epsilon} 
 \  
\frac{1}{ {\bf p}^{2} - {\bf q}^2 + i \epsilon} ~
 \ 
 f_S  (\spacevec{q}, \spacevec{p} )\,.
\label{eq:ACmsing}
\end{align}
\end{enumerate}

In the end, the difference between 
${\cal A}_C^{\rm sing} (\mu_\chi)$ and ${\cal A}_C^{<,{\rm sing}}$  
is quite  intuitive. 
As illustrated in Fig.~\ref{Fig:NN}, 
the integrand of ${\cal A}_C^{\rm sing}$  in Eq.~\eqref{eq:ACsing0} 
contains the LO nucleon weak current vertices 
and the leading contact $N\!N$ interaction.  
On the other hand,   the full-theory  integrand in ${\cal A}_C^{<,{\rm sing}}$  contains appropriate form factors in the 
nucleon  weak current vertices and in the $N\!N$ vertices 
($f_S  (\spacevec{q}, \spacevec{p})$). 
The latter parameterizes  the 
 off-shell  behavior of the short-range component of the $N\!N$ amplitude 
and  amounts to   changing  the  LO chiral EFT  function ${\cal I}_C (|\spacevec{k}|)$
 defined in Eq.~\eqref{eq:IC} to   ${\cal I}_C^< (|\spacevec{k}|)$  given in Eq.~\eqref{eq:ACmsing}, 
valid for a larger range of momenta $|\spacevec{k}|$ compared to the LO result.  
It will prove convenient to present results in terms of the ratio 
\be
r( |\spacevec{k}|)
 \equiv   \frac{{\rm Re} \, {\cal I}_C^< (| \spacevec{k}|)}{  {\rm Re} \, {\cal I}_C (| \spacevec{k}|) }\,. 
\label{eq:r0}
\ee

\begin{figure}[!t]
\centering
\includegraphics[width=0.5\textwidth]{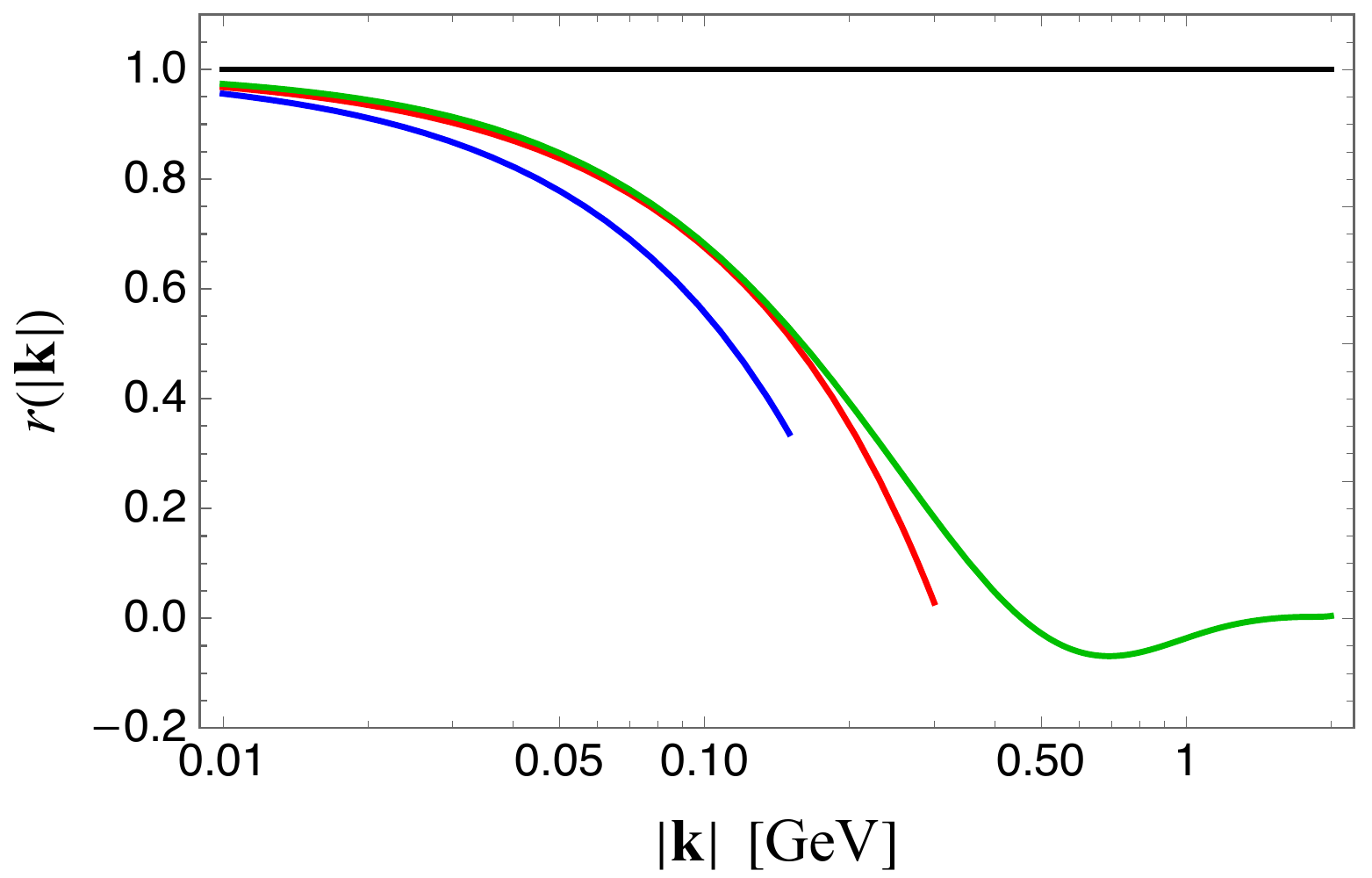}
\caption{
Ratio $r( |\spacevec{k}|)$  defined in Eq.~\eqref{eq:r0}. 
The various lines correspond to  the
NLO  $\slashpi$EFT result (blue), the NLO  chiral EFT result (red), 
and the three-Yukawa potential from Ref.~\cite{Kaplan:1999qa} (green), compared to the LO chiral EFT  result  $r( |\spacevec{k}|)=1$.
}
\label{Fig:r}
\end{figure}

We now describe our construction of ${\cal I}_C^< (|\spacevec{k}|)$ and $r (|\spacevec{k}|)$, based on 
the first two orders in chiral EFT and then extended to larger values of $|\spacevec{k}|$ using 
models  for the  short-range $N\!N$ interaction in the $^1S_0$ channel. For concreteness we work at the kinematic point $|\spacevec{p}|=|\spacevec{p}^\prime|=1\MeV$. 
\begin{enumerate}
\item At LO in chiral EFT one has $f_S  (\spacevec{q}, \spacevec{p}) = 1$ and the integrals simplify to \\
${\rm Re} \, {\cal I}_C^< (|\spacevec{k}|)  \to {\rm Re} \, {\cal I}_C (|\spacevec{k}|) =   \theta(|\spacevec{k}|-2|\spacevec{p}|)/(8 |\spacevec{k}|)$.

\item For $|\spacevec{k}| \leq M_\pi$ and 
$|\spacevec{k}| \leq \Lambda_\chi$  
we can evaluate the corrections to 
${\rm Re}  \ {\cal I}_C^< (| \spacevec{k}|)$ 
in $\slashpi$EFT  and chiral EFT, respectively, 
using $V_S (\spacevec{p}, \spacevec{p}^\prime)  = C + (C_2/2) (\spacevec{p}^2 +\spacevec{p}^{\prime 2})$. 
As discussed in Appendix~\ref{sect:HOST}, 
the results at next-to-leading order (NLO) in chiral EFT and $\slashpi$EFT 
are formally identical 
\begin{align}\label{eq:HOSfs}
f_S  (\spacevec{q}, \spacevec{p}) 
&= 1 -  \frac{C_2}{2 C}  (\spacevec{p}^2 - \spacevec{q}^2)\,, \notag\\
f_S  ( \spacevec{p}^\prime, \spacevec{q}+\spacevec k) 
&= 1 -  \frac{C_2}{2 C}  \left[\spacevec{p}^{\prime\,2} - (\spacevec{q}+\spacevec{k})^2\right]\,, 
\end{align}
with the identifications $C = C_0$  in $\slashpi$EFT and
 $C = C_0 + g_A^2/(4 F_\pi^2)$  in  chiral EFT.  
 Note that these HOS form factors are different depending on whether they involve the initial- or final-state on-shell momenta, so that $f_S  ( \spacevec{p}^\prime, \spacevec{q}) $ cannot be obtained from the expression of $f_S  (\spacevec{q}, \spacevec{p}) $ by interchanging $\spacevec{q}\leftrightarrow\spacevec{p}$ and subsequently replacing $\spacevec{p}\to\spacevec{p'}$.
 This leads to  (see Appendix~\ref{sect:HOST} for details)
\be
{\rm Re}  \ {\cal I}_C^< (| \spacevec{k}|) 
= \frac{1}{8 |\spacevec{k}|}   - \frac{C_2}{m_N C^2}\,.
\ee
In terms of the ratio  $r( |\spacevec{k}|)$
 introduced in Eq.~\eqref{eq:r0} 
the NLO analysis gives 
\be
r^{\rm NLO}( |\spacevec{k}|) = 1 - \frac{8 C_2}{m_N C^2}  |\spacevec{k}|\,.
\ee
In both EFTs the ratio $C_2/C^2$ 
is renormalization-scale independent. 
In $\slashpi$EFT $C_2/C^2 = m_N r_0/(8 \pi)$ is linked to the effective range, 
leading to 
\be
r^{\rm NLO}_\slashpi ( |\spacevec{k}|) = 1 - \frac{r_0}{\pi}  |\spacevec{k}|\,.
\ee
Since $r_0 \simeq 1/(72\MeV)$ (corresponding to $C_2/C^2 \simeq 0.52$), 
this produces sizable fractional deviations in ${\rm Re}  \ {\cal I}_C^< (| \spacevec{k}|)$ 
already for  $|\spacevec{k}| \leq M_\pi$. 
In chiral EFT one finds  a slightly reduced suppression $C_2/C^2 \simeq 0.38$~\cite{Kaplan:1996xu}, as part of the effective range 
is already captured by the pion-exchange contribution. 
In summary,  the NLO analysis shows that there is a sizable suppression of the LO result already for relatively low $|\spacevec{k}|$, 
linked to the large effective range in the $^1S_0$ channel.  

\item  To  extend ${\cal I}_C^< (| \spacevec{k}|) $ 
or equivalently $r( |\spacevec{k}|)$ to higher values of $|\spacevec{k}|$, 
we have computed  $f_S  (\spacevec{q}, \spacevec{p})$  using 
models of $V_S$, i.e., the  short-range $N\!N$ interaction in the $^1 S_0$ channel. 
We  have used the  three-Yukawa potential from Ref.~\cite{Kaplan:1999qa,Reid:1968sq}  
and the AV18  potential~\cite{Wiringa:1994wb},  all of which reproduce the $^1S_0$ phase shifts 
for $|\spacevec{p}|$ up to several hundred MeV. 
Using the Kaplan--Steel three-Yukawa  potential~\cite{Kaplan:1999qa}   as our baseline model, 
we  have  solved numerically  the Lippmann--Schwinger (LS) equation to obtain   
$f_S  (\spacevec{q}, \spacevec{p})$, which was then used to evaluate ${\cal I}_C^< (| \spacevec{k}|)$
(see   Appendix~\ref{sect:HOST} for details).
Reassuringly,  as illustrated in Fig.~\ref{Fig:r},  the behavior of  $r_{V_S}( |\spacevec{k}|)$  in the model calculation very closely tracks the 
NLO chiral EFT result  for  $|\spacevec{k}| < 200\MeV$, i.e., the region in which NLO  chiral EFT is expected to provide
 model-independent and accurate results. 
Using other $N\!N$ potentials~\cite{Reid:1968sq,Wiringa:1994wb}  does not change the qualitative picture, 
but induces small changes in the slope of $r (|\spacevec{k}|)$ for $|\spacevec{k}| < 400\MeV$ 
as well as the location and depth of the minimum for  $|\spacevec{k}| \sim 600\MeV$.  
This makes it clear that using   $r_{V_S}  (| \spacevec{k}|)$  
all the way to $|\spacevec{k}| \sim \Lambda \sim 1\GeV$  introduces model dependence. 
However,  in the integral~\eqref{eq:ACmsing} the 
region in which $r_{V_S}  (| \spacevec{k}|)$  has the largest model dependence 
is weighted considerably  less than the model-independent low-$|\spacevec{k}|$ region, 
because the integration kernel involves   $1/|\spacevec{k}|$ and the nucleon form factors, both  rapidly decreasing functions  of 
$| \spacevec{k}|$.  
In Section~\ref{sect:matchLL} we will quantify how this model dependence affects the extraction of the effective coupling.

\end{enumerate}

\subsection[High-momentum region: ${\cal A}^>$]{High-momentum region: $\boldsymbol{{\cal A}^>}$}
\label{sect:OPELL}

\begin{figure}[t]
\centering
\includegraphics[width=0.7\textwidth]{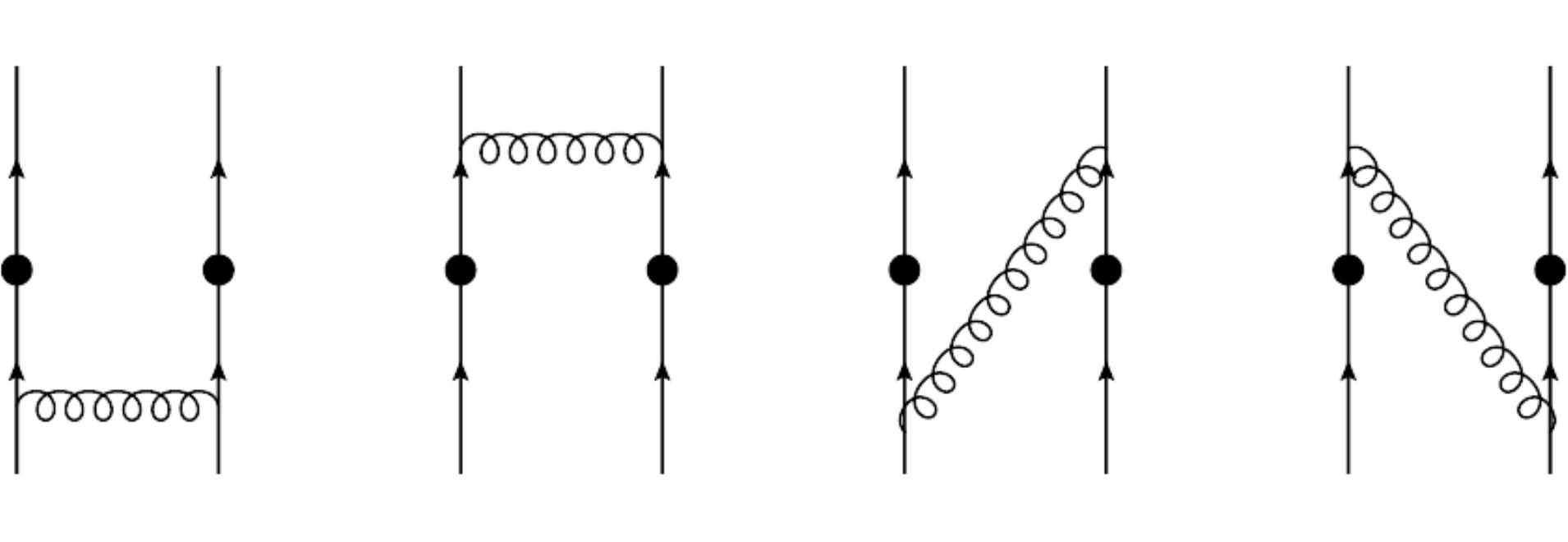}
\caption{Quark-level diagrams that determine the asymptotic behavior of 
$\hat \Pi^{LL}_{\mu \nu}$.
Solid lines represent quarks, curly lines represent gluons, and the black dots represent  
 insertions of the current $\bar{u}_L \gamma_\mu d_L$. 
}
\label{Fig:ope}
\end{figure}

At large Euclidean virtualities ($k_E^2 \gg \Lambda^2_\text{QCD}$)  the time-ordered product of 
currents $\hat{\Pi}_{\mu \nu}^{LL} (k,0)$ defined in Eq.~\eqref{eq:correlator}  admits an OPE. 
An analysis of the leading diagrams, depicted in Fig.~\ref{Fig:ope},   shows  that
\begin{align}
\hat{\Pi}_{\mu \nu}^{LL} (k,0) 
&= - \frac{4 i g_s^2}{(k^2 + i \epsilon)^3}  
\times \Big[
(k_\mu k_\nu - k^2 g_{\mu \nu} )  g^{\alpha \beta}  \, \hat{O}_{\alpha \beta}   (0)
\nonumber \\
&+  g_{\mu \nu} k_\alpha k_\beta \hat{O}^{\alpha \beta} (0) + k^2 \hat{O}_{\nu \mu}  (0)
-k_\nu k^\alpha \, \hat{O}_{\alpha \mu} (0)-  k_\mu k^\alpha \, \hat{O}_{\nu \alpha}  (0)
\Big]\,,
\label{eq:opeab}
\end{align}
where 
\be
\hat{O}_{\alpha \beta} (0) =  \bar{u}_L (0) \gamma_\alpha  T^a d_L (0)\ \bar{u}_L (0)\gamma_\beta  T^a d_L(0)  =  \hat{O}_{\beta \alpha} (0)\,,
\ee
and $T^a$ are $SU(3)$ color generators.

In the actual matching calculation, since we  contract $T_{\mu \nu}$ with $g_{\mu \nu}$,  we need
\be
g^{\mu \nu}   \, \hat{\Pi}_{\mu \nu}^{LL} (k,0)  
=  - \frac{4 i g_s^2}{(k^2  + i \epsilon)^3}  \ \Big( - 2 k^2 \hat{O}_\alpha^\alpha    + 2 k_\alpha k_\beta \hat{O}^{\alpha \beta} 
\Big)
\  \to  \  \frac{8 i g_s^2}{(k^2  + i \epsilon)^2}    \ \frac{3}{4}  \,  \hat{O}_\alpha^\alpha \,, 
\ee
where the last step holds under symmetric integration ($k_\alpha k_\beta \to  k^2 g_{\alpha \beta}/4$). 
Finally, we note that by using Fierz identities in color and then in flavor one obtains 
\be
\hat{O}_{\alpha \beta} =   \frac{1}{3}  \bar{u}_L \gamma_\alpha d_L \ \bar{u}_L \gamma_\beta d_L
\ \to \ 
\hat{O}_\alpha^\alpha  =   \frac{1}{3}  O_1 \,, 
\ee
where $O_1$ is defined in Eq.~\eqref{eq:O1def}.

Performing the $k^0$ integration in order to go from $g^{\mu \nu}   \, \hat{\Pi}_{\mu \nu}^{LL} (k,0)$
to $ \hat O^{LL} (\spacevec{k})$ we obtain 
\be
\hat O^{LL}_> (\spacevec{k}) = \frac{3 g_s^2}{4}  \frac{1}{|\spacevec{k}|^5} \ O_1 (0)\,, 
\ee
which is the expression to be used  to evaluate  the ${\cal A}^{>}$ component in Eq.~\eqref{eq:Mnew2}, 
to obtain 
\be\label{Alarger}
{\cal A}^> = \frac{3 \alpha_s}{2 \pi}\,  \langle f_- | \,  O_1(0)  \, | i_+ \rangle \, \int_{\Lambda}^{\infty} \ d |\spacevec{k}| \, \frac{1}{|\spacevec{k}|^3} 
\,.
\ee
This result  shows a factorization of the hard and soft contributions. The final step requires the evaluation of $ \langle f_- | \,  O_1(0)  \, | i_+ \rangle$, which can be 
done in chiral EFT to LO.

As discussed in Refs.~\cite{Prezeau:2003xn,Cirigliano:2018yza} the four-quark local operator $O_1$ 
admits a low-energy realization consistent with chiral symmetry and its breaking. To LO (and neglecting higher powers of  the pion fields) 
one has 
\be
O_1  \ \to  \  g_1^{NN} \, 
\bar p n \bar p n
\ + \ \frac{5}{6} g_1^{\pi \pi} \,  F_\pi^2 \, \partial_\mu \pi^- \partial^\mu \pi^- 
\ + \  \sqrt{2} g_A g_1^{\pi {N}} F_\pi  \, \bar{p} (S^\mu \partial_\mu \pi^-) n  + \ldots
\ee
The scaling of the non-perturbative parameters is $g_1^{NN}, g_1^{\pi { N}} \sim \Order(1)$, 
while a precise lattice calculation gives $g_1^{\pi \pi} = 0.36(2)$ at the renormalization scale $\mu=2\GeV$ 
in the $\overline{\rm MS}$ 
scheme~\cite{Nicholson:2018mwc}.\footnote{See  Refs.~\cite{Savage:1998yh,Cirigliano:2017ymo} for a determination 
of $g_1^{\pi \pi}$ based on chiral $SU(3)$, 
consistent with the direct lattice result. We have suppressed the dependence of the short-distance couplings on the QCD renormalization scale $\mu$, needed to cancel the scale dependence in $\alpha_s(\mu)$.}
The $\pi\pi$ and $\pi N$ couplings induce a pion-range transition operator, while the $N\!N$ 
couplings gives a short-range transition operator.  
The final result is given by 
\be
 \langle f_- | \,  O_1(0)  \, | i_+ \rangle 
= \tilde{\cal A }_A  + 
\chi^+_{\spacevec p^\prime}(\spacevec 0) \, K_{E^\prime} \, \tilde{\cal A}_B         
+ \   \tilde{\cal A}_{\bar B} \, K_E \, \chi^+_{\spacevec p}(\spacevec 0) 
+
\chi^+_{\spacevec p^\prime}(\spacevec 0) \, K_{E^\prime}  \, \left(  \tilde{\cal A}_C  + \frac{2 \hat{g}_1^{NN}}{C^2} \right) 
\,  K_E \, \chi^+_{\spacevec p}(\spacevec 0)\,, 
\label{eq:chiEFT3}
\ee
where 
\be
\hat g_1^{NN} =  g_1^{NN}+\frac{g_A^2}{2}\left(\frac{5}{6}g_1^{\pi\pi}-g_1^{\pi N}\right)
\ee
and  $\tilde {\cal A}_{A,B,C}$ have the same formal expression as 
${\cal A}_{A,B,C}$~\cite{Cirigliano:2019vdj} with the replacement~\cite{Cirigliano:2018yza}\footnote{Note that the mismatch in dimensions between $V^{^1S_0}_{\nu \, {\rm L}} (\spacevec{q})$  and   
$V^{^1S_0}_{O_1} (\spacevec{q}) $
is compensated by the additional powers of $1/|\spacevec k|$ in Eq.~\eqref{Alarger}.}  
\begin{align}
V^{^1S_0}_{\nu \, {\rm L}} (\spacevec{q})   
\ \to \ 
V^{^1S_0}_{O_1} (\spacevec{q})  =  
 \ g_A^2 
\, \left[
\frac{5 g_1^{\pi \pi}}{6} \, \frac{M_\pi^2 \, \spacevec{q}^2}{(\spacevec{q}^2 + M_\pi^2)^2}  
-   \left(  g_1^{\pi {N}} - \frac{5 g_1^{\pi \pi}}{6}   \right)
\, \frac{M_\pi^2}{\spacevec{q}^2 + M_\pi^2}
\right]\,.
\end{align}

These relations allow us to write ${\cal A}^>$ in a  form very useful for the matching to chiral EFT:
\be
{\cal A}^>  = \mathcal A^>_A  + 
\chi^+_{\spacevec p^\prime}(\spacevec 0) \, K_{E^\prime} \, \mathcal A^>_B         
+{\mathcal A}^>_{\bar B} \, K_E \, \chi^+_{\spacevec p}(\spacevec 0) 
+
\chi^+_{\spacevec p^\prime}(\spacevec 0) \, K_{E^\prime}  \, \mathcal A^>_C  \,  K_E \, \chi^+_{\spacevec p}(\spacevec 0)\,, 
\label{eq:Mp}
\ee
with
\begin{align}
{\cal A}_{A,B, \bar B}^> &=   \frac{3 \alpha_s}{2 \pi} \, \int_{\Lambda}^{\infty} \ d |\spacevec{k}| \ \frac{1}{|\spacevec{k}|^3}  \
\tilde{\cal A}_{A, B, \bar B}\,,\notag
\\
{\cal A}_C^> &=   \frac{3 \alpha_s}{2 \pi} \, \int_{\Lambda}^{\infty} \ d |\spacevec{k}| \ \frac{1}{|\spacevec{k}|^3}  \
\left( \frac{2 \hat g_1^{NN}}{C^2}   + \tilde{\cal A}_C \right)\,.
\label{eq:ACp}
\end{align}

The quantities ${\cal A}_{A, B, \bar B}^>$  are sub-leading compared to their finite chiral EFT counterparts 
${\cal A}_{A, B, \bar B}$, e.g.,  ${\cal A}_A^>/{\cal A}_A \sim Q^2/\Lambda^2$ and similarly 
for ${\cal A}_B$.  Therefore, as anticipated in the previous section,  these quantities 
do not enter the LO matching relation.

On the other hand, ${\cal A}_C^>$ has to be retained because  
its integrand 
provides the  QCD asymptotic behavior  
to which  the integrand in ${\cal A}_C^{<, \text{sing}}$   given in 
Eq.~\eqref{eq:ACmsing} has to tend for large $|\kk|$.
Given our limited knowledge of $g_1^{{NN}, \pi {N}}$ and the fact that  
the terms involving   $g_1^{{NN}}$,    $g_1^ {\pi {N}}$,  and $g_1^{\pi \pi}$ 
contribute at  the same order to $ \langle f_- | \,  O_1(0)  \, | i_+ \rangle$, 
we will retain for simplicity only the term proportional to $g_1^{ NN}$.  
For convenience,  in the matching analysis we express  $g_1^{ NN}$ as follows
\be
g_1^{NN} =  \left( \frac{m_N}{4 \pi} C \right)^2 \, F_\pi^2 \, \bar{g}_1^{NN}\,, \qquad \bar{g}_1^{NN} \sim \Order(1)\,.
\ee
In numerical estimates we will allow for $\bar g_1^{NN}$ as large as $\Order(10)$, given that the $\pi\pi$ short-distance coupling $\bar g_{LR}^{\pi\pi}=8.23$~\cite{Nicholson:2018mwc} relevant for the vector--vector correlator by far exceeds its $\Order(1)$ expectation.
However, the numerical impact of these poorly known short-distance parameters on the overall analysis is very minor, 
as  the main role of  ${\cal A}_C^>$  is to enforce the UV finiteness of the $nn \to pp$ amplitude.

\section{Matching and extraction of $\boldsymbol{{\cal C}_1}$}
\label{sect:matchLL}

\subsection{Results}

The matching condition~\eqref{eq:match0}, taking into account 
the form of ${\cal A}_\nu^{\chi {\rm EFT}}$ in Eq.~\eqref{eq:chiEFT2} 
and the structure of ${\cal A}^<$ (see Eq.~\eqref{eq:Mm} and following discussion)
and ${\cal A}^>$ (see Eq.~\eqref{eq:Mp} and following discussion) 
reduces to 
\be
{\cal A}_C^{< ,{\rm sing}} \  + \  {\cal A}_C^> \  = \   {\cal A}_C^{\rm sing} (\mu_\chi)  \ +  \ 
\frac{2 {\cal C}_1  (\mu_\chi)   }{C^2} \,. 
\label{eq:match2}
\ee
The amplitudes 
${\cal A}_C^{\rm sing} (\mu_\chi)$ and ${\cal A}_C^{<,{\rm sing}}$  are given in Eqs.~\eqref{eq:ACsing0} and  \eqref{eq:ACmsing},  respectively, 
and  ${\cal A}_C^>$ is given in Eq.~\eqref{eq:ACp}.

Defining  the dimensionless coupling $\tilde{\cal C}_1$  via
\be
{\cal C}_1 =   \left( \frac{m_N}{4 \pi} C \right)^2  \ \tilde{\cal C}_1
\label{eq:match3}
\ee
and the dimensionless quantities 
\be
\bar {\cal A}_C^X
 \equiv  \left( \frac{4 \pi}{m_N} \right)^2 \  {\cal A}^X_C\,, 
\ee
the matching condition reads
\be
\tilde{\cal C}_1  (\mu_\chi) 
 = \frac{1}{2}  \left[
\left( \bar{\cal A}_C^{<,{\rm sing}} \ - \bar{\cal A}_C^{\rm sing} (\mu_\chi) \right)   + \bar{\cal A}_C^{>}  
\right]\,.
 \label{eq:mc2}
\ee
Using our previous results, we can write the various quantities entering 
the matching condition in terms of the following integrals:
\begin{align}
\bar{\cal A}^{\rm sing}_{C}  (\mu_\chi) & =   - \frac{1+2 g_A^2}{2} + 
 \int_{0}^{\mu_\chi}   \  d |\spacevec{k}| \ a_\chi ( | \spacevec{k} |) \,,\notag 
 \\
\bar{\cal A}^{<.{\rm sing}}_{C}  &=  \int_0^{\Lambda} \  d |\spacevec{k}| \ a_< ( | \spacevec{k} |) \,,\notag
\\
\bar{\cal A}^{>}_{C} &=   \int_{\Lambda}^\infty   \  d |\spacevec{k}| \ a_> ( | \spacevec{k} |) \,,
\label{eq:integrals}
\end{align}
with\footnote{For the matching condition only the real part of $a_\chi$ matters.} 
\begin{align}
a_\chi ( | \spacevec{k} |)   &=  - (1+2 g_A^2)    \ 
\frac{1}{ |{\bf k}|} \,  \theta(|{\bf k}| - 2 |{\bf p}|)\,,\notag
\\
a_< ( | \spacevec{k} |)   &=  - 
\left(
g_V^2  (\spacevec{k}^2) + 2 g_A( \spacevec{k}^2 ) +   \frac{\spacevec{k}^2 g_M^2 (\spacevec{k}^2)}{2 m_N^2}
\right)
 \,  
\frac{r(|\spacevec{k}|)}{|\spacevec{k}|}\theta(|{\bf k}| - 2 |{\bf p}|)\,, 
\notag
\\
a_> ( | \spacevec{k} |)    &= 
 \frac{3 \alpha_s}{\pi} \ \bar{g}_1^{NN} \ \frac{F_\pi^2}{ |\spacevec{k}|^3} \,, 
 \label{eq:aLLm}
\end{align}
with  $r(|\spacevec{k}|)$ given in Eq.~\eqref{eq:r0}.
Putting all pieces together we can write 
\be
2\,  \tilde{\cal C}_1 (\mu_\chi) 
=  \frac{1+ 2 g_A^2}{2}   -  \int_{0}^{\mu_\chi}   \  d |\spacevec{k}| \ a_\chi ( | \spacevec{k} |) 
+ \int_0^{\Lambda} \  d |\spacevec{k}| \ a_< ( | \spacevec{k} |)  
+ \int_{\Lambda}^\infty   \  d |\spacevec{k}| \ a_> ( | \spacevec{k} |)\,.
\label{eq:match4}
\ee
By construction  the $\mu_\chi$ dependence is consistent with the renormalization group equation (RGE) for the rescaled coupling $\tilde{\cal C}_1(\mu_\chi)$~\cite{Cirigliano:2019vdj}. 
This representation shows that up to an additive constant the LEC 
$\tilde{\cal C}_1 (\mu_\chi)$ 
can be thought of as 
the difference between two integrals in $|\spacevec{k}|$, one in the full theory extending all the way to 
$|\spacevec{k}| \to \infty$,  and one in the EFT extending to  $|\spacevec{k}| = \mu_\chi$. 
Therefore, the  LEC 
$\tilde{\cal C}_1 (\mu_\chi)$ 
 corresponds to 
 (i) a possible mismatch between the LO chiral EFT and  the full amplitude at  $|\spacevec{k}| < \mu_\chi$;
 (ii) the component of the full amplitude  arising from  $|\spacevec{k}| > \mu_\chi$.

 \begin{figure}[t]
\begin{center}
\includegraphics[width=0.45\linewidth]{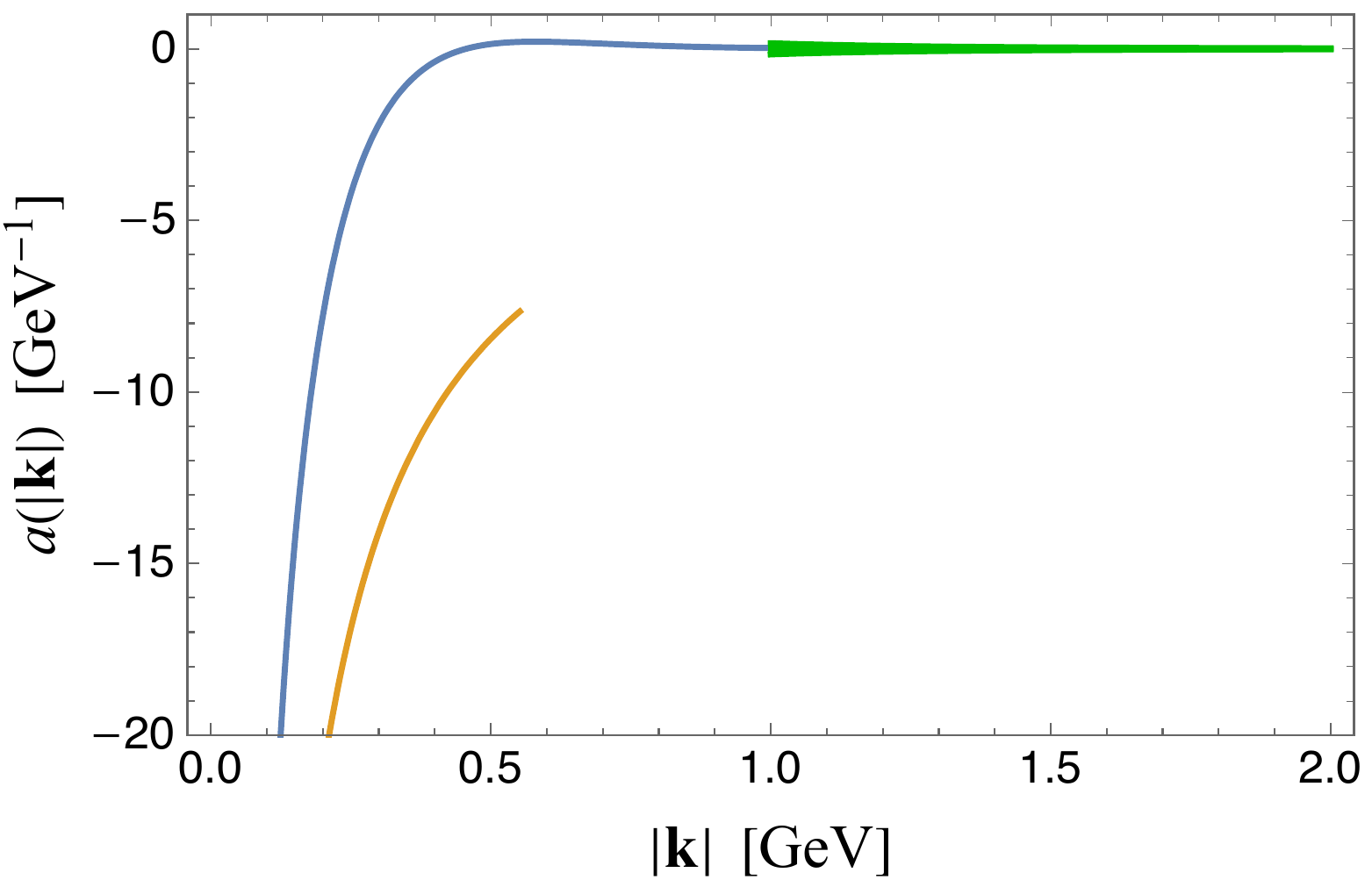}
\hspace{0.03\linewidth}
\includegraphics[width=0.45\linewidth]{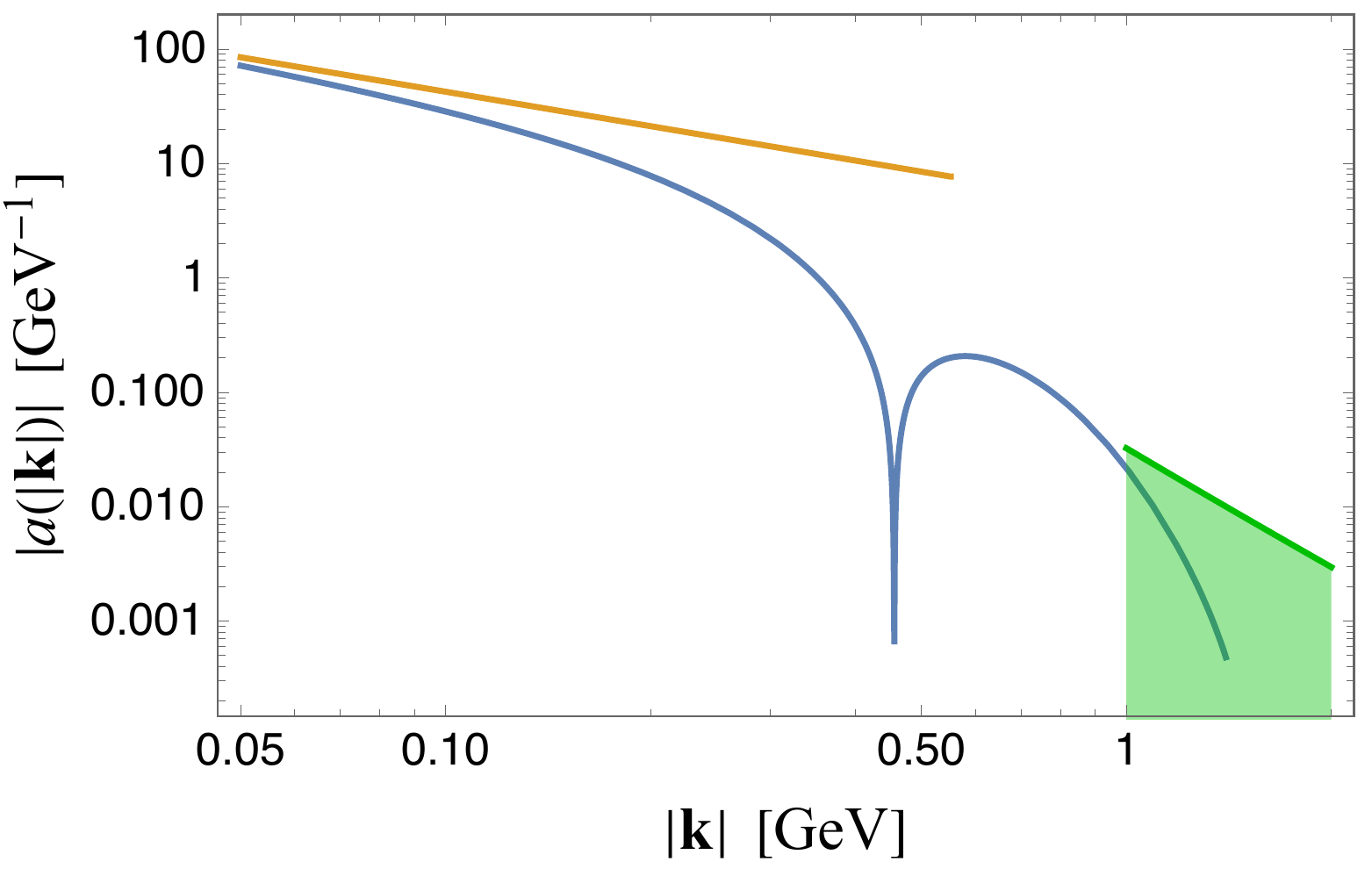}
\caption{ 
Left panel:  $a_\chi (|\spacevec{k}|)$  (yellow), 
$a_< (|\spacevec{k}|)$  (blue), 
and  $a_> (|\spacevec{k}|)$  (green band) assuming  $\bar g_1^{ NN}  \in [-10,10]$.
Right panel:  same plot for the absolute values in logarithmic scale. 
\label{fig:integrandLL}
}
\end{center}
\end{figure}
 
\begin{figure}[t]
\begin{center}
\includegraphics[width=0.45\linewidth]{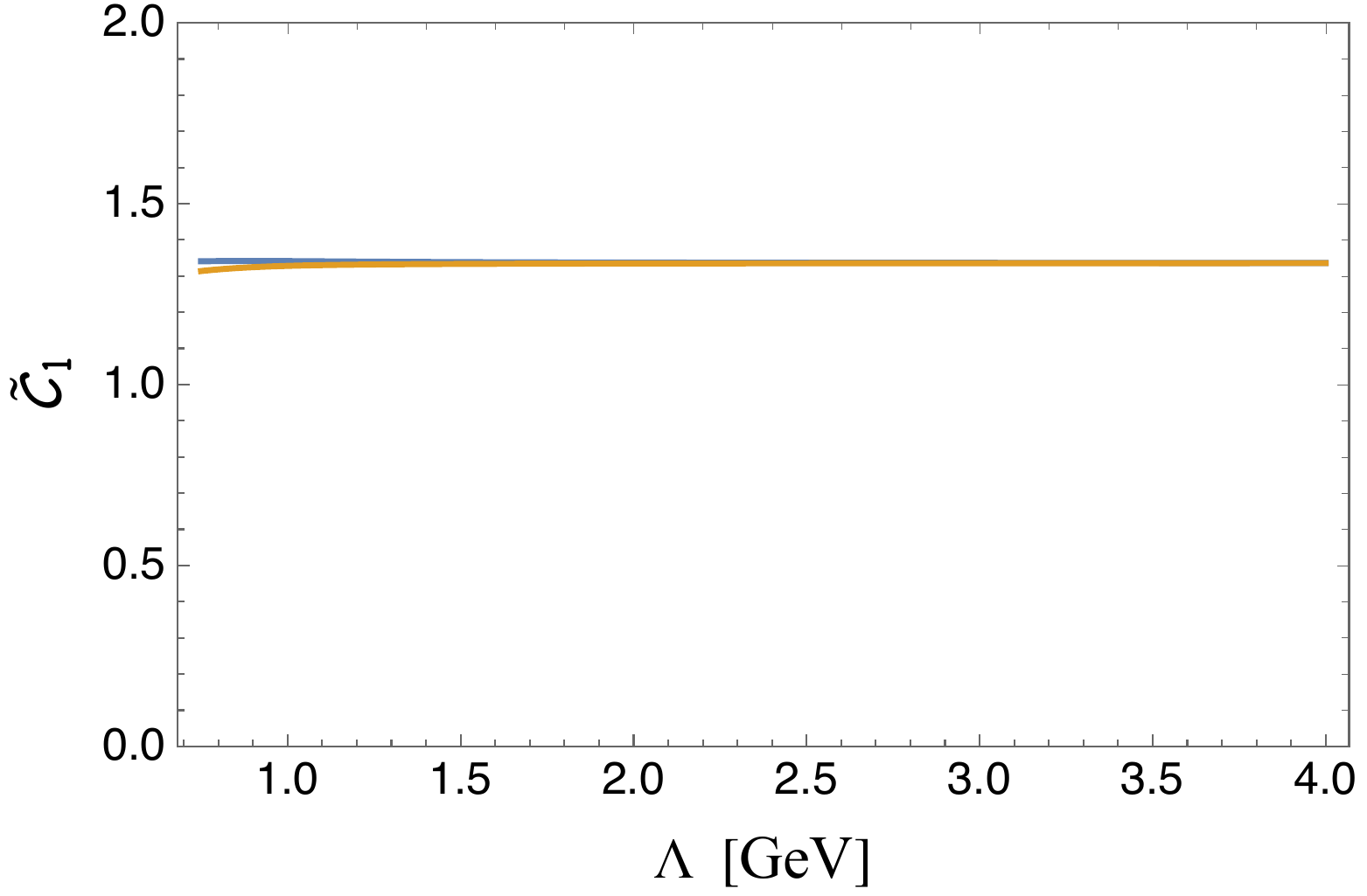}
\hspace{0.03\linewidth}
\includegraphics[width=0.45\linewidth]{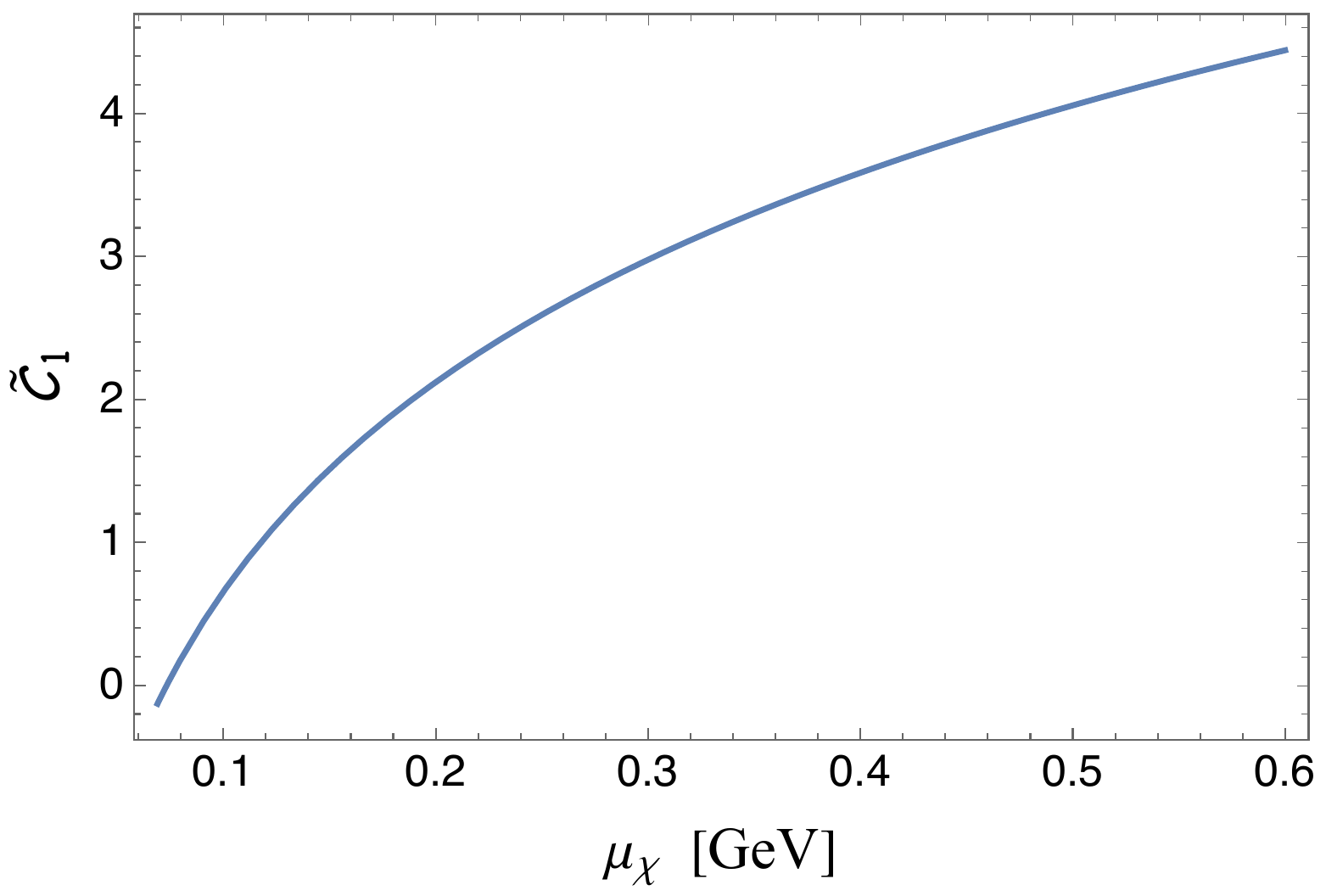}
\caption{ 
Left panel:  dependence of $\tilde{\cal C}_1 (\mu_\chi =M_\pi)$  on the  matching scale $\Lambda$ using 
$\bar g_1^{NN} = \pm 10$. 
Right panel:  dependence of $\tilde{\cal C}_1$ on the chiral renormalization scale $\mu_\chi$. 
\label{fig:C1LL}
}
\end{center}
\end{figure}

In Fig.~\ref{fig:integrandLL} we show    $a_\chi (|\spacevec{k}|)$  (yellow), 
$a_< (|\spacevec{k}|)$  (blue),   and  $a_> (|\spacevec{k}|)$  (green) assuming $\bar{g}_1^{NN} \in [-10,+10]$. 
The behavior of the integrand indicates that 
the LEC is dominated by the low- and intermediate-momentum regions. 
In the left panel of  Fig.~\ref{fig:C1LL}  we show the  dependence of $\tilde{\cal C}_1 (\mu_\chi =M_\pi)$  on the separation scale  $\Lambda$, which proves to be relatively mild.  The right panel shows the dependence of $\tilde{\cal C}_1$ on the chiral renormalization scale $\mu_\chi$. 
The impact of varying $\bar g_1^{NN}  \in [-10,+10]$ is illustrated by the two different curves in the left panel of  Fig.~\ref{fig:C1LL}  and is very small. 

Overall, the result  for the LEC 
$\tilde{\cal C}_1 (\mu_\chi = M_\pi)  \simeq +1.32$ 
 is relatively stable and confirms the expected scaling of the contact term, i.e., 
 $\tilde{\cal C}_1 \sim \Order(1)$~\cite{Cirigliano:2018hja}. 
This matching analysis implies that the contribution from 
$\tilde{\cal C}_1$ and the long-range neutrino exchange encoded in ${\cal A}_C$ interfere {\it destructively}, 
at least when one uses dimensional regularization with minimal subtraction. 
Another way to see this is as follows:  the $^1S_0$ long-range  neutrino potential  
$V_{\nu {\rm L}} = (1+2g_A^2)/\spacevec{q}^2 >0$  
has the opposite sign compared to the short-range potential 
$V_{\nu {\rm S}} = -2 {\cal C}_1   <0$,   since  ${\cal C}_1 (\mu_\chi) >0$  for 
values of $\mu_\chi$ appropriate for the  chiral EFT analysis 
(see the right panel of Fig.~\ref{fig:C1LL}  for a  quantitative statement). 

The origin of the  sign can be traced back to the mismatch between the LO chiral EFT and  the full amplitude for  $|\spacevec{k}| < \mu_\chi$, 
an effect  quite visible in the doubly-logarithmic plot in the right panel of Fig.~\ref{fig:integrandLL}.
In turn, the sizable deviation of $a_< (|\spacevec{k}|)$  (blue)  from   $a_\chi (|\spacevec{k}|)$  (yellow) at $|\spacevec{k}| < \mu_\chi = M_\pi$ is due to the 
behavior of $r(|\spacevec{k}|)$ (see Fig.~\ref{Fig:r} and discussion in Section~\ref{sect:MLLm}), which enters multiplicatively in 
$a_< (|\spacevec{k}|)$,   as per Eq.~\eqref{eq:aLLm}, 
and  effectively encodes higher-order corrections to the forward Compton amplitude. 
As discussed in Section~\ref{sect:MLLm}, 
the all-important negative slope of $r (|\spacevec{k}|)$  for $|\spacevec{k}|  < 300\MeV$   
is controlled by the large   $^1S_0$  effective range $r_0 \simeq 2.7\fm$ and  is not a model artifact. 

While these arguments allow us to understand the origin of the sign of the contact term at a given scale in the \MS scheme, we stress that no general statement is possible, see Section~\ref{sect:synthetic} for more details. For instance, in the cutoff schemes discussed there, both destructive and constructive effects are possible depending on the scale of the regulator.

\subsection{Discussion of uncertainties}

Our baseline result $\tilde {\cal C}_1 (\mu_\chi = M_\pi)  \simeq +1.32$  
is obtained using the Kaplan--Steele~\cite{Kaplan:1999qa}  $V_S$ potential. 
This result  is relatively stable with respect to input parameters in the form factors, the local matrix element $\bar{g}_1^{NN}$ 
controlling the high-$|\spacevec{k}|$ tail, and  the matching scale $\Lambda$, as long as  $\Lambda \geq 1\GeV$.  
The coupling $\bar{g}_1^{NN}(\mu)$, expected to be $\Order(1)$, is presently unknown, but  in view of the large 
value of the similar  two-pion matrix element   $\bar g_{LR}^{\pi \pi}=8.23$ 
(see Appendix~\ref{sec:LROPE} for details), we take the range  $\bar{g}_1^{NN}\in [-10,+10]$, with minor impact on the final result. 
These parametric uncertainties do not exceed $\delta \tilde{\cal C}_1 \simeq \pm 0.05_{\rm par}$ and are 
dominated by the input $\Lambda_A = 1.0(2)\GeV$.

The main systematics of our approach are due to the following effects:
\begin{enumerate}

\item   We have effectively truncated the spectral 
representation~\eqref{eq:spectral2} to keep only the elastic $N\!N$ intermediate state. 
In order to assess the size of the neglected terms, 
we provide below   a rough estimate of the simplest inelastic channel,  namely $N\!N\pi$,\footnote{In the discussion 
of higher-order  EFT contributions to  the $nn \to pp$ amplitude given in   Ref.~\cite{Cirigliano:2017tvr}, 
these contributions were  called ``non-factorizable.'' }  see Fig.~\ref{Fig:inel}.
As discussed in Appendix~\ref{sect:inel}, we find  contributions to $\tilde{\cal C}_1$ from the $N\!N\pi$ intermediate states  on the order of 
$|\delta \tilde{\cal C}_1| = 0.1\text{--}0.35$ 
and due to this we assign an uncertainty of  $\delta \tilde{\cal C}_1 \vert_{\rm inelastic}= \pm 0.5$ .

\item Within the $N\!N$ channel,  we have built our integrand in $|\spacevec{k}|$ starting 
from the low-$|\spacevec{k}|$ end.   To extend the form of the integrand to $|\spacevec{k}| \sim \Lambda \sim 1\GeV$,  
we have taken two main steps: (i) We have included nucleon form factors of the weak currents. 
This is known to saturate the elastic contribution to the electromagnetic mass difference of both the nucleon and pion 
in the Cottingham approach.   (ii) We have included HOS  form factors $f_S (\spacevec{q}, \spacevec{p})$ 
that encode the higher-momentum behavior of  the $N\!N$ scattering amplitude. 
This effect, parameterized by the ratio $r(|\spacevec{k}|)$ defined in Eq.~\eqref{eq:r0}, 
has no analog in the Cottingham literature, which usually deals with two  electroweak current insertions 
on a single hadron.  
As discussed in Section~\ref{sect:MLLm}, 
we have performed the modeling of  $r(|\spacevec{k}|)$  
using  a three-Yukawa potential and tested against the  NLO chiral EFT analysis.   
Comparison with the  chiral EFT analysis shows that inclusion of this effect is essential to reproduce the correct 
Compton amplitude, and hence the integrand, already at $|\spacevec{k}| \leq M_\pi$. 
On the other hand, the extension of   $r(|\spacevec{k}|)$  to  $|\spacevec{k}| \sim \Lambda \sim 1\GeV$  introduces model dependence. 
To quantify this model dependence  we have employed both the 
simple  three-resonance model and the AV18 model for 
the short-range $N\!N$ interactions in the $^1 S_0$ channel. 
In both cases the behavior of   $r(|\spacevec{k}|)$  
shows the same features, see Fig.~\ref{fig:host}:
a rapid drop controlled by the large effective range, a zero for $|\spacevec{k}|$ between 400 and 500 MeV, 
and then a minimum around  $(600\text{--}700)\MeV$.   However, the key point is that 
because of the fall-off of the integrand multiplying  $r(|\spacevec{k}|)$, 
any feature above 500 MeV is washed out, and the differences between AV18 and Kaplan--Steele potentials 
have little impact on the extracted LEC. 
 Numerically, compared  to our  baseline Kaplan--Steele model, 
 we find a variation on $\tilde{\cal C}_1$  of $\approx+ 0.20$   and $\approx- 0.20$  
when using the Reid or AV18 potential, respectively. 
This difference correlates with the  different slopes of   
$r(|\spacevec{k})$  for $|\spacevec{k}|<300\MeV$, see Fig.~\ref{fig:host}. 
Based on this, we will assign an error of $\delta \tilde{\cal C}_1$  of $\pm 0.2$  
due to the choice of the  short-range potential $V_S$.

\item 
The above discussion and Fig.~\ref{fig:integrandLL} 
both point to the fact that the bulk of  $\tilde{\cal C}_1$  
is controlled by the behavior of $a_<(|\spacevec{k}|)$ for 
$|\spacevec{k}|<300\MeV$, where our integrand can be 
linked to the   model-independent  chiral EFT  behavior. 
Quantitatively, we find that the region  
$|\spacevec{k}| \in [0.4, 1.5]\GeV$, entailing the largest unknowns, 
 contributes 
$\Delta |\tilde{\cal C}_1 |\leq 0.05$, which is a reassuring result.

\item 
Finally, we note that our approach can be used to estimate 
the  combination of LECs $\tilde{\cal C}_1 + \tilde{\cal C}_2$,  
which corresponds to two insertions of the electromagnetic current and can  be extracted from data. 
As discussed in Section~\ref{sect:VV},  our matching calculation for   $\tilde{\cal C}_1 + \tilde{\cal C}_2$ 
compares within uncertainties fairly well with the value extracted  from  the CIB combination of  $N\! N$ scattering lengths. 
This is a non-trivial validation of the method and 
gives us confidence that our results provide a realistic estimate of these couplings and their uncertainties. 

\end{enumerate}

\begin{figure}[t]
\centering
\includegraphics[width=0.5\textwidth]{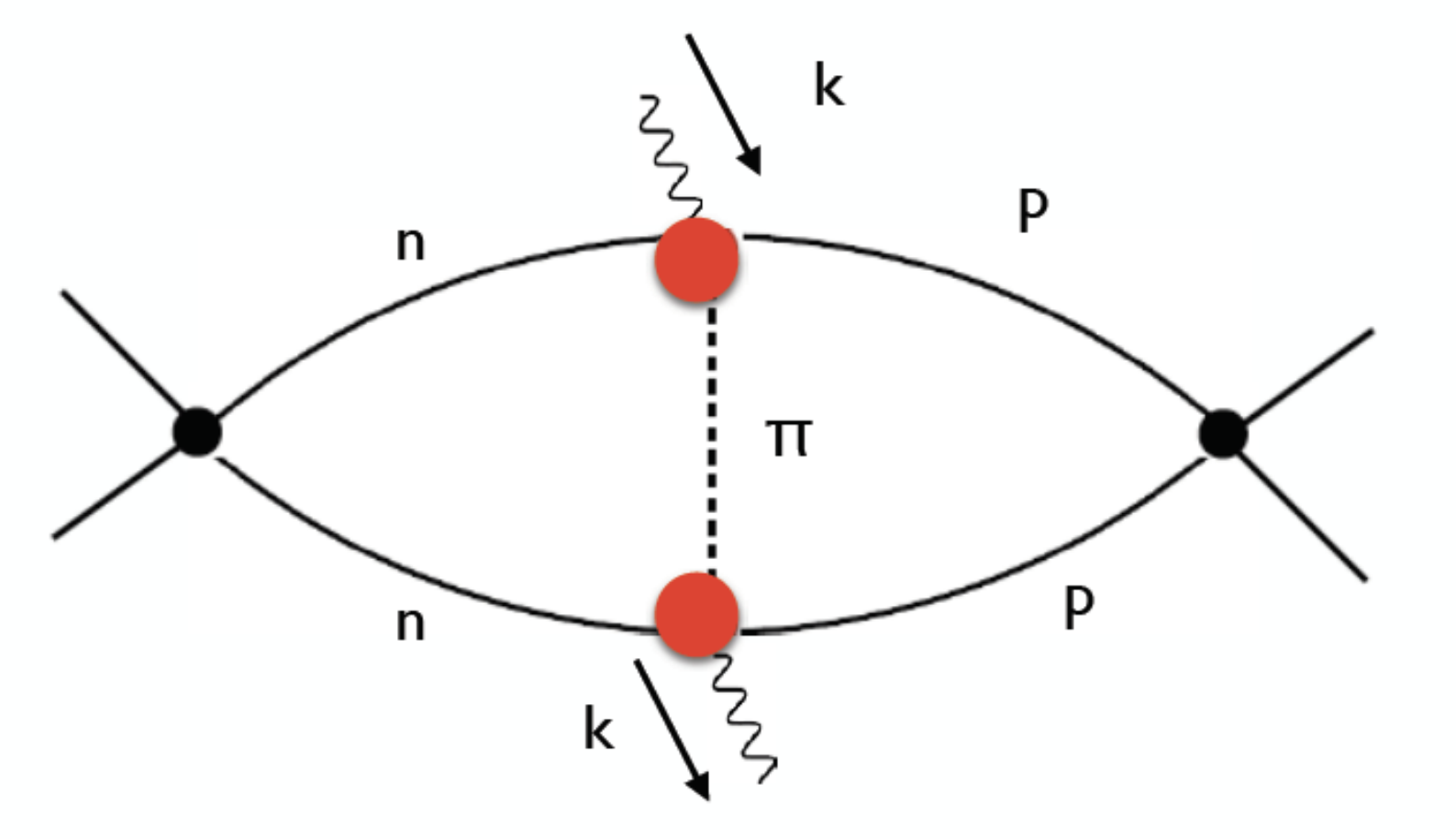}
\caption{
Representative inelastic contribution to the generalized forward Compton amplitude.
Solid lines represent nucleons, dashed lines pions, and wavy lines the  current insertions. 
}
\label{Fig:inel}
\end{figure}

In light of the above discussion, we quote  the following  estimate for the LEC 
\be
 \tilde {\cal C}_1 (\mu_\chi = M_\pi) 
 \simeq 1.32(50)_{\rm inel}(20)_{V_S}(5)_{\rm par}
 =  1.3(6)\,,
\ee
which becomes 
\be
 \tilde {\cal C}_1 (\mu_\chi = 4M_\pi) 
 =  4.2(6) 
\ee
at a renormalization scale that corresponds more closely to cutoffs used in \textit{ab-initio} many-body calculations (albeit in a different scheme and thus not in direct correspondence, see Section~\ref{sect:synthetic} for more details). 

The final uncertainty is dominated by the missing 
inelastic contributions and implies a relative precision of  (20--30)\%  on  the renormalized  singular  amplitude  $\bar \A_C^\text{sing} + 2\, \tilde{\cal C}_1$ 
 at $|\pp|\sim(20\text{--}30)\MeV$---in line with the expectation from the Cottingham analyses of pion and nucleon masses. 
Note that  this  translates into a smaller relative error on the total amplitude $\A_\nu^{\chi\text{EFT}}$, as we will discuss  in Section~\ref{sect:synthetic}.

\section{Vector--vector  amplitude and  $\boldsymbol{{\cal C}_1 +{ \cal  C}_2}$}
\label{sect:VV}

In  analogy to the matching for the purely left-handed coupling $ {\cal C}_1 = g_\nu^{NN}$, 
one can determine the left--right coupling ${\cal C}_2$ and the vector combination ${\cal C}_1 + {\cal C}_2$. 
Since  ${\cal C}_1 + {\cal C}_2$  can be extracted from a fit to  the $I=2$ $N\!N$ scattering amplitudes,  this offers a way to validate the matching approach 
through experimental data. 
The matching calculation is most easily done by considering the $nn \to pp$ transition in the unphysical theory in which 
the $W$ boson couples to both the left-handed  quark current $\bar{u}_L \gamma_\mu d_L$ (as in the Standard Model) and the right-handed 
current $\bar{u}_R \gamma_\mu d_R$ with equal couplings.

\subsection{Effective Lagrangians}

Let us briefly recall the basic elements of this extended analysis~\cite{Cirigliano:2018hja,Cirigliano:2019vdj}.
The starting point is the quark-level electromagnetic and weak Lagrangian in the Standard Model
\begin{equation}\label{quark}
\mathcal L = \bar q_L \gamma^\mu \left(l_\mu + \hat{l}_\mu \right) q_L  
+ \bar q_R \gamma^\mu \left(r_\mu + \hat{r}_\mu\right) q_R  
\,,
\end{equation}
where $q$ denotes the quark doublet 
$q = (u,\; d)^T$ and we defined
\begin{align}
l_\mu &= \frac{e}{2} A_\mu \, \tau^3 
- 2 \sqrt{2}\, G_F \left[V_{ud} \, \bar e_L \gamma_\mu  \nu_L \, \tau^+  + {\rm h.c.}\right]\,,  
& \hat{l}_\mu &= \frac{e}{6} A_\mu\,, \notag
\\
r_\mu &= \frac{e}{2} A_\mu \, \tau^3\,,  
& \hat{r}_\mu &= \frac{e}{6} A_\mu\, , 
\end{align}
where $\tau^+ = (\tau_1 + i \tau_2)/2$. 
We are interested in the isovector components of the interactions, involving $l_\mu$ and $r_\mu$. 
We will  consider the unphysical theory in which the $W$ boson has vector-like couplings 
and hence couples with same strength to both  $\bar{u}_L \gamma_\mu d_L$  and  $\bar{u}_R \gamma_\mu d_R$. 
This amounts to $r_\mu = l_\mu$, namely 
\be
r_\mu = \frac{e}{2} A_\mu \, \tau^3   - 2 \sqrt{2}\, G_F \left[V_{ud} \, \bar e_L \gamma_\mu  \nu_L \, \tau^+  + {\rm h.c.}\right]\,.  
\ee
Double insertions of the isovector component of Eq.~\eqref{quark}  give rise 
to $0\nu\beta\beta$ amplitudes and $I=2$ electromagnetic effects through the effective actions:
\be
S_{\rm LNV} = S^{++}\,,\qquad 
S_{\rm EM, I=2} = S^{33}\,, 
\ee
with 
\be
S^{aa} =F^{aa}  \, \int d^4x d^4y \   g^{\mu \nu} S(x-y)  \ T \Big( 
J^a_{\mu L}(x)   \,  J^a_{\nu L}(y)   + 
J^a_{\mu R}(x)   \, J^a_{\nu R}(y)   + 
2  J^a_{\mu L}(x) \,  J^a_{\nu R}(y)   
\Big)\,,
\ee
and
\be
J^a_{\mu L} (x)  =  \bar{q}_L (x) \gamma_\mu t^a  q_L(x)\,,\qquad 
J^a_{\mu R} (x)  =  \bar{q}_R (x) \gamma_\mu t^a  q_R(x)\,.
\ee
The interaction-specific factors are\footnote{For simplicity we neglect the electron four-momenta, so we can 
replace $e_L(x), e_L(y)  \to  e_L (0)$ and factor the electron fields out of the convolution integral.}
\begin{align}
F^{++} &=  \frac{8 G_F^2 V_{ud}^2 m_{\beta \beta}}{2!} \   \bar{e}_L e_L^c\,,    & t^+ &= \tau^+   = \frac{\tau_1 + i \tau_2}{2}\,,\notag
\\
F^{33} &=  \frac{e^2}{2!}\,, & t^3 &= \frac{\tau^3}{2}\,, 
\end{align}
and $S(x-y)$ is the massless scalar propagator defined in 
Eq.~\eqref{eq:EffProp}, which arises from the neutrino or photon propagator in the $0\nu\beta\beta$ and $I=2$ cases, respectively.

At low energy, the above effective actions  manifest themselves through both (i)  long-distance effects with exchange 
of soft, potential, and ultrasoft  neutrinos (photons)  between the pion and nucleon realization of the electroweak currents; (ii) local interactions, 
which can be thought of as arising from the exchange of hard neutrino (photon) modes. 
Since the amplitudes transform according  to the same irreducible representation of the isospin group, 
in the isospin symmetry limit   the exchange of hard neutrinos  leads to identical contributions 
as photon exchange, up to the overall factors $F^{++}$ and $F^{33}$. 

The above considerations imply  the following relations between low-energy chiral Lagrangians. 
In the nucleon sector we have:
\begin{align}
\mathcal L_{e^2}^{NN}  &=   \frac{e^2}{4} \ \Big[ {\cal C}_1  \, O^{33}_1 +  {\cal C}_2  \, O^{33}_2   \Big]\,,\notag 
\\
\mathcal L_{|\Delta L| =2}^{NN}  &= 2 \,  G_F^2 V_{ud}^2 m_{\beta \beta}  \  \bar{e}_L e_L^c    \ 
\Big[ {\cal C}_1  \, O^{++}_1 +  {\cal C}_2  \, O^{++}_2   \Big]\,, 
\end{align}
with 
\begin{align}
O_1^{aa}  &= \bar N \mathcal Q^a_L N \, \bar N \mathcal Q^a_L N  
-\frac{\textrm{Tr}[\mathcal Q^a_L   \cq^a_L ]}{6} \bar N \boldtau N \cdot \bar N \boldtau N
+ \{ L \leftrightarrow R \} \, ,
\nonumber \\
O_2^{aa} &=  2 \left( \bar N \mathcal Q^a_L N \bar N \mathcal  Q^a_R N  
-\frac{\textrm{Tr}[\mathcal Q^a_L \mathcal Q^a_R] }{6} \bar N \boldtau N \cdot 
\bar N \boldtau N \right) \, ,
\end{align}
where  
\be
\mathcal Q_L^a  = u^\dagger t^a u\,, 
\qquad 
\mathcal Q_R^a  = u t^a u^\dagger\,,
\ee
and  $u^2 = U = \exp(i \boldtau \cdot \boldsymbol{\pi}/F_\pi)$
incorporates the pion fields.

In the pion sector we have non-derivative LO local operators arising from the LR current correlator:
\begin{align}
\mathcal L_{e^2}^{\pi \pi}  &=  e^2 Z F_\pi^4  \  \textrm{Tr} [\mathcal Q^{3}_L \mathcal Q^{3}_R ]\,,\notag
\\
\mathcal L_{|\Delta L| =2}^{\pi \pi }  &=  \left( 8  G_F^2 V_{ud}^2 m_{\beta \beta}  \  \bar{e}_L e_L^c   \right) \ 
Z F_\pi^4 \ 
 \textrm{Tr} [\mathcal Q^{+}_L \mathcal Q^{+}_R ]\,, 
\label{eq:ZLag}
\end{align}
where, at LO in chiral perturbation theory (ChPT), $Z$ is related to the pion-mass splitting by\footnote{Note that the 
operator in the first line of Eq.~\eqref{eq:ZLag} differs from 
the usual definition in ChPT~\cite{Bijnens:2014lea}  by an inessential  constant.} 
\begin{equation}
Z e^2 F_\pi^2 = \frac{1}{2}\delta M^2_\pi  = \frac{1}{2} \left( M_{\pi^\pm}^2 - M_{\pi^0}^2 \right).
\end{equation}

 \subsection[Pion two-point function and  the low-energy constant $Z$]{Pion two-point function and  the low-energy constant $\boldsymbol{Z}$}
 \label{sect:pipi}
 
 Before discussing the $nn \to pp$ amplitude,  it is  instructive  to consider  the simpler  
 $\pi^- \to \pi^+$ transition.   
 In complete analogy with Eq.~\eqref{eq:correlator} one can define a vector--vector correlator via the replacement 
 $J_\mu^L  =  \bar u_L \gamma_\mu  d_L \to   V_\mu = J_\mu^L +J_\mu^R =  \bar u \gamma_\mu d$. 
 The object of interest for our matching calculation is 
 \be
T_{\mu \nu}^{\pi \pi}  (k,p)  =  \langle \pi^+ (p) |  \  \hat \Pi^{VV}_{\mu \nu}  (k,0)  \  | \pi^- (p) \rangle\,,   \qquad
 {\cal T}^{\pi \pi} (k,p) 
\equiv g^{\mu \nu} T_{\mu \nu}^{\pi \pi}  (k,p)\,.
 \ee
${\cal T}^{\pi \pi} (k,0)$  determines the LEC $Z$    defined 
in Eq.~\eqref{eq:ZLag}, while 
${\cal T}^{\pi \pi} (k, p)$ (where $p$ is an off-shell four-momentum with potential scaling)  
contributes to $nn \to pp$  as discussed below.

In the small $p,k$ regime $ {\cal T}^{\pi \pi} (k,p)$  can be computed reliably in  ChPT. 
At LO one has\footnote{Up to a proportionality factor due to isospin, this expression agrees with the
forward Compton amplitude, see for example Ref.~\cite{Donoghue:1996zn}.}
 \be
 {\cal T}_\chi^{\pi \pi} (k,p) 
  = 2 i \left[ 2 g_{\mu \nu}
  - \frac{ (2 p +k)_\mu (2 p+ k)_\nu}{(p+k)^2 - M_\pi^2 + i \epsilon} 
   - \frac{ (2 p -k)_\mu (2 p- k)_\nu}{(p-k)^2 - M_\pi^2 + i \epsilon}
 \right] \ g^{\mu \nu}\,,
  \label{eq:tpipim0}
 \ee
 which in the chiral limit and for $p \to 0$ reduces to 
  \be
 {\cal T}_\chi^{\pi \pi} (k,0)  \Big \vert_{M_\pi= 0}  
  = 4 i \left[  g_{\mu \nu} - \frac{k_\mu k_\nu}{k^2} \right] \, g^{\mu \nu} = 4 i (d-1)\,.
  \label{eq:tpipim}
 \ee
The expression for ${\cal T}^{\pi \pi} (k,p)$ can be extended to the intermediate $k$ momentum region by 
using resonance models that work reasonably well in the meson sector~\cite{Ecker:1988te} 
and automatically respect the constraints of chiral symmetry beyond the LO amplitude~\eqref{eq:tpipim0}.
We will not follow this route but we will rather use the following prescription
\be
 {\cal T}_<^{\pi \pi} (k,p)  =  {\cal T}_\chi^{\pi \pi} (k,p)  \times \left( F_\pi^V (k^2) \right)^2\,, 
\label{eq:Tpipim}
\ee
which captures the full pion-pole contribution to the on-shell Compton amplitude~\cite{Colangelo:2015ama,Colangelo:2017qdm,Colangelo:2017fiz} 
in terms of the pion vector form factor  $F_\pi^V(k^2)$,  for which we will take the simple monopole form 
\be
F_\pi^V ({k}^2) =  \frac{M_V^2}{M_V^2 - k^2}\,,
\label{eq:pionFF}
\ee
with $M_V=M_\rho$.
While the form~\eqref{eq:Tpipim} does not capture the contributions form intermediate-state axial-vector  mesons~\cite{Bardeen:1988zw,Ecker:1988te,Donoghue:1993hj,Baur:1995ig,Donoghue:1996zn}, 
the resulting numerical impact on the LEC $Z$ is below the 20\% level~\cite{Donoghue:1993hj}.

 In the  large $k$  regime, the form of  $ {\cal T}^{\pi \pi} (k,p)$ is dictated by the OPE, see Section~\ref{sect:OPELL} 
 and  Appendix~\ref{sec:LROPE}. It is given by 
 \be
 {\cal T}_>^{\pi \pi} (k,p)  = \frac{4 i g_s^2}{\left(k^2 + i \epsilon \right)^2} \,   \langle \pi^+ (p) |  \  \hat  O^{VV}   \  | \pi^- (p) \rangle \,,
 \qquad  \hat  O^{VV} = O_1 + \frac{1}{2}  \left(O_4 - 3 O_5 \right)\,.
\label{eq:tpipiM}
 \ee
To LO in ChPT, the pion matrix elements are given by 
\be
  \langle \pi^+ (p) |  \  \hat  O^{VV}   \  | \pi^- (p) \rangle =  F_0^2 \ \left[ \frac{5}{3}  g_1^{\pi \pi}  p^2 + \frac{1}{2}  \left( g_4^{\pi \pi}  - 3 g_5^{\pi \pi} \right) \right]\,, 
  \label{eq:OLR}
\ee
 where the LECs $g^{\pi \pi}_{1,4,5}$ are known from lattice QCD~\cite{Cirigliano:2017ymo,Nicholson:2018mwc} 
 and scale as $g_1^{\pi \pi} \sim \Order(1)$ and $g_{4,5}^{\pi \pi} \sim \Order(\Lambda_\chi^2)$. 
 From this scaling it is clear that the dominant contribution to the two-pion matrix element is proportional to the combination 
 $g_{LR}^{\pi \pi} \equiv (1/2) (g_4^{\pi \pi} - 3 g_5^{\pi \pi})$. 

We are now in a position to write down a matching relation for the LEC $Z$ introduced in Eq.~\eqref{eq:ZLag}.
Writing the pion two-point function at zero momentum in ChPT and in full QCD, one derives the relation 
\be
8 Z F_0^2 =  \int \frac{d^4k}{(2 \pi)^4} \ \frac{1}{k^2 + i\epsilon} 
\, {\cal T}^{\pi \pi} (k,0)\,.
\ee
 In the spirit of the matching strategy used in this work, we split up the integration over three-momentum $|\spacevec{k}|$ in 
 two regions and by using ${\cal T}_<^{\pi \pi} (k,0)$  
 from Eq.~\eqref{eq:tpipim}
 and ${\cal T}_>^{\pi \pi} (k,0)$ from Eqs.~\eqref{eq:tpipiM}--\eqref{eq:OLR} we arrive at (see Appendix~\ref {app:Zm} for details) 
\begin{align}
 Z &= Z^< + Z^>\,,\notag
\\
 Z^< &=  \frac{6}{(4 \pi F_0)^2} \, \int_0^\Lambda  \, d |\spacevec{k}|  \, |\spacevec{k}|    \times 
 \left(  1 - \frac{|\spacevec{k}|}{\omega_V} \left( 1 + \frac{M_V^2}{2 \omega_V^2}\right)  \right)\,,\notag
 \\
 Z^> &=  \frac{3 \alpha_s (\mu) g_{LR}^{\pi \pi} (\mu)}{16 \pi} \, \int_{\Lambda}^{\infty}   d |\spacevec{k}|  \, \frac{1}{|\spacevec{k}|^3}\,, 
 \label{eq:Zmatch}
 \end{align}
where  $\omega_V = \sqrt{\spacevec{k}^2 + M_V^2}$ and  $\mu$ is the QCD renormalization scale. 
 The above result is in agreement with estimates of the pion electromagnetic mass splitting~\cite{Bardeen:1988zw,Ecker:1988te,Donoghue:1993hj,Baur:1995ig,Donoghue:1996zn}.
Taking $M_V = 775\MeV$,  $F_\pi = 92.28\MeV$, 
and ${g}_{LR}^{\pi \pi}=8.23$  from lattice QCD~\cite{Nicholson:2018mwc},
the sum of low- and high-momentum components   $Z^<(\Lambda) + Z^> (\Lambda)$   equals 0.63  at $\Lambda = 2\GeV$ 
and reaches the asymptotic value  $Z^<  (\Lambda \to \infty)= 0.67$. 
The  deficit compared to   the experimental value $Z \simeq 0.8$  is understood in terms of 
the neglected inelastic contributions from the axial-vector resonances~\cite{Bardeen:1988zw,Ecker:1988te,Donoghue:1993hj}, thus providing another estimate of the error due to neglecting inelastic corrections.

\subsection[{$ n n \to p p$ vector-like amplitude  in chiral EFT and full theory}]{$\boldsymbol{ n n \to p p}$ vector-like amplitude  in chiral EFT and full theory}

Let us now consider the amplitude ${\cal A}_{VV}$  in the unphysical theory in which the $W$ boson has vector-like couplings to quarks. 
Denoting by ${\cal A}_{LL,LR}$ the  $nn \to pp$ amplitudes generated by $W$-exchange between two left-handed quark currents 
and a left-handed and a right-handed quark current, respectively (see Eqs.~\eqref{eq:correlator} and~\eqref{eq:M}), we have 
\be
{\cal A}_{VV}   =  2 {\cal A}_{LL} + 2 {\cal A}_{LR}\,. 
\label{eq:MV}
\ee
Note that ${\cal A}_{LL} \equiv {\cal A}_\nu$, i.e., the amplitude in the physical theory,  defined in Eq.~\eqref{eq:M}, 
and   the factor of 2 in front of  ${\cal A}_{LL}$ arises from the fact that  $LL$ and $RR$ products, both present 
due to vector-like couplings,  give the same result by parity. 

\subsubsection[{${\cal A}_{VV}$ in chiral EFT}]{$\boldsymbol{{\cal A}_{VV}}$ in chiral EFT}

At LO in chiral EFT,  the $nn \to pp$  vector amplitude takes the form
\be
{\cal A}_{VV}^{\chi {\rm EFT}} = 2 {\cal A}_{LL}^{NN} +  2 {\cal A}_{LR}^{NN} +  {\cal A}_{VV}^{\pi \pi}\,, 
\label{eq:MVEFT}
\ee
where  ${\cal A}_{LL}^{NN}$ is given in Eq.~\eqref{eq:chiEFT2}.
${\cal A}^{NN}_{LR}$  can be obtained  from ${\cal A}^{NN}_{LL}$ by flipping the sign of the  ``$A\times A$'' axial contribution, 
which in practice amounts to setting  $g_A^2 \to - g_A^2$ everywhere and replacing ${\cal C}_1 \to {\cal C}_2$ in the 
counter term amplitude. 

\begin{figure}[t]
\centering
\includegraphics[width=0.7\textwidth]{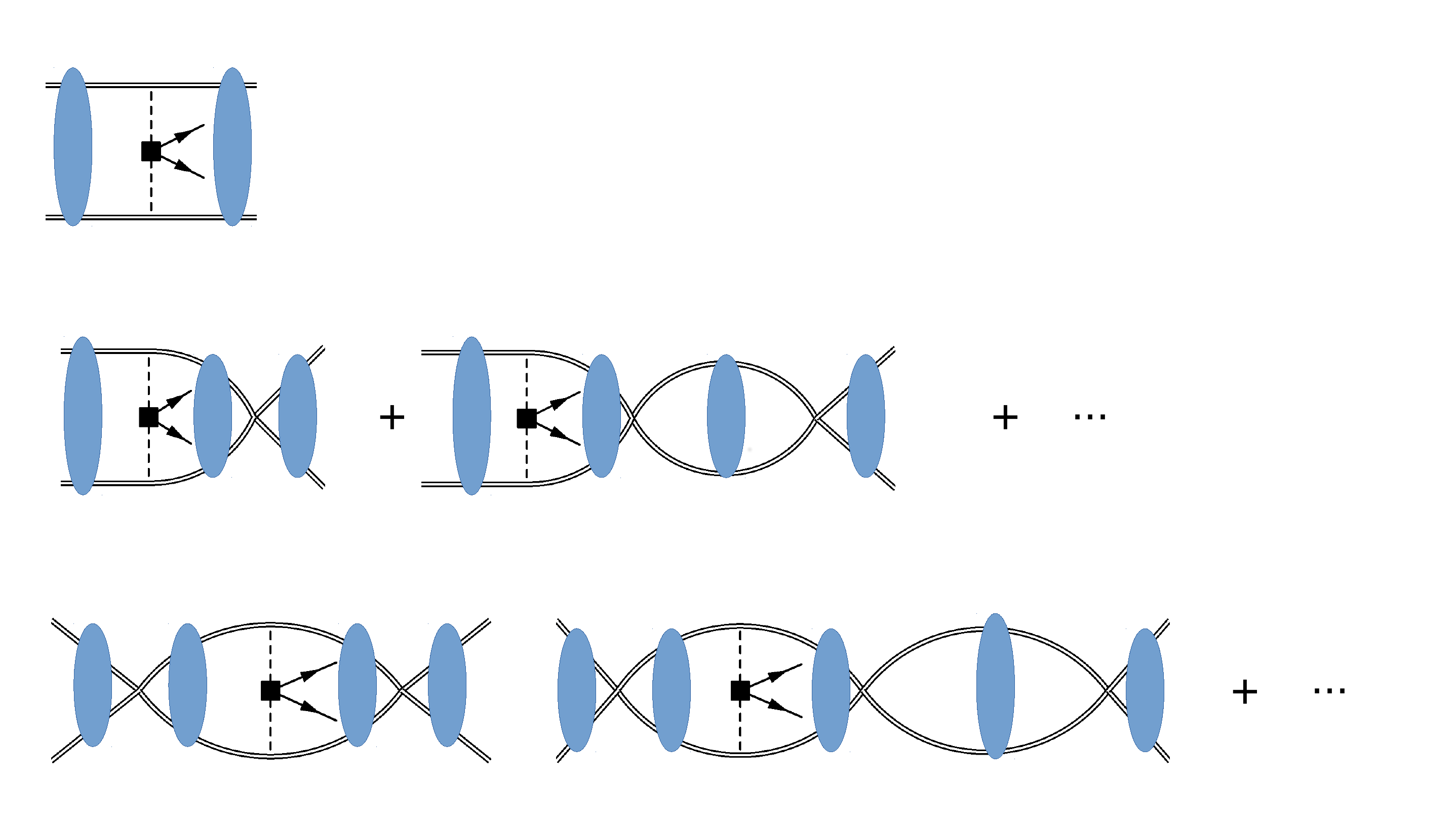}
\caption{Diagrammatic representation of  ${\cal A}_{VV}^{\pi \pi}$, 
the pion  contributions to  the LO
$n n \rightarrow p p ee$ amplitude in the vector-like unphysical theory.  
Double, dashed, and plain lines denote nucleons, 
pions, and electrons, respectively. 
The black square denotes an insertion of  the non-derivative coupling $Z$, see Eq.~\eqref{eq:ZLag}. 
The blue ellipse represents iteration of 
$\hat V_\pi$, as in Fig.~\ref{Fig:chiEFT}.
The ellipses in the second and third lines denote diagrams with 
arbitrary numbers of $N\!N$ bubble insertions.  
}
\label{Fig:chiEFTpi}
\end{figure}

In the  vector theory, the main new effect compared to the physical amplitude is the presence of a LO contribution induced by pion exchange,  
denoted by $ {\cal A}_{VV}^{\pi \pi}$ in Eq.~\eqref{eq:MVEFT},  
induced by  the non-derivative operator in Eq.~\eqref{eq:ZLag} with coupling constant $Z$. 
The amplitude  ${\cal A}_{VV}^{\pi \pi}$ has the same form of  
$ {\cal A}_\nu^{\chi {\rm EFT}}$ in  Eq.~\eqref{eq:Mmchi}, except for the fact that
in each diagram the neutrino propagator is replaced by a pion propagator with one insertion of $Z$ that converts a $\pi^-$ into a $\pi^+$, 
see Fig.~\ref{Fig:chiEFTpi}. 
This implies that in the  computation  of ${\cal A}_{A,B,C}$  in Eqs.~\eqref{eq:mcc}--\eqref{eq:chiEFT2} 
one needs to make the replacement:
\be
V^{^1S_0}_{\nu \, {\rm L}} (\spacevec{k})   = 
\frac{1}{\spacevec{k}^2} \left[
1+ 2 g_A^2  + \frac{ g_A^2 M_\pi^4}{ (\spacevec k^2 + M_\pi^2)^2} \right]
\ \to \ 
V^{^1S_0}_{Z} (\spacevec{k}) \ = 
\frac{8 Z g_A^2}{\spacevec{k}^2} \ \left( 1   - \frac{M_\pi^2}{\spacevec{k}^2 + M_\pi^2}   \right)^2\,.
\ee
As discussed in Ref.~\cite{Cirigliano:2019vdj}, the UV  behavior of this potential induces additional divergences reabsorbed by the coupling ${\cal C}_2$.  

Finally, we note that in complete analogy to the physical  $LL$ case, the matching condition for the $VV$ amplitude will involve only the 
real part  of the singular component of the  amplitude ${\cal A}_C$ associated with Eq.~\eqref {eq:MVEFT}. 
Combining the $NN$ and $\pi \pi$ contributions, 
we write it in the following  way that will 
prove useful in the matching procedure:
\begin{align}
{\cal A}_C^{\rm sing}  (\mu_\chi) 
\Big \vert_{VV}
&=
-   2   \frac{m_N^2}{(4 \pi)^2}    \, 
  \left( 1 + 2 Z g_A^2   \right)  \,
\left[
 \log \frac{\mu_\chi^2}{4 |{\bf p}|^2 }    
+1 
\right]\notag
\\
&=  2  \frac{m_N^2}{(4 \pi)^2}    \,  
\left[ - ( 1 + 2 Z g_A^2) - 4 Z g_A^2 \, \log \frac{\mu_\chi}{ 2  |{\bf p}|}
+  \int_0^{\mu_\chi} \ 
 d |\spacevec{k}| \ a^{VV}_\chi ( | \spacevec{k} |)  
\right]\,,\notag
\\
a^{VV}_\chi ( | \spacevec{k} |)   &=  - 2  \ 
\frac{1}{ |{\bf k}|} \,  \theta(|{\bf k}| - 2 |{\bf p}|)\,.
\label{eq:ACsingVV}
\end{align}

 \subsubsection[{${\cal A}_{VV}$ in the full theory}]{$\boldsymbol{{\cal A}_{VV}}$ in the full theory}

As before, we split the full amplitude into two terms, capturing the contributions from small plus intermediate and hard 
neutrino momenta:   ${\cal A}_{VV}^{\rm full} = {\cal A}_{VV}^< + {\cal A}_{VV}^>$. 
One can again organize the full theory calculation of ${\cal A}_{VV}^<$ in  close analogy to the EFT expression given in Eq.~\eqref{eq:MVEFT}.
${\cal A}_{LL}^{NN,<}$ is discussed in  Section~\ref{sect:MLLm} and the corresponding  ${\cal A}_{LR}^{NN,<}$ is obtained  
by defining $\hat O^{LR} (\spacevec{k})$ in complete analogy to $\hat O^{LL} (\spacevec{k})$ 
in Eq.~\eqref{eq:OLLdef} and then following through.  In practice,  this again amounts to flipping the sign of the axial 
contribution in ${\cal A}_{LL}^{NN,<}$.  As discussed in Section~\ref{sect:matchLL}, for matching purposes 
one only needs the singular parts of  the amplitude ${\cal A}_C^<$, which can be written as 
\be
{\cal A}_C^{<,{\rm sing}} \Big \vert_{VV}
 = 2 \left(  {\cal A}_C^{<,{\rm sing}} \Big \vert_{LL}
 + {\cal A}_C^{<,{\rm sing}} \Big \vert_{LR}
\right) + 
 {\cal A}_C^{\pi \pi,<, {\rm sing}}\,.
 \label{eq:ACVVm}
\ee
${\cal A}_C^{<,{\rm sing}}  \vert_{LL}$ is given in Eqs.~\eqref{eq:ACmsing}, \eqref{eq:integrals}, and  \eqref{eq:aLLm}. 
${\cal A}_C^{<,{\rm sing}}  \vert_{LR}$ is obtained by flipping the sign of the axial terms in 
${\cal A}_C^{<,{\rm sing}}  \vert_{LL}$, and the sum of the $N\!N$ intermediate-state contributions is 
\begin{align}
2 \left(  {\cal A}_C^{<,{\rm sing}} \Big \vert_{LL}
 + {\cal A}_C^{<,{\rm sing}} \Big \vert_{LR} \right) 
 &= 2 \,  \frac{m_N^2}{(4 \pi)^2} \,    \int_0^{\Lambda} \  d |\spacevec{k}| \ a^{NN}_< ( | \spacevec{k} |) \,, 
 \notag\\
  a^{NN}_< ( | \spacevec{k} |)   &=  - 
\left(
2 g_V^2  (\spacevec{k}^2) +   \frac{\spacevec{k}^2 g_M^2 (\spacevec{k}^2)}{m_N^2}
\right)
 \,  
8 \,  {\cal I}_C^< ( |\spacevec{k}|)\,. 
\label{eq:ACNNm}
\end{align}

\begin{figure}[t]
\centering
\includegraphics[width=0.7\textwidth]{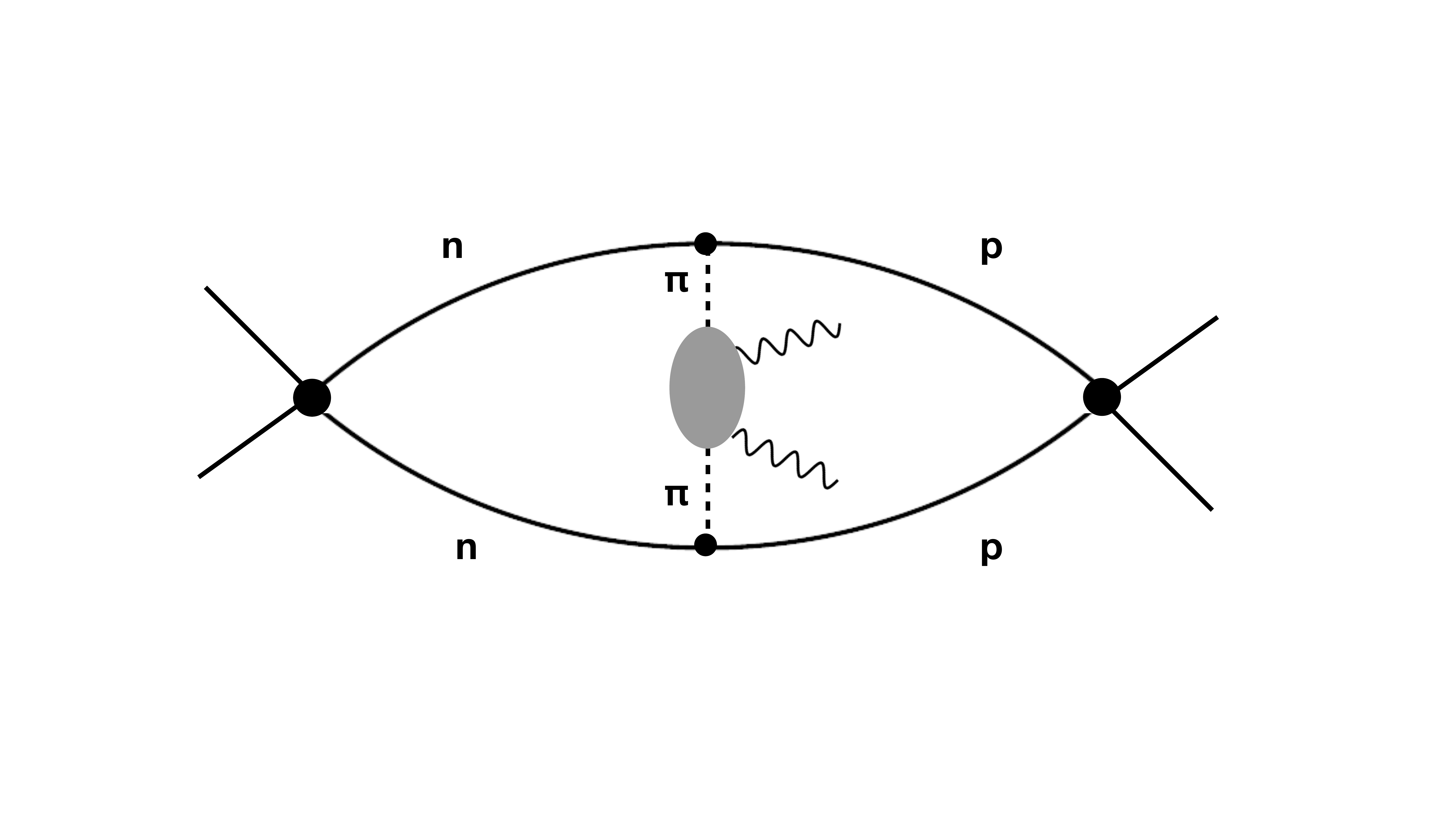}
\vspace{-1.5cm}
\caption{
Contribution to ${\cal A}_{VV}$  due to the insertion of two vector currents on a potential pion line.
In the small $|\spacevec{k}|$ region this diagram represents   ${\cal A}_C^{\pi \pi,<, {\rm sing}}$.
In the large $|\spacevec{k}|$ region it represents $\hat{\cal A}_C^{\rm sing}$. See main text for notation. }
\label{Fig:pipi}
\end{figure}

In the full theory analysis,  the counterpart of the EFT terms proportional to $Z$  (called $ {\cal A}_{VV}^{\pi \pi} $ 
in Eq.~\eqref{eq:MVEFT})  is due  to the insertion of  $\hat O^{VV} (\spacevec{k}) \equiv ( \hat O^{LL} + \hat O^{RR} +  2 \hat O^{LR}) (\spacevec{k})$ on a  pion   
exchanged between the two nucleons, as depicted in Fig.~\ref{Fig:pipi}. 
The  corresponding   ${\cal A}_C^{\pi \pi,<, {\rm sing}}$  is given by 
\be
{\cal A}_C^{\pi \pi,<, {\rm sing}}  = 
- m_N^2 \, \frac{g_A^2}{F_0^2}
\int \frac{d^4k}{(2 \pi)^4} \ \frac{1}{k^2 + i\epsilon}  \,  
\int \frac{d^3 \spacevec{q}}{(2 \pi)^3}  \
\frac{1}{\spacevec{q}^2}
    \,  {\cal I}_C^< ( |\spacevec{q}|) 
     \ {\cal T}_<^{\pi \pi} (k,q)\,.
    \label{eq:ACpipim}
\ee 
The integral in $d^3 \spacevec{q}$ is actually convergent.  
This can be seen by rewriting   ${\cal T}_<^{\pi \pi} (k,p)$ as follows: 
\begin{align}
{\cal T}_<^{\pi \pi} (k,q) &= 2 i  \bigg[  2 (d -4)  - F_+ (k,q) - F_-(k,q)  \bigg]   \times \left( F_\pi^V (k^2) \right)^2\,,\notag 
\\
F_\pm (k,q) &= \frac{4 M_\pi^2 - 3 k^2 \mp 4 k \cdot q}{(q \pm k)^2 - M_\pi^2 + i \epsilon}\,.
\end{align}
Note that terms involving $q^0 \sim |\spacevec{q}|^2/m_N$ are neglected at the order at which we work. 
Up to a  finite piece coming from the term proportional to $d-4$  in ${\cal T}_<^{\pi \pi}$,  
the integral  in Eq.~\eqref{eq:ACpipim} can be evaluated as follows:  first performing the integration in $k^0$ using the residues' theorem;
then performing the angular integral in $d^3 \spacevec{q}$;   finally  evaluating the integral in $d | \spacevec{q}|$.  
We will present our results in order of complexity:
\begin{enumerate}
\item First, we will take $F_\pi^V(k^2) =1$ and $ {\cal I}_C^< ( |\spacevec{q}|) = \theta( |\spacevec{q}| - 2 |\spacevec{p}|)/(8 |\spacevec{q}|)$. 
\item Next, we will use the realistic pion electromagnetic form factor  $F_\pi^V(k^2)$ from Eq.~\eqref{eq:pionFF}.
\item  
Finally, we will  include higher-momentum effects in the $N\!N$ interactions by writing 
\be {\cal I}_C^< ( |\spacevec{q}|) =  
\theta( |\spacevec{q}| - 2 |\spacevec{p}|)\frac{ r ( |\spacevec{q}|) }{8 |\spacevec{q}|}\,,\ee 
with $r ( |\spacevec{q}|) $ defined in Eq.~\eqref{eq:r0} and illustrated in Fig.~\ref{Fig:r}.  
\end{enumerate}
Throughout, we work in the chiral limit ($M_\pi \to 0$). 

The result is quite simple 
when using  $F_\pi^V(k^2) =1$    (corresponding to the limit $M_V \to \infty$)  and $ {\cal I}_C^< ( |\spacevec{q}|) = \theta( |\spacevec{q}| - 2 |\spacevec{p}|)/(8 |\spacevec{q}|)$:
\be
\label{eq:ACsing}
{\cal A}_C^{\pi \pi,<, {\rm sing}}  =
- \frac{m_N^2}{(4 \pi)^2}  \, 
\frac{16 g_A^2}{(4 \pi F_0)^2}  \ \int_0^\Lambda d |\spacevec{k}|  | \spacevec{k}| \left( 4 \log 2 - 2  + 3 \log \frac{|\spacevec{k}|}{2 |\spacevec{p}|} \right)
\,.
\ee
This expression can be recast into the following useful form using Eq.~\eqref{eq:Zmatch}  
\begin{align}
{\cal A}_C^{\pi \pi,<, {\rm sing}}   
&=  2 \,   \frac{m_N^2}{(4 \pi)^2}  \, \left[
- 4 g_A^2   Z^< \, \log \frac{\nu_\chi}{ {2 |\spacevec{p}|}}\ + \  \ \int_0^\Lambda d |\spacevec{k}|   \  a_<^{\pi \pi} (|\spacevec{k}|, \nu_\chi) 
\right]\,,
\notag
\\
a_<^{\pi \pi} (|\spacevec{k}|, \nu_\chi)  &=  - 8 g_A^2 \  \frac{ | \spacevec{k}|}{(4 \pi F_0)^2}  \left( 4 \log 2 - 2  + 3 \log \frac{|\spacevec{k}|}{ \nu_\chi } \right) 
\,, 
\label{eq:ACppm}
\end{align}
where we have extracted a factor of two for later convenience, have  introduced the arbitrary scale $\nu_\chi$, and isolated the dependence on the IR physics (external momentum $|\spacevec{p}|$), which 
must reproduce the same dependence in the EFT amplitude (cf.\ the term in Eq.~\eqref{eq:ACsingVV} proportional to $Z$). 

For a realistic pion form factor~\eqref{eq:pionFF},  the residues from the $\rho$ propagators need to be added, which changes the bracket in Eq.~\eqref{eq:ACsing} to 
\begin{align}
&4\log 2 -\frac{(\omega_V-|\spacevec{k}|)\big(2\omega_V^2-3|\spacevec{k}|\omega_V-|\spacevec{k}|^2\big)}{2\omega_V^3}+\frac{(\omega_V-|\spacevec{k}|)^2(2\omega_V+|\spacevec{k}|)}{2\omega_V^3}\bigg(3\log\frac{\omega_V+|\spacevec{k}|}{2|\spacevec{p}|}-1\bigg)\notag\\
&-3\log\bigg(1+\frac{\omega_V}{|\spacevec{k}|}\bigg)-\log\bigg(1+\frac{|\spacevec{k}|}{\omega_V}\bigg)\,.
\end{align}
In the limit $\omega_V\to\infty$ this reduces to Eq.~\eqref{eq:ACsing}, and for $|\spacevec{k}|\to\infty$ this expression displays the 
$1/|\spacevec{k}|^4$ fall-off that makes the integrand 
in Eq.~\eqref{eq:ACsing} consistent with the OPE behavior at large $|\kk|$. 
To extract the new form of the integrand, we first express $Z^<$, see Eq.~\eqref{eq:Zmatch}, as
\begin{align}
 Z^<=\frac{3}{16\pi^2 F_0^2}  \int_0^\Lambda d |\spacevec{k}|   |\spacevec{k}| \frac{(\omega_V-|\spacevec{k}|)^2(2\omega_V+|\spacevec{k}|)}{\omega_V^3}\,.
\end{align}
In this way, Eq.~\eqref{eq:ACppm} still applies upon the replacement
\begin{align}
\label{eq:apipi2}
 a_<^{\pi \pi} (|\spacevec{k}|, \nu_\chi)  
& =  - 4 g_A^2 \  \frac{ | \spacevec{k}|}{(4 \pi F_0)^2}  \Bigg\{ 8 \log 2 -  \frac{(\omega_V-|\spacevec{k}|)\big(2\omega_V^2-3|\spacevec{k}|\omega_V-|\spacevec{k}|^2\big)}{\omega_V^3}\notag\\
&+\frac{(\omega_V-|\spacevec{k}|)^2(2\omega_V+|\spacevec{k}|)}{\omega_V^3}\bigg(3\log\frac{\omega_V+|\spacevec{k}|}{\nu_\chi}-1\bigg)\notag\\
&-6\log\bigg(1+\frac{\omega_V}{|\spacevec{k}|}\bigg)-2\log\bigg(1+\frac{|\spacevec{k}|}{\omega_V}\bigg) \Bigg\}\,.
\end{align}

Finally, we include the HOS form factor effects by 
rewriting  in Eq.~\eqref{eq:ACpipim}
\be
 {\cal I}_C^< ( |\spacevec{q}|) =   \theta( |\spacevec{q}| - 2 |\spacevec{p}|) \times
 \left[  \frac{1  }{8 |\spacevec{q}| }+
  \frac{r ( |\spacevec{q}|)-1  }{8 |\spacevec{q}| }
  \right]\,.
\ee
The first term gives the analytic result in Eqs.~\eqref{eq:ACppm} and \eqref{eq:apipi2}.
The second term gives a correction to $a_<^{\pi \pi} (|\spacevec{k}|, \nu_\chi) $, 
which we denote by $\delta a_<^{\pi \pi} (|\spacevec{k}|)$, 
  \be
\label{eq:dapipi}
  \delta a_<^{\pi \pi} (|\spacevec{k}|) 
  =  4 g_A^2 \  \frac{ | \spacevec{k}|}{(4 \pi F_0)^2}  
\ \int_{2 |\spacevec{p}|}^\infty d |\spacevec{q}| \, \frac{1 - r (|\spacevec{q}|)}{|\spacevec{q}|} \, g (|\spacevec{k}|, |\spacevec{q}|) \,, 
  \ee
where  $ g (|\spacevec{k}|, |\spacevec{q}|) $ is given by
\begin{align}
 g (|\spacevec{k}|, |\spacevec{q}|)&=\begin{cases}
                                      g_- (|\spacevec{k}|, |\spacevec{q}|)\qquad \text{if} \qquad |\spacevec{q}|<|\spacevec{k}|\\
                                      g_+ (|\spacevec{k}|, |\spacevec{q}|)\qquad \text{if} \qquad |\spacevec{q}|>|\spacevec{k}|
                                     \end{cases},
                                     \qquad \omega=\omega_V=\sqrt{\kk^2+M_V^2}\,,\\
g_- (|\spacevec{k}|, |\spacevec{q}|)&=\frac{\omega-|\spacevec{k}|}{\omega^2}\bigg[\frac{(\omega+|\spacevec{k}|)^2(|\spacevec{k}|+3\omega)}{(\omega+|\spacevec{k}|)^2-|\spacevec{q}|^2}
-\frac{2(|\spacevec{k}|^2+2|\spacevec{k}|\omega-\omega^2)}{\omega}\bigg]\notag\\
&-\frac{2|\spacevec{q}|}{|\spacevec{k}|}\log\frac{2|\spacevec{k}|+|\spacevec{q}|}{2|\spacevec{k}|-|\spacevec{q}|}-\frac{|\spacevec{k}|^4+6|\spacevec{q}|^2\omega^2+\omega^4-2|\spacevec{k}|^2(|\spacevec{q}|^2+\omega^2)}{2|\spacevec{q}|\omega^3}\log\frac{\omega+|\spacevec{k}|-|\spacevec{q}|}{\omega+|\spacevec{k}|+|\spacevec{q}|}\,,\notag\\ 
g_+ (|\spacevec{k}|, |\spacevec{q}|)&=\frac{\omega-|\spacevec{k}|}{\omega^2}\bigg[
\frac{(|\spacevec{q}|-2\omega)(|\spacevec{k}|\omega+(|\spacevec{q}|+\omega)^2)}{(\omega+|\spacevec{q}|)^2-|\spacevec{k}|^2}-
\frac{\omega |\spacevec{q}|(|\spacevec{q}|-6\omega)+|\spacevec{k}|(\omega+|\spacevec{k}|)(2|\spacevec{q}|-\omega)}{\omega |\spacevec{q}|}\bigg]
\notag\\
&-\frac{2|\spacevec{q}|}{|\spacevec{k}|}\log\frac{2|\spacevec{k}|+|\spacevec{q}|}{|\spacevec{q}|}-\frac{|\spacevec{k}|^4+6|\spacevec{q}|^2\omega^2+\omega^4-2|\spacevec{k}|^2(|\spacevec{q}|^2+\omega^2)}{2|\spacevec{q}|\omega^3}\log\frac{\omega-|\spacevec{k}|+|\spacevec{q}|}{\omega+|\spacevec{k}|+|\spacevec{q}|}\,.\notag
\end{align} 
The integration in Eq.~\eqref{eq:dapipi} will be done numerically.

Putting together the $N\!N$ and $\pi \pi$ contributions, the full low- and intermediate-$|\spacevec{k}|$ 
integrand in  ${\cal A}_C^{<, \rm sing} \vert_{VV}$ is given by  
\be
  a^{VV}_< ( | \spacevec{k} |, \nu_\chi)  = 
  a^{NN}_< ( | \spacevec{k} |)   + a^{\pi \pi}_< ( | \spacevec{k} |, \nu_\chi)   + \delta a^{\pi \pi}_< ( | \spacevec{k} |) \,,
\label{eq:aVVm2}
\ee
where the input functions are given in Eqs.~\eqref{eq:ACNNm}, \eqref{eq:apipi2}, and \eqref{eq:dapipi}.

Concerning  ${\cal A}_{VV}^>$,  we use the OPE results  given in Section~\ref{sect:OPELL} 
 and  Appendix~\ref{sec:LROPE},  assembled  to produce the vector--vector  linear combination, as done in  Section~\ref{sect:pipi}.
This leads to
\be
{\cal A}_C^{>,{\rm sing}} \Big \vert_{VV}
 =   \frac{3 \alpha_s}{ \pi} \, \int_{\Lambda}^{\infty} \ d |\spacevec{k}| \ \frac{1}{|\spacevec{k}|^3}  \
\left( \frac{2 (g_1^{NN} +  g_{LR}^{NN}) (\nu_\chi) }{C^2}   + \tilde{\cal A}_C^{\rm sing} (\nu_\chi) + \hat{\cal A}_C^{\rm sing}  (\nu_\chi) \right)\,,
\ee
where $\tilde{\cal A}_C$ is defined in Section~\ref{sect:OPELL},  
$g_{LR}^{NN} (\nu_\chi)$ and 
$\hat{\cal A}_C$ are defined in Appendix~\ref{sec:LROPE}, and $\nu_\chi$ is the chiral EFT renormalization scale 
used in evaluating $\langle pp | \hat O^{VV} | nn \rangle$.  We chose to keep $\nu_\chi \neq \mu_\chi$ for clarity. 
However, to simplify the analysis of the IR divergences, we use here  the same scale $\nu_\chi$ introduced in Eq.~\eqref{eq:ACppm} to separate 
out the $\log|\pp|$ term from the low-$|\kk|$ amplitude ${\cal A}_C^{\pi \pi,<, {\rm sing}} $. 
Given the scaling of the $N\!N$ couplings and the potentials that determine  $\tilde{\cal A}_C$  and  $\hat{\cal A}_C$, 
discussed in Appendix~\ref{sec:LROPE}, 
in the above expression we  keep only the leading terms $g_{LR}^{NN}$ and $\hat{\cal A}_C$. 
The real part of the singular component of $\hat{\cal A}_C$ is generated by pion exchange and is given by 
\be
\hat {\cal A}_C^{\rm sing}  (\nu_\chi) 
=
-    \frac{m_N^2}{(4 \pi)^2}    \, 
  \frac{g_{LR}^{\pi \pi} \,  g_A^2 }{4}   \,
\left[   
\log \frac{\nu_\chi^2}{4 |{\bf p}|^2 }    
+ 1
\right]\,,
\label{eq:ACsingVVsd}
\ee
so that we can write\footnote{Here and below the dependence on the QCD short-distance renormalization scale  $\mu$ in $\alpha_s$, $g_{LR}^{\pi \pi}$, 
and $g_{LR}^{NN}$ is suppressed for simplicity. Note that, in contrast to $g_1^{NN}$, $g_{LR}^{NN}$ carries an additional dependence on the chiral EFT scale $\nu_\chi$.} 
\begin{align}
{\cal A}_C^{>,{\rm sing}} \Big \vert_{VV}
 &=   \frac{3 \alpha_s}{ \pi} \, \int_{\Lambda}^{\infty} \ d |\spacevec{k}| \ \frac{1}{|\spacevec{k}|^3}  \
\left( \frac{2   g_{LR}^{NN} (\nu_\chi) }{C^2}  -   \frac{m_N^2}{(4 \pi)^2}    \,   \frac{  g_{LR}^{\pi \pi} \,  g_A^2 }{4}   \, 
 \right)\notag \\
 &-     \frac{m_N^2}{(4 \pi)^2}   \,  8 g_A^2 \, 
 \left[ \frac{3  \, \alpha_s  \,  g_{LR}^{\pi \pi} }{16 \pi} \, \int_{\Lambda}^{\infty}   d |\spacevec{k}|  \, \frac{1}{|\spacevec{k}|^3} \right] 
 \, \log \frac{\nu_\chi}{2 | \spacevec{p}|}\,. 
\end{align}
In the second line of the above equation, the expression within square brackets is identified as  $Z^>$, cf.\ Eq.~\eqref{eq:Zmatch}.
Moreover, using the definitions~\eqref{eq:gLRNN}  we can further simplify this expression to obtain:
\begin{align}
{\cal A}_C^{>,{\rm sing}} \Big \vert_{VV}
 &=    2 \,  \frac{m_N^2}{(4 \pi)^2}  \,  \left[
-  4 g_A^2  \  Z^> \ \log \frac{\nu_\chi}{2 | \spacevec{p}|} 
+   \ \int_\Lambda^\infty d |\spacevec{k}|   \  a_>^{VV} (|\spacevec{k}|, \nu_\chi) 
 \right]\,,\notag
\\
a_>^{VV} (|\spacevec{k}|, \nu_\chi)  &= 
   \frac{3 \alpha_s}{ 2\pi} \,  \ \frac{(4 \pi F_\pi)^2}{|\spacevec{k}|^3}  \  
\left(2   \,  \bar g_{LR}^{NN} (\nu_\chi)    -     \frac{  \bar g_{LR}^{\pi \pi} \,  g_A^2 }{4}  
 \right)\,.
\label{eq:ACVVp}
\end{align}
Using the lattice QCD results of Ref.~\cite{Nicholson:2018mwc},
 we find $\bar g_{LR}^{\pi \pi}  = 8.23$ 
in the $\overline{\rm MS}$ scheme at $\mu = 2\GeV$. 
While both $\bar g_{LR}^{\pi \pi} $
 and  $\bar{g}_{LR}^{NN} $ are expected to be $\Order(1)$, 
given the large numerical value of $\bar g_{LR}^{\pi \pi} $,  for the nucleon coupling we will assume  $\bar{g}_{LR}^{NN} \in [-10,+10]$.

\subsection{Matching}

The matching condition for the vector amplitude ${\cal A}_{VV}$ leads to (compare to Eq.~\eqref{eq:match2} for the $LL$ amplitude) 
\be
{\cal A}_C^{< ,{\rm sing}}\Big \vert_{VV}  \  + \  {\cal A}_C^{>,{\rm sing}} \Big \vert_{VV} \  = \   {\cal A}_C^{\rm sing} (\mu_\chi)  \Big \vert_{VV} \ +  \ 
\frac{4 ( {\cal C}_1 (\mu_\chi) + {\cal C}_2 (\mu_\chi)) }{C^2}\,. 
\label{eq:match33}
\ee
The  input needed in Eq.~\eqref{eq:match33} can be found in Eqs.~\eqref{eq:ACsingVV},  \eqref{eq:ACVVm},  \eqref{eq:ACNNm}, 
\eqref{eq:ACppm},   \eqref{eq:apipi2}, and \eqref{eq:ACVVp}.
Rescaling the couplings and integrals as in Section~\ref{sect:matchLL} and using $Z = Z^< + Z^>$,  one arrives at 
\begin{align}
 2\,  \bigg( \tilde C_1 (\mu_\chi)   &+  \tilde C_2 (\mu_\chi) \bigg)
=  1 + 2 \, Z \, g_A^2  - 4 Z g_A^2  \, \log \frac{\nu_\chi}{\mu_\chi} 
-  \int_{0}^{\mu_\chi}   d |\spacevec{k}| \    a^{VV}_\chi ( | \spacevec{k} |)   
\nonumber  \\
&+ 
 \int_0^{\Lambda}   d |\spacevec{k}| 
\   a^{VV}_< ( | \spacevec{k} |, \nu_\chi)  
 + \int_{\Lambda}^\infty     d |\spacevec{k}| \  a^{VV}_> ( | \spacevec{k} |, \nu_\chi) \,, 
\label{eq:match4V}
\end{align}
with $a^{VV}_< ( | \spacevec{k} |, \nu_\chi)$  given 
in Eq.~\eqref{eq:aVVm2}. 
Note that the logarithmic IR dependence on the external momentum $|\pp|$ has disappeared 
in the matching relation, providing a strong consistency check on the calculation.  
Except for $\delta a_<^{\pi \pi} (|\spacevec{k}|) $, which is  given in Eq.~\eqref{eq:dapipi} and 
has to be obtained via numerical integration, 
the  quantities relevant to evaluate the matching 
condition~\eqref{eq:match4V} are
\begin{align}
a^{VV}_\chi ( | \spacevec{k} |)   &=  - 2  \ 
\frac{1}{ |{\bf k}|} \,  \theta(|{\bf k}| - 2 |{\bf p}|)\,,\notag
\\
a^{NN}_< ( | \spacevec{k} |)   &=  - 
\left(
2 g_V^2  (\spacevec{k}^2) +   \frac{\spacevec{k}^2 g_M^2 (\spacevec{k}^2)}{m_N^2}
\right)
 \,  
8 \,  {\cal I}_C^< ( |\spacevec{k}|) \,,\notag
\\
a_<^{\pi \pi} (|\spacevec{k}|, \nu_\chi)  &=  - 4 g_A^2 \  \frac{ | \spacevec{k}|}{(4 \pi F_0)^2}  \Bigg\{ 8 \log 2 -\frac{(\omega_V-|\spacevec{k}|)\big(2\omega_V^2-3|\spacevec{k}|\omega_V-|\spacevec{k}|^2\big)}{\omega_V^3}\notag\\
&+\frac{(\omega_V-|\spacevec{k}|)^2(2\omega_V+|\spacevec{k}|)}{\omega_V^3}\bigg(3\log\frac{\omega_V+|\spacevec{k}|}{\nu_\chi}-1\bigg)\notag\\
&-6\log\bigg(1+\frac{\omega_V}{|\spacevec{k}|}\bigg)-2\log\bigg(1+\frac{|\spacevec{k}|}{\omega_V}\bigg) \Bigg\}\,,  \notag
\\
a_>^{VV} (|\spacevec{k}|, \nu_\chi)  &= 
   \frac{3 \alpha_s}{ 2\pi} \,  \ \frac{1}{|\spacevec{k}|^3}  \
\left(2   \bar g_{LR}^{NN} (\nu_\chi)   -     \frac{  \bar g_{LR}^{\pi \pi} \,  g_A^2 }{4}  
 \right)\,. 
\end{align}

\begin{figure}[t]
\begin{center}
\includegraphics[width=0.45\linewidth]{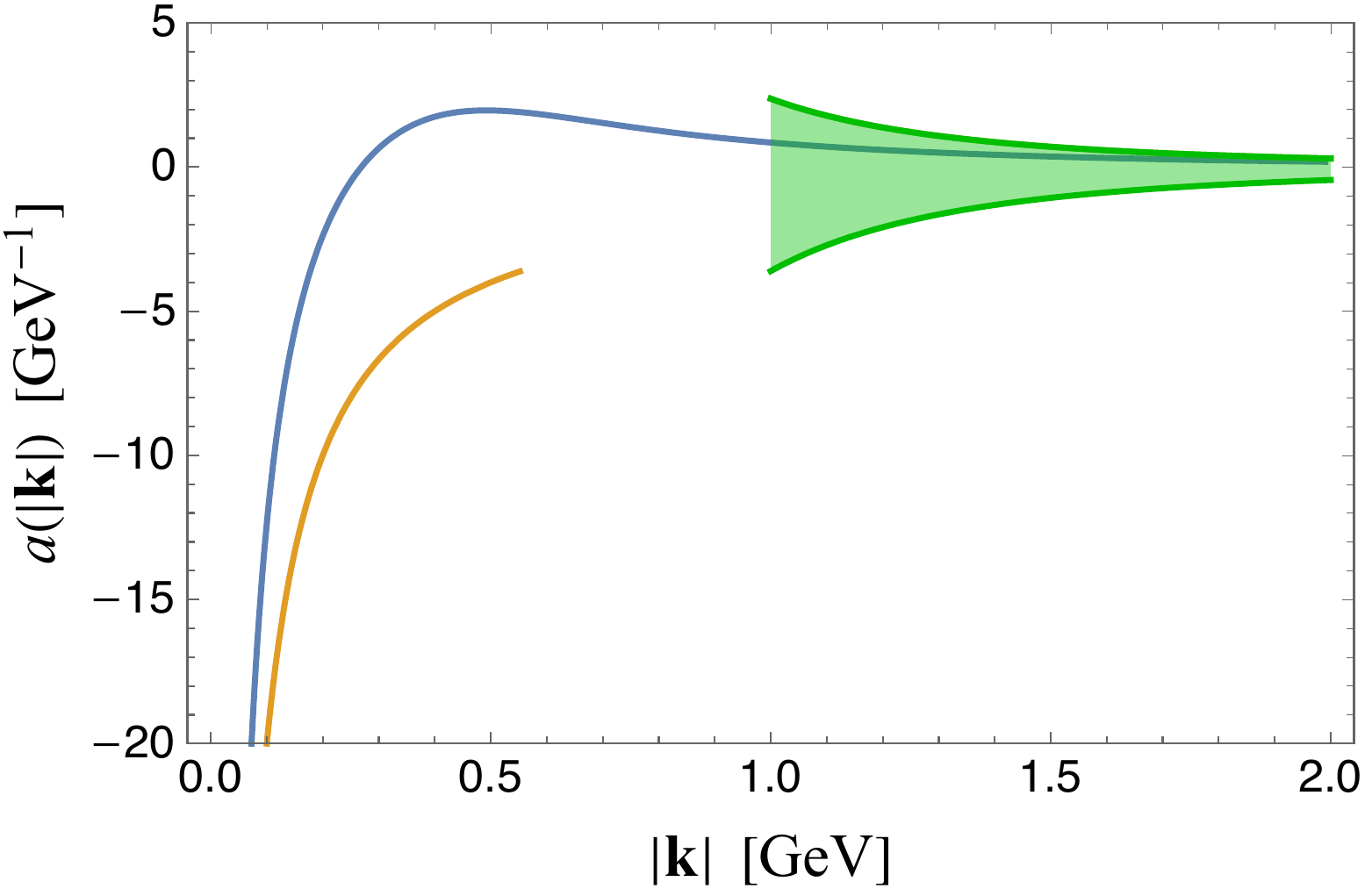}
\hspace{0.03\linewidth}
\includegraphics[width=0.45\linewidth]{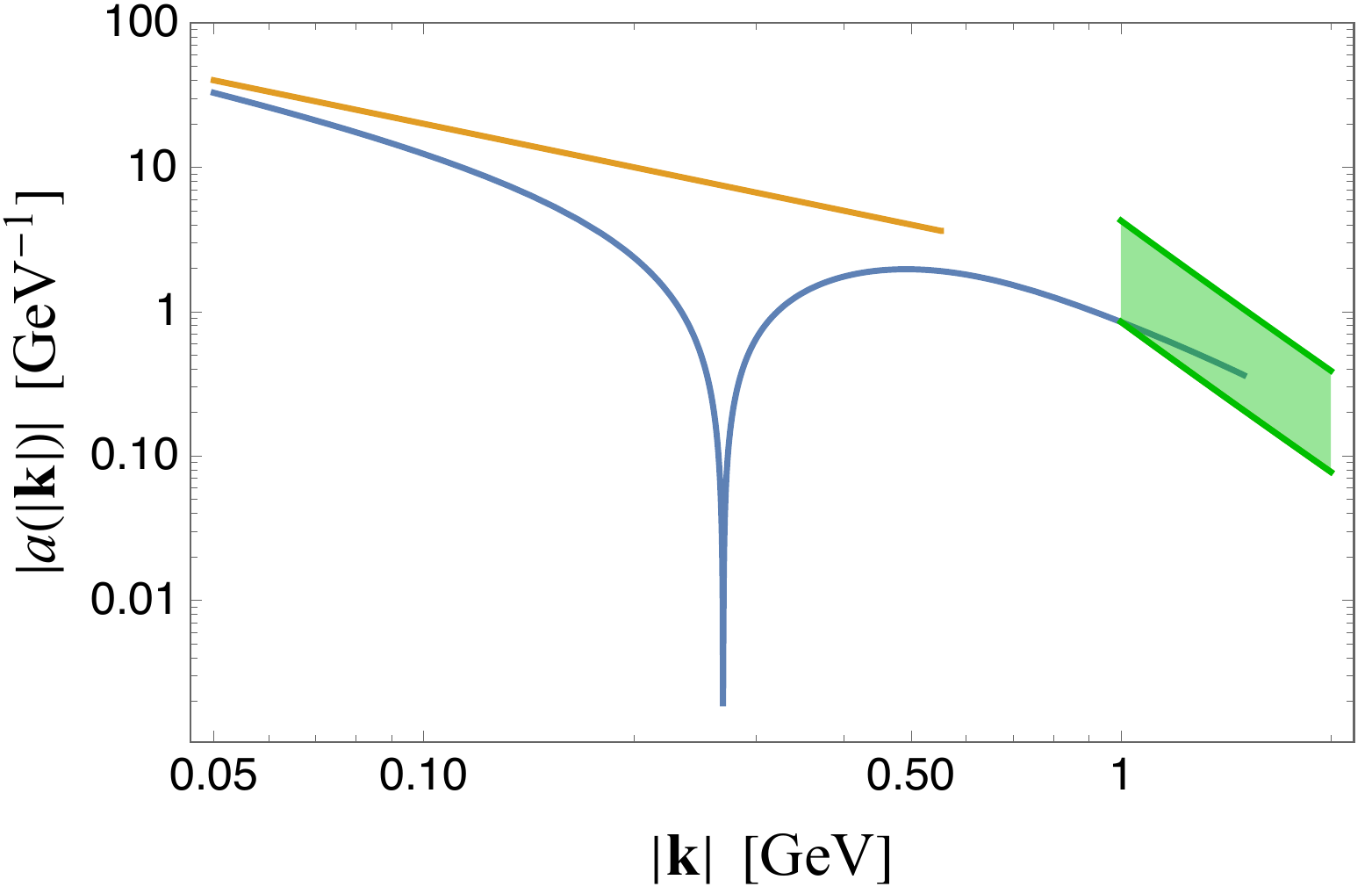}
\caption{ 
Left panel:  
$a^{VV}_\chi (|\spacevec{k}|)$  (yellow),
$a^{VV}_< ( | \spacevec{k} |,M_\pi) $  
(blue),  and  $a^{VV}_> (|\spacevec{k}|, M_\pi)$  (green) assuming   $\bar{g}_{LR}^{NN} \in [-10,10]$ (lower to upper green curves and shaded band).
Right panel:  same plot for the absolute values in logarithmic scale. 
\label{fig:integrandVVNN}
}
\end{center}
\end{figure}

\begin{figure}[t]
\centering
\includegraphics[width=0.5\textwidth]{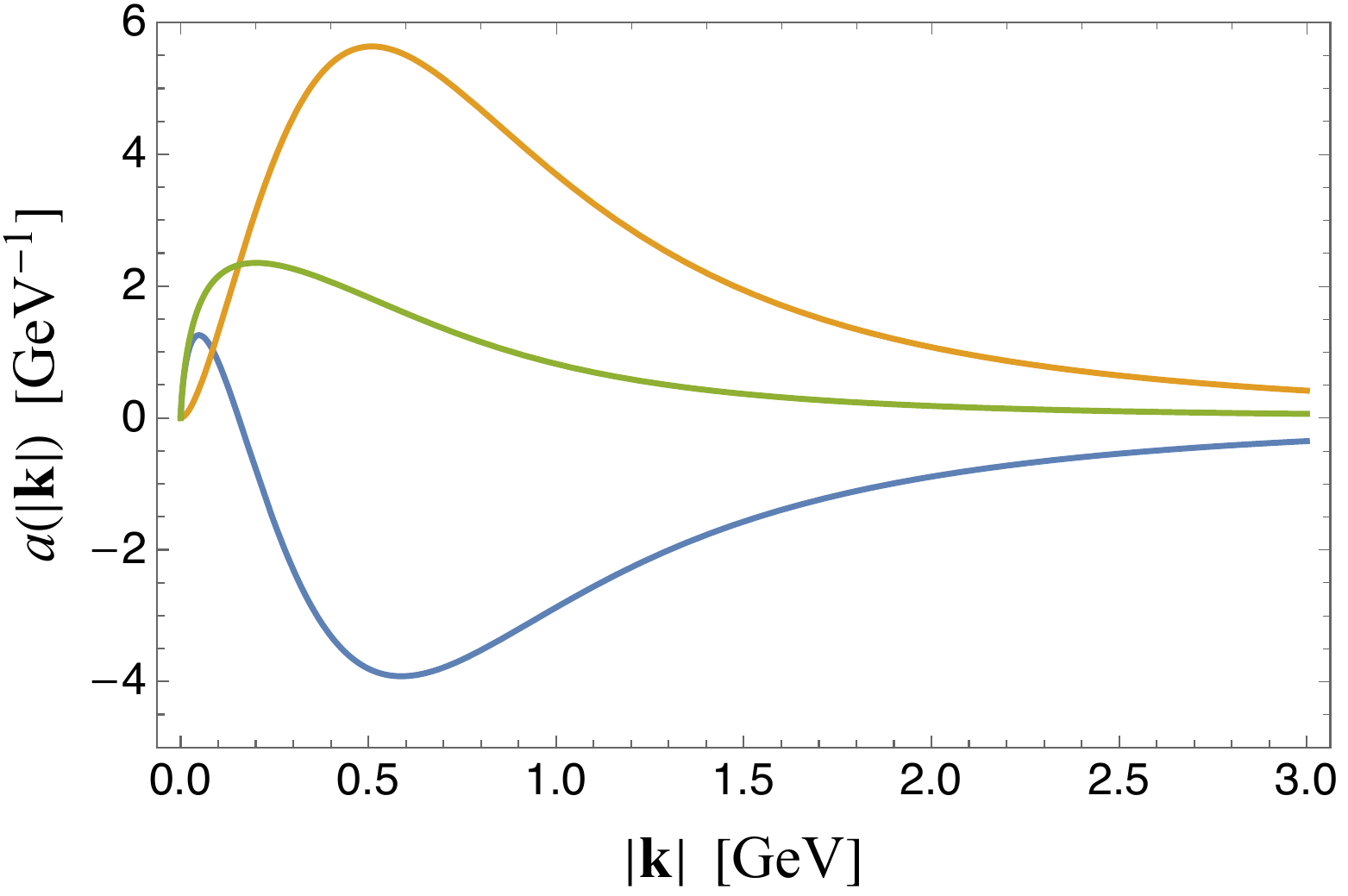}
\caption{
$a^{\pi \pi}_< (|\spacevec{k}|,M_\pi)$  (blue), 
$\delta a^{\pi \pi}_< (|\spacevec{k}|)$  (yellow), 
and their sum (green).
}
\label{Fig:integrandpi}
\end{figure}

\begin{figure}[t]
\begin{center}
\includegraphics[width=0.45\linewidth]{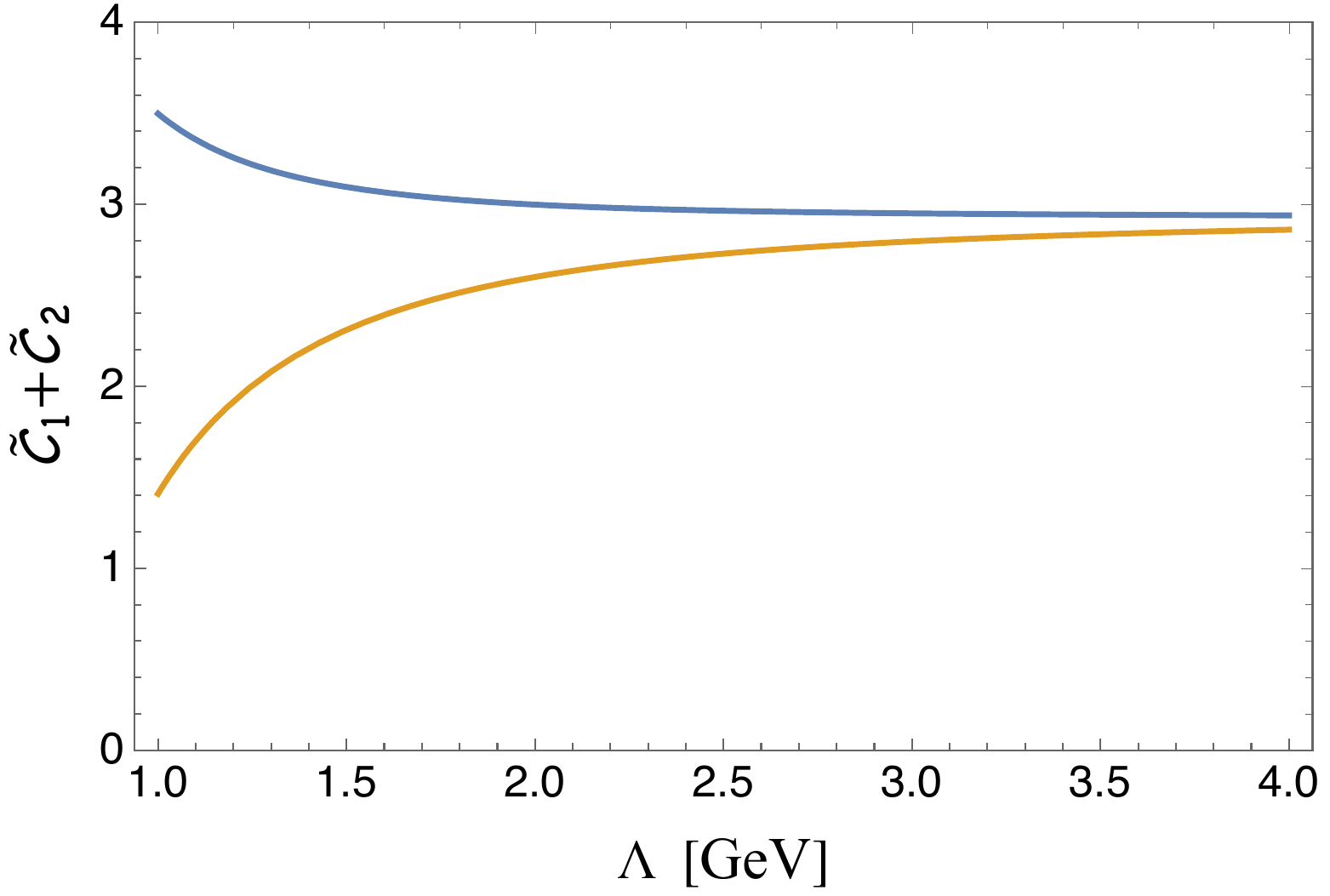}
\hspace{0.03\linewidth}
\includegraphics[width=0.45\linewidth]{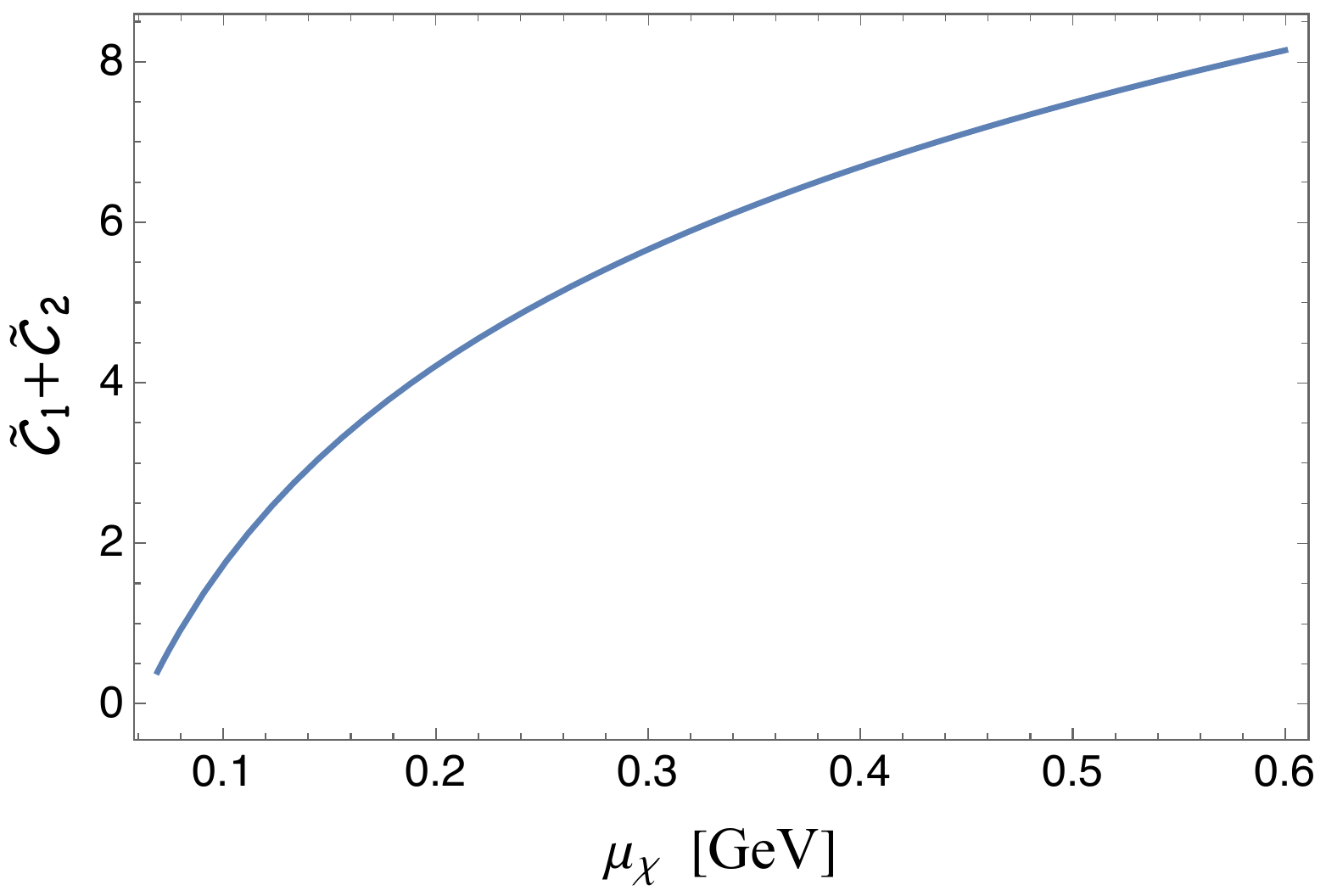}
\caption{ 
Left panel:  dependence of $(\tilde{\cal C}_1 + \tilde{\cal C}_2) (\mu_\chi =M_\pi)$  on the  matching scale $\Lambda$ using 
$\bar  g_{LR}^{NN}  =  \pm 10$. 
Right panel:  dependence of $\tilde{\cal C}_1+\tilde{\cal C}_2$ on the chiral renormalization scale $\mu_\chi$. 
\label{fig:CVV}
}
\end{center}
\end{figure}

Without loss of generality, 
in the numerical results we set $\nu_\chi = M_\pi$. 
The various components of the integrand are shown in 
Figs.~\ref{fig:integrandVVNN}   and \ref{Fig:integrandpi}.
Fig.~\ref{fig:integrandVVNN}   displays the chiral,  low-  and high-$|\spacevec{k}|$ components of the integrand, 
namely $a^{VV}_\chi (|\spacevec{k}|)$ (yellow), 
$  a^{VV}_< ( | \spacevec{k} |,M_\pi) $  
   (blue),   and  $a^{VV}_> (|\spacevec{k}|, M_\pi)$, 
 in both linear scale  (left panel) and logarithmic scale (right panel).
Fig.~\ref{Fig:integrandpi}  shows   $a^{\pi\pi}_< (|\spacevec{k}|)$  (blue), 
 $\delta a^{\pi\pi}_< (|\spacevec{k}|)$  (yellow),  and their sum in green. 
 Fig.~\ref{fig:CVV}   shows the dependence of  $(\tilde{\cal C}_1 + \tilde{\cal C}_2) $ 
 on the matching scale $\Lambda$ (left panel) and the chiral 
 renormalization scale $\mu_\chi$ (right panel).
  
We now discuss the uncertainties associated with our estimate of  $ (\tilde{\cal C}_1 + \tilde{\cal C}_2)$: 
(i) Concerning the parametric uncertainties,  the dependence on  the vector form factor parameterization is mild: we find an upward shift of $\pm 0.03$ 
when using the Arrington--Sick fit~\cite{Arrington:2006hm}, compared to our baseline dipole parameterization. 
On the other hand, the  uncertainty due to the choice of  $\Lambda \in [2\GeV, 4\GeV]$, coupled to the 
range $\bar g_{LR}^{NN} | \in [-10,+10]$ (motivated by the large value of the corresponding two-pion matrix element,  see
Appendix~\ref{sec:LROPE}),  is  at the level of  $\pm 0.3$, larger than in the case of $\tilde{\cal C}_1$, see Fig.~\ref{fig:CVV} (left panel). 
(ii) To estimate  the systematic uncertainty  due to the choice of the short-range $N\!N$ interaction $V_S$, 
we repeat the analysis using the Reid  and  AV18 potentials,  finding shifts in $(\tilde{\cal C}_1 + \tilde{\cal C}_2) $ 
 of $\approx+0.25$ and $\approx -0.3$, respectively.  
(iii)  For the systematic effect due to  the neglected inelastic channels we take a range
 $\delta (\tilde{\cal C}_1 + \tilde{\cal C}_2)\approx \pm 1.1$, which   leads to 
 a relative error of about 50\% in the singular $N\!N$ electromagnetic amplitude 
at $|\pp| \sim 25\MeV$,  larger than the 30\% in the weak amplitude controlled by $\tilde{\cal C}_1$. 
In addition to the new class of pion-exchange diagrams, this larger assignment is motivated as follows: in contrast to the $\tilde{\cal C}_1$ analysis, 
the parametric error  is now more sizable, mainly driven by the $N\!N$ short-distance coupling $\bar g_{LR}^{NN}$. In fact,  
reducing $\Lambda$ to values as low as $1\GeV$ and thus into the energy region where the applicability of the OPE becomes questionable and inelastic effects important, leads to a variation  $\delta (\tilde{\cal C}_1 + \tilde{\cal C}_2) \simeq \pm 1.0$. Since this effect concerns the intermediate-momentum region which is most uncertain in our analysis, the resulting variation could be either booked as an uncertainty obtained by extrapolating the OPE expression beyond its region of validity, or in terms of neglected intermediate states. We prefer to keep the OPE scale $\Lambda\gtrsim 2\GeV$, and thus increase the estimate of inelastic contributions accordingly to account for the more prominent effect of the short-distance coupling for $\tilde{\cal C}_1 + \tilde{\cal C}_2$.

Altogether, 
our final result is 
 \be
(\tilde{\cal C}_1 + \tilde{\cal C}_2)  (\mu_\chi =M_\pi) \simeq   2.9  (1.1)_{\rm inel}(0.3)_{V_S}(0.3)_{\rm par}
= 2.9(1.2)\,.
\label{C1C2}
\ee
Concerning the central value,    the pion contributions 
($2 Z g_A^2$ plus the  integral of   $a^{\pi \pi}_< (|\spacevec{k}|,M_\pi )+  \delta a^{\pi \pi}_< (|\spacevec{k}|)$) 
amount to  $+2.4$.   Moreover, the most uncertain intermediate momentum region 
$|\spacevec{k}| \in [0.4, 1.5]\GeV$ contributes 
$\Delta  ( \tilde{\cal C}_1  + \tilde{\cal C}_2)  \simeq 0.55$,  safely below our  uncertainty estimate. 
At  the renormalization scale $\mu=4M_\pi$ our result becomes 
 \be
(\tilde{\cal C}_1 + \tilde{\cal C}_2)  (\mu_\chi =4 M_\pi)   = 7.8(1.2)\,.
\label{C1C2v2}
\ee

\subsection[Charge-independence-breaking contribution to $N\!N$ scattering]{Charge-independence-breaking contribution to $\boldsymbol{N\!N}$ scattering}

The result~\eqref{C1C2} already compares quite well to the phenomenological determination $(\tilde{\cal C}_1 + \tilde{\cal C}_2) (\mu_\chi =\mpi)=5.0$ from Ref.~\cite{Cirigliano:2019vdj}. However, since the contact term is scale and scheme dependent,
 it is more appropriate to compare directly observables calculated based on Eq.~\eqref{C1C2}. 
We therefore focus on the CIB contribution to $N\!N$ scattering.   
We summarize here our main findings and report  
technical details of the analysis in Appendix~\ref{sect:AppCIB}.
 
We first note that within LO chiral EFT the scattering lengths $a_{nn}$, $a_{np}$, and $a_{pp}^C$ (the latter defined in the modified effective range expansion to account for Coulomb effects~\cite{Bethe:1949yr,Jackson:1950zz,Kong:1999sf}) can be mapped onto contact terms for each channel
\be
\tilde C_{np} =  \tilde C + \frac{e^2}{3} \left( \tilde {\cal C}_1 + \tilde {\cal C}_2 \right)\,,\qquad 
\tilde C_{nn/pp} = \tilde C - \frac{e^2}{6} \left( \tilde{\cal C}_1 + \tilde {\cal C}_2 \right)  \pm \frac{1}{2}  \tilde {\cal C}_\text{CSB}\,,
\ee
where $\tilde C$ denotes the isospin-symmetric combination, $\tilde {\cal C}_1 + \tilde {\cal C}_2$ the CIB contribution, and $\tilde {\cal C}_\text{CSB}$ a charge-symmetry-breaking (CSB) term~\cite{Epelbaum:1999zn}. To test our prediction for $\tilde {\cal C}_1 + \tilde {\cal C}_2$, we can thus use two observables to determine $\tilde C$ and $\tilde {\cal C}_\text{CSB}$, and then predict the third based on Eq.~\eqref{C1C2}. 
Several choices are possible (see Appendix~\ref{sect:AppCIB}), of which arguably the simplest is
\be
\label{aCIB}
a_\text{CIB}=\frac{a_{nn}+a_{pp}^C}{2}-a_{np}=10.4(2) \fm\,.
\ee
This combination of scattering lengths 
would isolate the CIB contribution if $N\!N$ scattering were perturbative and Coulomb interactions absent, but in practice depends on all couplings in a complicated manner. 
To obtain the numerical result  we have used the empirical values $a_{pp}^C=-7.817(4)\fm$~\cite{Bergervoet:1988zz,Reinert:2017usi}, $a_{np}=-23.74(2)\fm$~\cite{Klarsfeld:1984es,Machleidt:2000ge}, $a_{nn}=-18.9(4)\fm$~\cite{Chen:2008zzj}. From Eq.~\eqref{C1C2} we find $a_\text{CIB}=15.9^{+4.5}_{-4.0}\fm$, in good agreement with Eq.~\eqref{aCIB}, given that additional uncertainties from higher chiral orders could be attached. The comparison to the phenomenology of CIB in $N\!N$ scattering thus validates our approach at the level of (30--50)\% and shows that our uncertainty estimates are realistic.

\section{Synthetic data for $\boldsymbol{nn \to pp}$ near threshold}
\label{sect:synthetic}

Having determined the LEC  $\tilde {\cal C}_1 $  in the \MS scheme, we can now compute the low-energy $nn\rightarrow ppe^-e^-$ amplitude $\A_\nu$. 
Our matching strategy was based on dimensional regularization with minimal subtraction as it provides convenient and factorized expressions for the amplitude.
While dimensional regularization is rarely used by nuclear practitioners, this is no obstacle to applying our results in nuclear-structure calculations. Observables, such as $\A_\nu$, are scheme independent so that the LNV contact 
term can by obtained in any scheme through a fit to our synthetic data for the amplitude 
\begin{equation}
\mathcal A_\nu(|\spacevec{p}|,|\spacevec{p}^\prime|) = 
-\langle \Psi_{pp}(|\spacevec{p}^\prime|) | \ V_{\nu\, \rm L}^{{}^1S_0}  + V_{\nu\, \rm S}^{{}^1S_0}  \ 
| \Psi_{nn}(|\spacevec{p}|) \rangle\,,
\label{Anu}
\end{equation}
where both initial $|\Psi_{nn}(|\spacevec{p}|)\rangle$ and 
final $|\Psi_{pp}(|\spacevec{p}^\prime|)\rangle$ states are in the ${}^1S_0$ channel, 
 $V_{\nu\, \rm L}^{{}^1S_0}$ denotes the  usual long-range neutrino potential, and 
 $V_{\nu\, \rm S}^{{}^1S_0}$ is  the short-range interaction proportional to $\tilde{\cal C}_{1}$.
We denote by $E= \spacevec p^2/m_n$ and $E' = \spacevec p'^2/m_p$ the 
center-of-mass energies of the incoming neutrons and outgoing protons
of masses $m_n$ and $m_p$, respectively, and by
$\spacevec p$ and $\spacevec p'$ the corresponding relative momenta. 
Assuming the outgoing electrons to be at rest one can determine the maximum momentum carried by 
the outgoing protons,  given the incoming momentum of the neutrons $|\spacevec{p}|$,  via 
\begin{equation}
E' = E + 2(m_n-m_p - m_e)\,, \qquad |\spacevec p^\prime |  = 
\sqrt{\spacevec p^2 + 2 m_N (m_n - m_p - m_e)}\,,
\label{kin}
\end{equation}
with $m_e$ the electron mass and $2 m_N = m_n + m_p$.
If the electrons are not at rest but carry total zero momentum in the incoming neutrons' rest frame, 
the outgoing protons fly back-to-back with momentum  $|\spacevec{p}^\prime|$ ranging between zero 
and the maximum value given in Eq.~\eqref{kin}. 

In Ref.~\cite{Cirigliano:2020dmx}
we chose the momenta  $|\spacevec{p}|=25\MeV$ and  $|\spacevec{p}^\prime|=30\MeV$, which proved an advantageous kinematic point for which   
the LO chiral amplitude provides an accurate prediction, 
while staying away form the sizable isospin-breaking effects at threshold. 
In practice we compute the amplitude in coordinate space\footnote{The relation between 
${\cal C}_1$ and $\tilde{\cal C}_1$ is given in Eq.~\eqref{eq:match3}.}
\begin{equation}
\mathcal A_\nu (|\spacevec{p}|, |\spacevec{p}^\prime|) 
= - \int d^3 \spacevec r \,  
\psi^{-\, *}_{\spacevec p^\prime}(\spacevec r)  
\left(V^{{}^1S_0} _{\nu\,\rm L}(\spacevec r) - 2 \,  {\cal C}_1   \, \delta^{(3)} (\spacevec r) 
\right)
\psi^+_{\spacevec p}(\spacevec r)\,, 
\label{ALNV}
\end{equation}
with the long-range neutrino potential given by 
\begin{equation}
V^{{}^1S_0} _{\nu {\rm L}} (r) = \frac{1 + 2 g_A^2}{4\pi r}   + \frac{g_A^2}{4\pi r} 
\left[ 1 -  e^{- M_\pi r}  \left( 1 + \frac{M_\pi r}{2}       \right) \right]\,,
\label{FGTT}
\end{equation}
and  wave functions for the scattering states 
$\psi^+_{\spacevec{p}} (\spacevec{r})$ obtained by solving the Schr\"odinger equation with  LO isospin-symmetric chiral potential  
\be
V_{^1S_0} (\spacevec{r}) =   - \alpha_\pi \frac{e^{- M_\pi r}}{r} + C \, \delta^{(3)} (\spacevec{r})\,, 
\qquad  \alpha_\pi = \frac{g_A^2 M_\pi^2}{4 \pi F_\pi^2}\,, \qquad M_\pi = \frac{M_{\pi^0} + 2 M_{\pi^\pm}}{3}\,, 
\ee
with  $C$ tuned to reproduce 
either the $np$ or $nn$ scattering length (the difference being negligible at the chosen kinematic point). 
To solve for the wave functions  we have used the methods described in Ref.~\cite{Kaplan:1996xu}, and 
to compute the amplitude, isolating its singular component, we have  used the approach described  in detail in Ref.~\cite{Cirigliano:2019vdj}.

Our amplitude  satisfies Watson's theorem~\cite{Watson:1954uc} and at 
 the kinematic point  $|\spacevec{p} |=25\MeV$,  $|\spacevec{p}^\prime |=30\MeV$ 
we find\footnote{The amplitude $\mathcal{A}_{\nu}$ is related to the $S$-matrix element 
for the process 
$n( \pp) \ n (- \pp) \to p ( \pp^\prime) \ p (-  \pp^\prime)  \ e (   \pp_{e}) \ e ( -\pp_{e} ) $
by  
$S_{\nu} = i  (2 \pi)^4 \, \delta^{(4)}(p_f - p_i) \,  ( 4  G_F^2 V_{ud}^2  m_{\beta \beta} \   \bar{u}_L   (\pp_e)  u_L^c (-\pp_e)  )    \, {\mathcal A}_{\nu}$\,. 
} 
\begin{align}
 \A_\nu (|\spacevec{p}|, |\spacevec{p}^\prime|)  \times 
 e^{-i (\delta_{^1S_0}(|\spacevec{p}|) + \delta_{^1S_0}( |\spacevec{p}^\prime|))}
&=   - \left( 2.271 - 0.075  \ \tilde{\cal C}_1 (4\, M_\pi) \right)  \times 10^{-2} \MeV^{-2}\notag
\\
&=   - 1.95(5)_{\tilde{\cal C}_1}  \times 10^{-2} \MeV^{-2}\,,
\label{eq:sdata}
\end{align}
where in the second line we have used our prediction   
$\tilde{\cal C}_1 (4 M_\pi) = 4.2(6)$. 
At the chosen kinematic point and chiral renormalization scale 
we find that the contact-term contribution interferes destructively with the long-range neutrino exchange 
and reduces the amplitude by about $15\%$. Since, 
as discussed in Refs.~\cite{Cirigliano:2018hja,Cirigliano:2019vdj}, 
the effect of a contact term of natural size that affects $\Delta I=0$ transitions such as $nn \to pp$ at the $(10\text{--}20)\%$ level, becomes amplified to the level of 
$(50\text{--}70)\%$ in  $\Delta I=2$ nuclear transitions due to a node in the matrix element density, it is possible that a more pronounced effect occurs in nuclei of interest for $0\nu\beta\beta$ searches, but we stress that our observation is based on the \MS scheme at $\mu=4M_\pi$ and thus need not carry over to other scales and schemes.

\begin{figure}[t]
\includegraphics[width=0.49\textwidth]{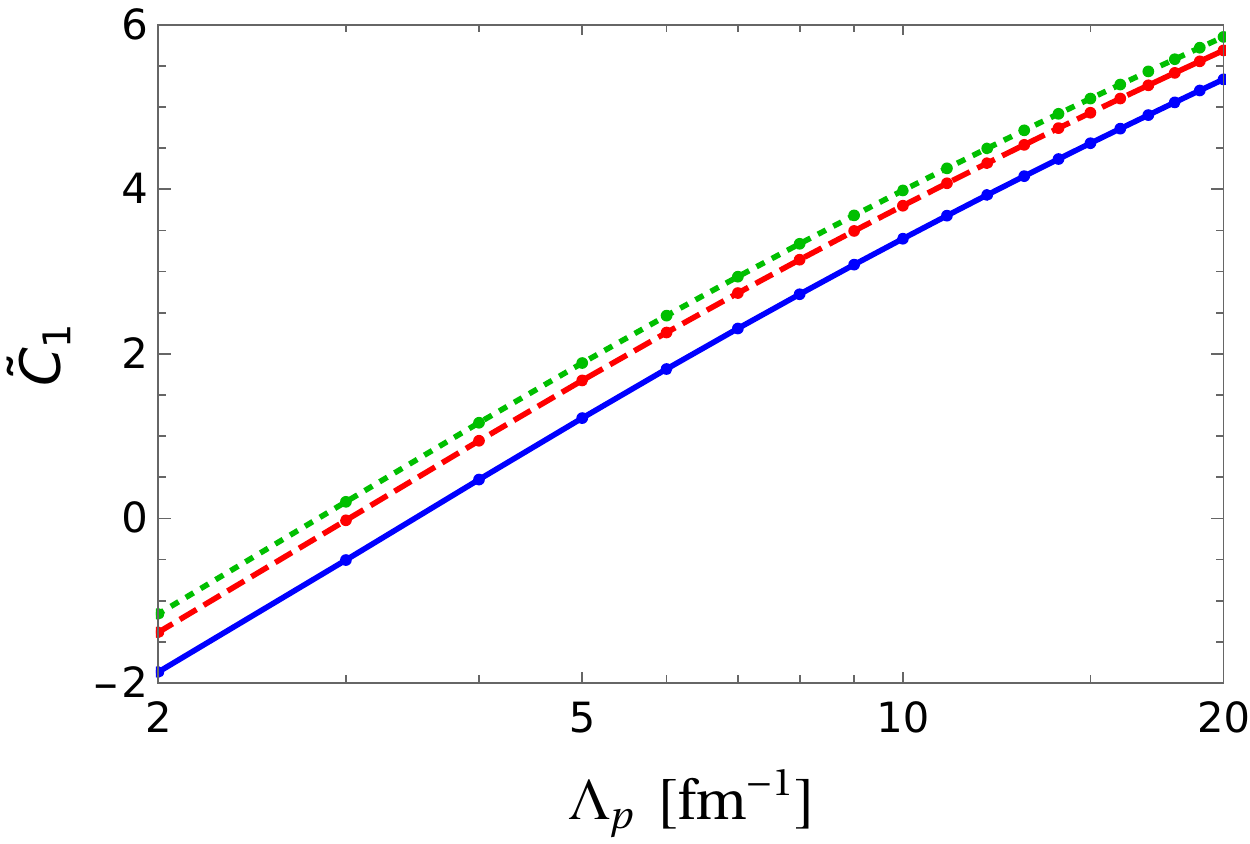}
\includegraphics[width=0.49\textwidth]{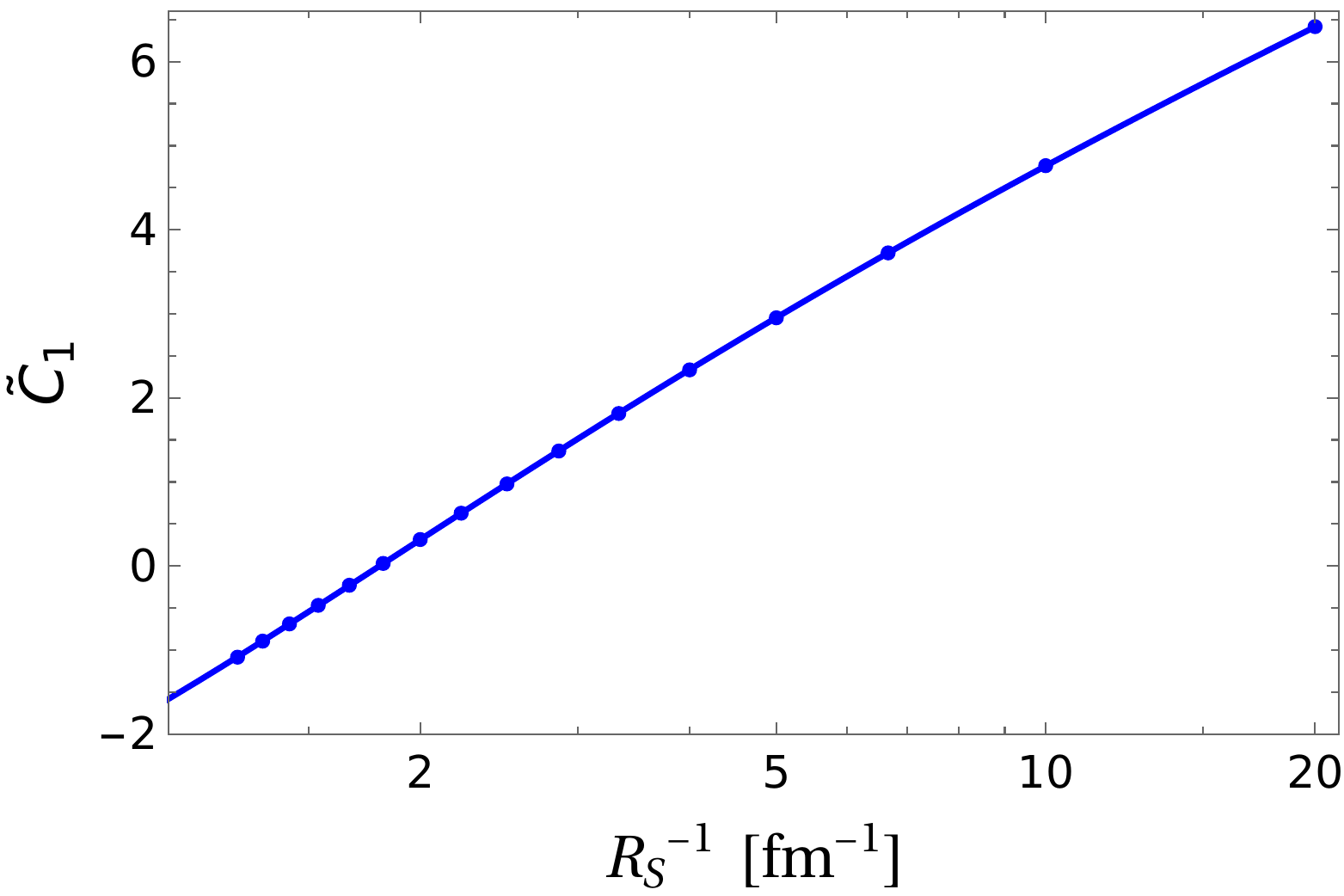}
\caption{Left: values of $\tilde{\cal  C}_1(\Lambda_p,n)$ as function of $\Lambda_p$ for 3 choices of the regulator $n=2$ (solid), $n=3$ (dashed), and $n=4$ (dotted).
Right: values of $\tilde {\cal C}_1(R_S)$ as function of the coordinate-space regulator $R_S$.
}
\label{gnuNN}
\end{figure}

We advocate using the synthetic data   \eqref{eq:sdata}  
to determine the LEC $\tilde {\cal C}_1$  in any  regularization/renormalization scheme 
applicable in many-body nuclear calculations~\cite{Epelbaum:2008ga,Machleidt:2011zz,Hammer:2012id,Hammer:2019poc,Hebeler:2020ocj}. As an example, we show here how to perform such a determination for 
often-applied momentum- and coordinate-space regulators. In momentum space, the strong potential is regulated through an exponential cutoff of the form
\begin{equation}
V_{\mathrm{strong}}(\pp^\prime, \pp) \rightarrow 
\mathrm{exp}\left[-\left(\frac{\pp^{\prime\,2}}{\Lambda_p^2}\right)^n\right]\, 
V_{\mathrm{strong}}(\pp^\prime, \pp)\, 
\mathrm{exp}\left[-\left(\frac{\pp^{2}}{\Lambda_p^2} \right)^n\right]\,,
\label{Lambda1b}
\end{equation}  
in terms of a momentum cutoff $\Lambda_p$. The LS equation is solved numerically for different values of $\Lambda_p$ and $n$ and the strong counter term, $C(\Lambda_p,n)$ in the ${}^1S_0$ channel, is determined from a fit to the $N\!N$ scattering lengths. We then insert the long- and short-range neutrino potential between initial- and final-state scattering states and determine $\tilde C_1(\Lambda_p,n)$ by fitting to Eq.~\eqref{eq:sdata}.
An analogous procedure is followed in coordinate space. In this case, contact interactions are regulated by a local Gaussian regulator
\begin{equation}
\delta^{(3)}(\spacevec r) \rightarrow \frac{1}{\left(\sqrt{\pi} R_S\right)^3} \exp \left(-\frac{r^2}{R_S^2}\right).
\end{equation}
The $pp$ and $nn$ wave functions obtained for a fixed value of $R_S$ are then used to compute the amplitude  \eqref{ALNV}.

\begin{figure}[t]
\includegraphics[width=0.49\textwidth]{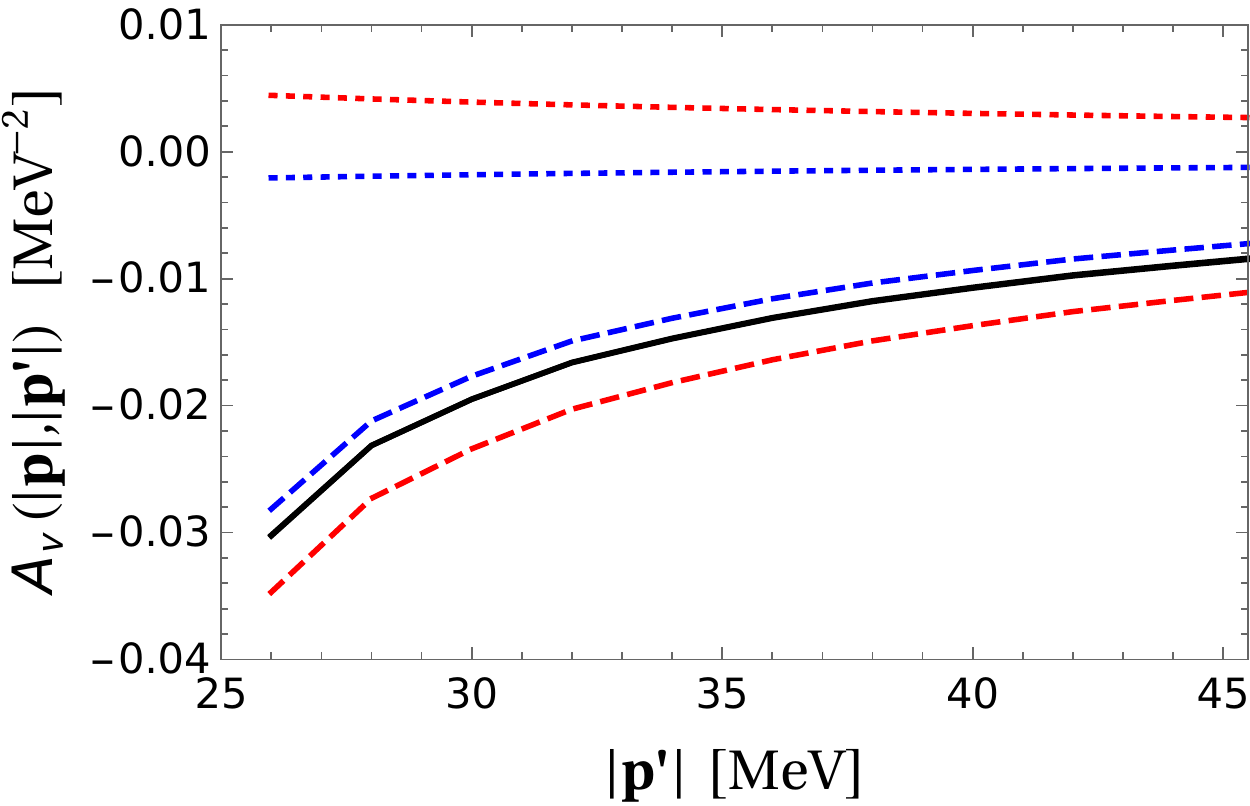}
\includegraphics[width=0.49\textwidth]{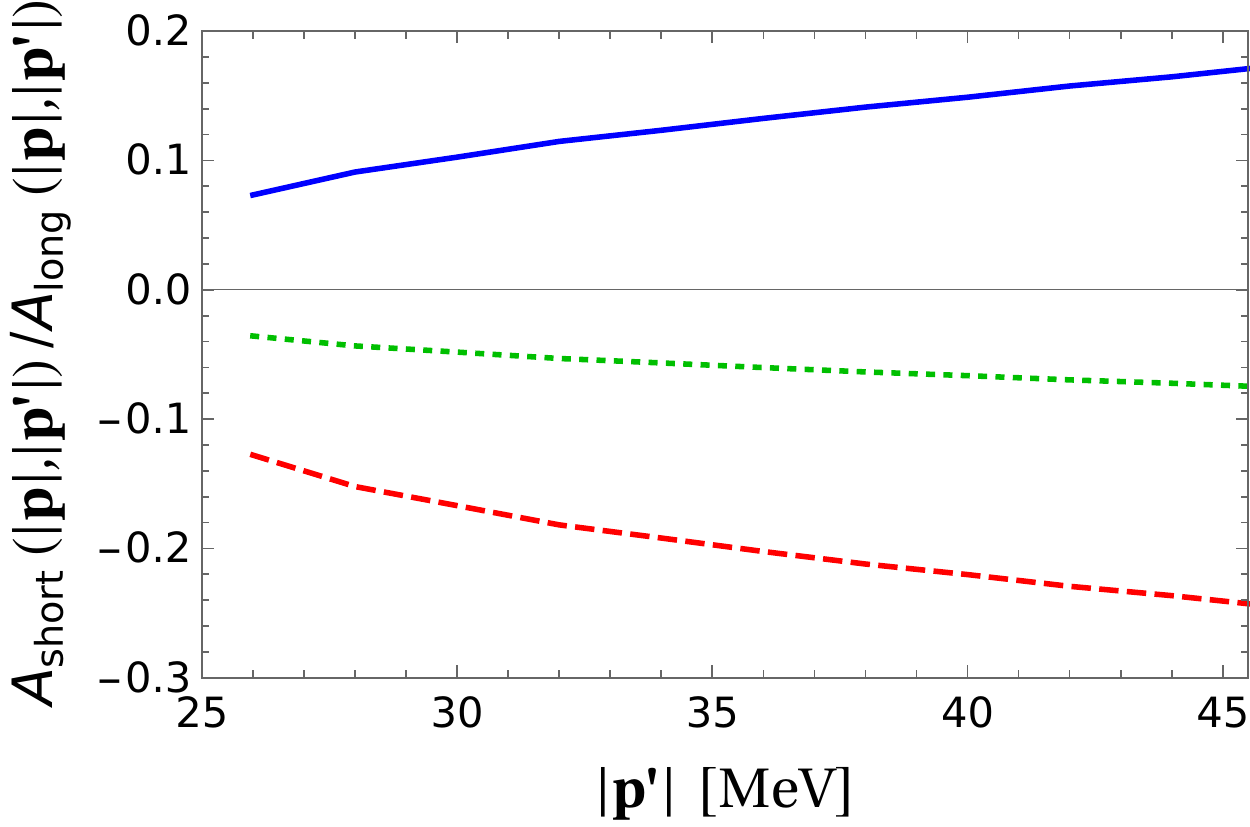}
\includegraphics[width=0.49\textwidth]{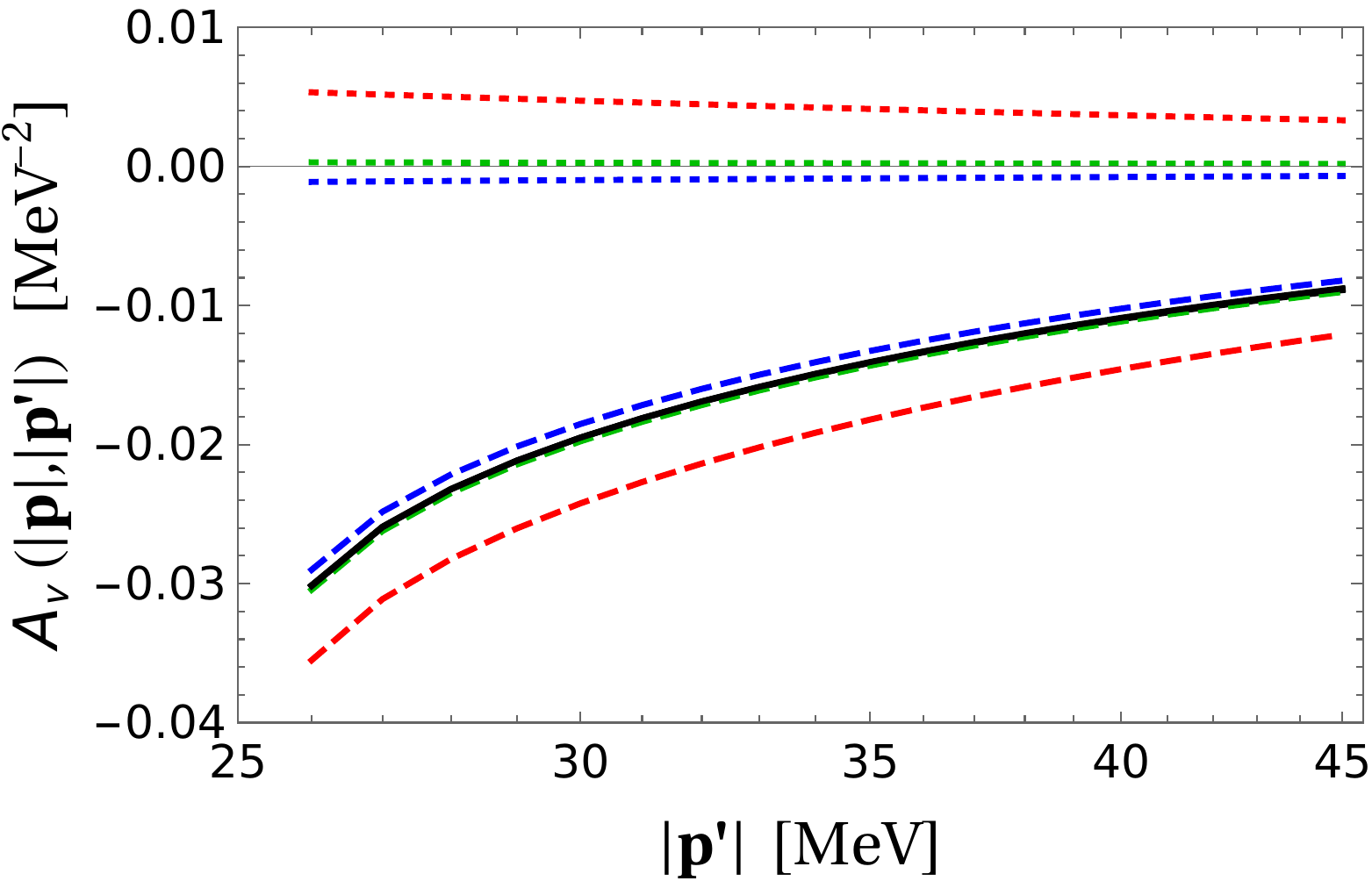}
\includegraphics[width=0.49\textwidth]{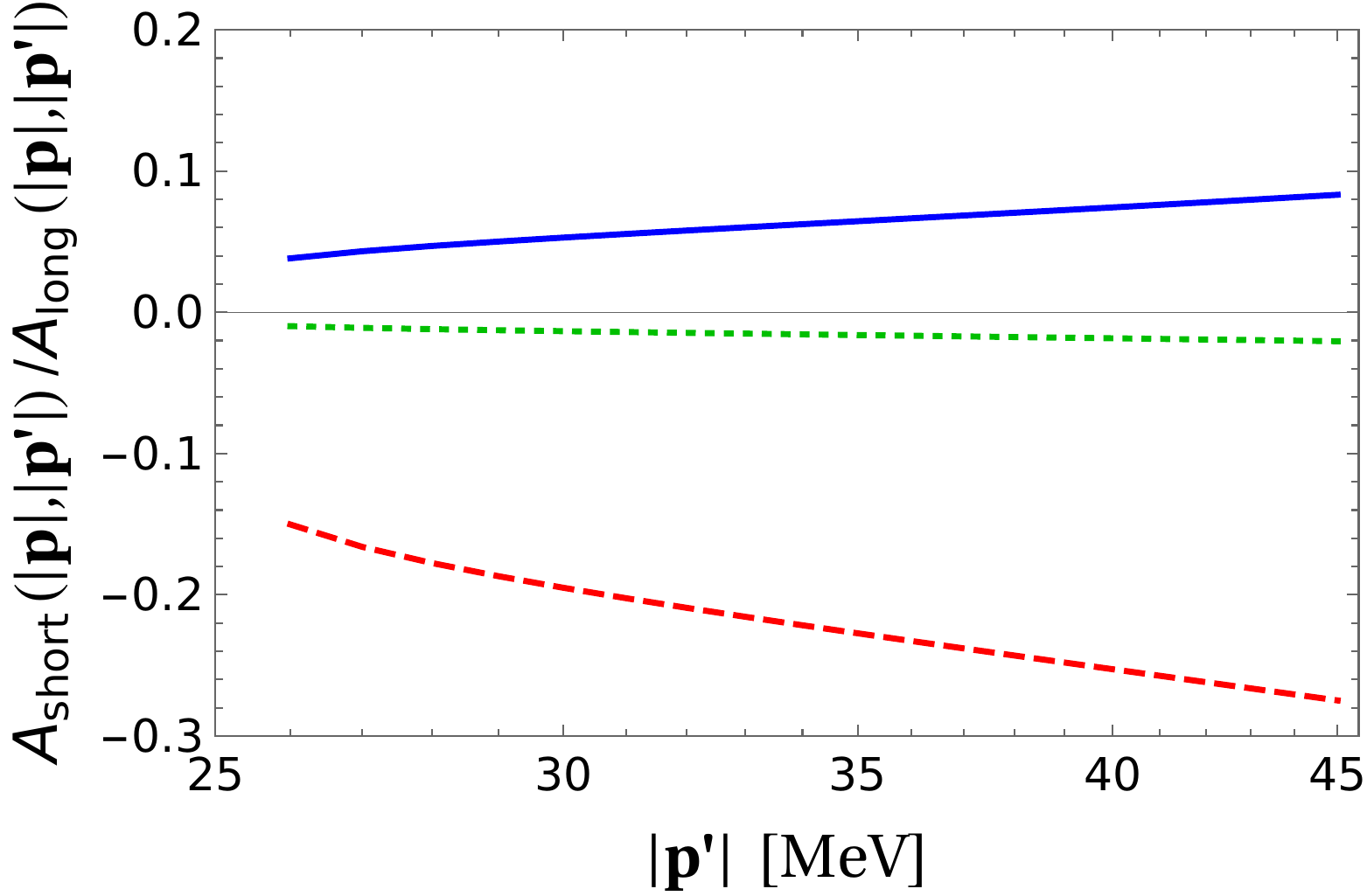}
\caption{Top left:
Long-distance (dashed), short-distance (dotted), and total (solid)  contributions to 
$ \A_\nu (|\spacevec{p}|= 25\MeV, |\spacevec{p}^\prime|)$   for $\Lambda_p =2\fm^{-1}$ (blue) and $\Lambda_p =20\fm^{-1}$ (red) (the total contribution is shown in black, as it is identical for the two cutoffs). 
Top right: ratio of short-to-long-distance contributions for $\Lambda_p =2\fm^{-1}$ (blue),  $\Lambda_p =5\fm^{-1}$ (green, dotted), and $\Lambda_p =20\fm^{-1}$ (red, dashed). Bottom: same as top, but with the coordinate space regulator $R_S$, with $R^{-1}_S = 1.25\fm^{-1}$ (blue), 
$R^{-1}_S = 2\fm^{-1}$ (green), and $R^{-1}_S = 20\fm^{-1}$ (red).}
\label{Anp_pout}
\end{figure}

In Fig.~\ref{gnuNN} we depict the values of the counter term $\tilde{\cal C}_1$ in the two schemes.
In the left panel, we show $\tilde{\cal C}_1(\Lambda_p,n)$ for a wide range of $\Lambda_p$ and three choices of $n$, showing a clear logarithmic dependence of $\Lambda_p$ and a mild dependence on $n$. In the right panel, we show the same plot in the $R_S$ scheme, which again exhibits a logarithmic dependence on $R_S$, 
with power corrections becoming important at small $R^{-1}_S$. 

The relative importance of the short-distance contribution depends on the regulator and on the kinematics. 
In Fig.~\ref{Anp_pout} we show the amplitude $\mathcal A_\nu$ for $|\pp| = 25$ MeV, while varying $|\pp'|$ between $26$ and $46$ MeV.
The region includes the fitting point $|\pp^\prime| = 30$ MeV, and at these small momenta the LO chiral wave functions provide a good description of the ${}^1S_0$ phase shift.
The plots in the top panel use the momentum-space regulator $\Lambda_p$, while the bottom plots use the coordinate-space $R_S$ scheme.
In the top-left panel, we have chosen two specific regulators $\Lambda_p =2\fm^{-1}$ (blue) and $\Lambda_p =20\fm^{-1}$ (red) and kept $n=2$ as the dependence on $n$ is mild. We show individually the long-distance (dashed), short-distance (dotted), and sum (solid) contributions. The sum does not depend on $\Lambda_p$, but the counter term goes from constructive ($+15\%$) to destructive ($-20\%$) between the two choices of the regulators. This is shown more clearly in the right panel of Fig.~\ref{Anp_pout},  where the ratio of short-to-long-distance contributions is given for three choices of regulators $\Lambda_p =\{2,\,5,\,20\}\,\fm^{-1}$ in blue, green, and red respectively. This confirms explicitly that the separation of long- and short-distance contributions is not physical. 

The bottom panel shows the same plots in the $R_S$ scheme, using $R^{-1}_S = 1.25\fm^{-1}$ (green), 
$R^{-1}_S = 2\fm^{-1}$ (blue), and $R^{-1}_S = 20\fm^{-1}$ (red). Also in this case, the interference goes from destructive at large cutoff to constructive at small cutoff, with the short-range amplitude being approximately zero at $2\fm^{-1}$. Of course, the total amplitude agrees between the $\Lambda_p$ and $R_S$ regularization schemes as it should.
These plots show that, as already emphasized above, we  cannot say whether the new short-distance contribution will add constructively or destructively in \textit{ab-initio} calculations of nuclear transitions. This question can only be answered within a specific regularization scheme and choice for the strong $N\!N$ potential.

\section{Conclusions}
\label{sect:conclusion}

In this work we have presented the details of the new method developed in Ref.~\cite{Cirigliano:2020dmx} to estimate 
the  LO contact-term contribution
to the amplitude $nn \to pp e^-e^-$  through the exchange of light Majorana neutrinos. 
This is currently the missing ingredient in order to construct  complete LO $nn \to pp$ transition operators in chiral EFT, 
to be used in {\it ab-initio} nuclear-structure calculations of matrix elements in nuclei of experimental interest for $0\nu\beta\beta$ searches. 

Our approach  to estimate the contact term  is based on a representation of the 
$nn \to pp e^-e^-$  amplitude as the momentum integral of the neutrino propagator ($1/k^2$)  times the generalized forward 
Compton scattering amplitude  $n(p_1)   n(p_2)   W^+ (k)  \to  \ p(p_1^\prime)  p(p_2^\prime)  W^- (k)$, 
in close analogy to the Cottingham formula for the electromagnetic mass splittings of pions and nucleons. To extract the contact term from this integral representation,
we have constructed  model-independent descriptions of the integrand in the low-momentum 
region, using  chiral EFT, and  in the high-momentum region,  using the OPE. 
In the low- and intermediate-momentum region we have kept  only the elastic contribution, i.e.,  the effect of two-nucleon intermediate states, with the most important momentum dependence generated by single-nucleon form factors as well as the $N\!N$ amplitude itself. While we do not have a strict dispersive derivation as in the case of hadron mass splittings, this approach has the potential to match the $(20\text{--}30)\%$ accuracy of the elastic approximation observed there, an expectation that we verified by studying CIB in low-energy $N\!N$ scattering within the same framework and reproducing, within uncertainties,  the CIB contribution to the $^1S_0$  $N\!N$ scattering lengths. 
This phenomenological success gives us confidence that our method is sound and the uncertainty estimate realistic. 

The phenomenological validation is particularly important given that 
our  method does introduce   model-dependent input in the intermediate-momentum region, albeit
anchored to known constraints from QCD at low and high momenta. The extraction of the LO contact term, in a given scheme, then proceeds  
by matching the full amplitude obtained in this way to the LO  chiral EFT amplitude. We considered several sources of uncertainty,    
chief among them the missing contributions from inelastic intermediate states, such as $N\!N \pi$. 
The intermediate steps of our analysis have been performed in dimensional regularization  with the  \MS scheme,  while  the final result can be expressed  in terms of  the scheme-independent  renormalized amplitude $\A_\nu(|\pp|,|\pp^\prime|)$ 
at a set of kinematic points near threshold, where LO chiral EFT is expected to give an excellent approximation. 
Using our  synthetic data, as discussed in Section~\ref{sect:synthetic},  one  can then determine the  contact term  in any 
regularization and renormalization scheme, in particular the ones  employed in nuclear-structure calculations 
for isotopes of experimental interest for $0\nu\beta\beta$ searches. This application is timely in view of the remarkable progress in \textit{ab-initio} calculations of  $0\nu\beta\beta$ decay rates of light and intermediate-mass nuclei~\cite{Cirigliano:2019vdj,Yao:2019rck, Belley:2020ejd, Novario:2020dmr, Yao:2020olm}, ranging from ${}^6$He to ${}^{48}$Ca and  ${}^{76}$Ge, starting from microscopic nuclear forces obtained from chiral EFT. So far these decay rates include the long-distance neutrino-exchange contributions while omitting the contact term, which can now be remedied using our synthetic data for $nn\to pp e^-e^-$, allowing, for the first time, for  
complete LO calculations of nuclear $0\nu\beta\beta$ decay rates. 
For  heavier nuclei such as ${}^{136}$Xe, which are still beyond the capabilities of \textit{ab-initio} techniques, the impact of the short-range term could be studied indirectly, e.g., by matching \textit{ab-initio} results and nuclear models for nuclei accessible to both approaches (this strategy was used in Ref.~\cite{Hoferichter:2020osn} for the axial-vector current).

As a benchmark point, we find
 for the amplitude  ${\cal A}_\nu$ in the \MS scheme at $\mu_\chi=4 \mpi$, with initial- and final-state nucleon momenta  $|\pp| = 25\MeV$ and $|\pp^\prime|=30\MeV$,  
  that the contact-term contribution adds destructively to the neutrino exchange at the 15\%  level. However, as discussed in detail in Section~\ref{sect:synthetic}, this statement depends on the renormalization scheme and scale, e.g., the two cutoff schemes studied there can lead to both constructive and destructive effects depending on the choice of scale. This illustrates that the separation into short- and long-distance contributions is 
unphysical, 
see Fig.~\ref{Anp_pout} for the decomposition in the $\Lambda_p$ and $R_S$ schemes as a function of the final-state momentum and the cutoff scales.   
  Further, as discussed in Refs.~\cite{Cirigliano:2018hja,Cirigliano:2019vdj}, 
while a contact term of natural size affects $\Delta I=0$ transitions such as $nn \to ppe^-e^-$ at the (10--20)\% level, its  effect is amplified to the level of 
(50--70)\% in  $\Delta I=2$ nuclear transitions due to a node in the matrix element density. 
The size of the effect in realistic $0\nu\beta\beta$ 
nuclear transitions can now be addressed, greatly reducing a 
crucial uncertainty in the interpretation of future experimental searches~\cite{LeNoblet:2020efd,Zsigmond:2020bfx,Schmidt:2019gre,Pocar:2020zqz,Tetsuno:2020ngo,Lozza:2020xig,Gando:2020cxo}.

Improving the accuracy of the results presented here would require at least a thorough study of the inelastic $N\!N\pi$ channel, but at the same time a more systematic connection to the Cottingham formula would likely become necessary as well: while the identification of the leading elastic effects is rather intuitive, the extension to subleading corrections is not. 
Additional experimental input could, in principle,  separate ${\cal C}_1$ from ${\cal C}_2$, e.g., via CIB in nuclei, but such an extraction also requires the development of a suitable theoretical framework. 
While our results allow for first phenomenological estimates of the impact of the contact term on $0\nu\beta\beta$ decay rates, they thus also define a benchmark for future lattice-QCD calculations~\cite{Feng:2018pdq,Tuo:2019bue,Cirigliano:2020yhp,Detmold:2020jqv,Feng:2020nqj,Davoudi:2020ngi,Davoudi:2020gxs}. In addition to comparing the final result for ${\cal A}_\nu$, there could also be aspects of the matching strategy and spectral representation, as described in Section~\ref{sect:integral}, that might prove synergistic between the two  approaches. Finally, having concentrated on the LO contact term for light Majorana exchange in this paper, we remark that contact-term contributions arise at LO for other operators mediating $0\nu\beta\beta$ decay as well, both at dimension $7$ and dimension $9$~\cite{Cirigliano:2017djv}, or through exchange of massive sterile neutrinos \cite{Dekens:2020ttz}, and generalizations of the strategies presented here could help constrain the contact terms in these mechanisms.

\medskip

\noindent {\bf Acknowledgments}---We thank Jon Engel,  Evgeny Epelbaum, Michael  Graesser,  Bira van Kolck, and Andr\'e Walker-Loud for discussions at various stages of this work.  The work of VC and EM is supported  by the US Department of Energy through the Los Alamos National Laboratory. Los Alamos National Laboratory is operated by Triad National Security, LLC, for the National Nuclear Security Administration of U.S.\ Department of Energy (Contract No.\ 89233218CNA000001).
WD is supported by  U.S.\ Department of Energy Office of Science, under contract  DE-SC0009919. 
MH is supported by an Eccellenza Grant (Project No.\ PCEFP2\_181117) of the Swiss National Science Foundation. JdV is supported by the 
RHIC Physics Fellow Program of the RIKEN BNL Research Center.
The DOE DBD Topical Nuclear Theory Collaboration inspired  this work and partially supported some of us during its genesis. 
MH thanks the T-2 group at LANL for their hospitality and support during a visit when this project was initiated.

\appendix

\section{Half-off-shell $\boldsymbol{T}$ matrix}
\label{sect:HOST}

In this appendix we describe the calculation of the HOS  $T$-matrix elements
defined in Eq.~\eqref{eq:host1}  and the associated form factors  in Eq.~\eqref{eq:fsbarnopi}.
These quantities affect the   $nn \to pp$  amplitude  through 
the ratio $r(|\spacevec{k}|)$, introduced in Eq.~\eqref{eq:r0}, 
 parameterizing the higher-order  corrections to the forward Compton amplitude. 
We present the calculations  in order of increasing difficulty, 
starting from   $\slashpi$EFT, in which analytic results can be obtained, 
moving to chiral EFT, and finally using $N\!N$ interactions from potential models that successfully 
fit the $N\!N$ scattering data.

\subsection{The half-off-shell form factor in pionless EFT}
\label{sect:HOSTpi}

In this appendix  we discuss  the HOS form factors   
in  the framework of $\slashpi$EFT. Since analytic results are readily available, 
the $\slashpi$EFT  analysis will provide several insights on the general problem at hand.

\subsubsection[Half-off-shell  $T$ matrix  in pionless EFT at NLO]{Half-off-shell  $\boldsymbol{T}$ matrix  in pionless EFT at NLO}

Working in dimensional regularization with power divergence subtraction~\cite{Kaplan:1998we}, the resummation of bubble diagrams with $N\!N$ vertices involving $C_0$ and $C_2$ leads to 
($E = \spacevec{p}^2/m_N$,  $\spacevec{q}^2 \neq \spacevec{p}^2 $)
\be
 \langle \spacevec{q} |  \hat T_S (E)  | \spacevec{p}  \rangle =   \frac{\tilde{C} (E) + \frac{C_2}{2} (\spacevec{q}^2 - \spacevec{p}^2)}{1 - \tilde C (E) I_0}\,, 
 \qquad  \tilde{C} (E) = C_0  + C_2 \, \spacevec{p}^2\,,
\ee
where  we used the definitions~\eqref{eq:TS} with $\hat V_\pi \to 0$ and 
\be
I_0   \left( E \right) = \Big(\frac{\mu}{2}\Big)^{4-d}
\int \frac{d^{d-1} q }{(2 \pi)^{d-1}}  
\, \frac{ m_N }{\spacevec{p}^{2} - \spacevec{q}^2 + i \epsilon} 
= - \frac{m_N}{4 \pi} \, (\mu + i |\spacevec{p}|)\,.
\label{eq:I0v2}
\ee 
This  leads to 
\be
 f_S  (\spacevec{q}, \spacevec{p}  )
  =  1 + \frac{C_2}{2 C_0}  \frac{\spacevec{q}^2 - \spacevec{p}^2}{1 + (C_2/C_0) \spacevec{p}^2} 
 \simeq    1 + \frac{C_2}{2 C_0}  \big( \spacevec{q}^2 - \spacevec{p}^2 \big)\,.
\label{eq:HOSTffspionless} 
\ee
As mentioned before, these HOS form factors are different depending on whether they involve the initial- or final-state on-shell momenta, cf.\ Eq.~\eqref{eq:HOSfs}.
By themselves these form factors are clearly not physical. 
First, one notes that they depend on the renormalization scale $\mu$, as $C_2/C_0$ is $\mu$-dependent. 
The form factors vary  significantly  with $\mu$,  while having the expected qualitative features 
at small $|\spacevec{k}|$.
Second, as shown below,  $f_S$  can be changed by performing  field redefinitions, as expected for any off-shell quantity. 
However, in the actual application  insertions of $f_S$   combine into a physical contribution to the $nnW^+(k) \to pp W^-(k)$    amplitude, 
or equivalently to the  integrand in the sum of  ${\cal M}^<_{B + \bar B}$ and ${\cal M}^<_C$  in  Eq.~\eqref{eq:ABCfull}. 
In the end, the net effect is a modification of the function  ${\cal I}_C^< (| \spacevec{k}|)$ and hence a change in 
$r (|\spacevec{k}|)$.

\subsubsection[Behavior of $f_S$ and ${\cal I}_C^< (| \spacevec{k}|)$ under field redefinitions]{Behavior of $\boldsymbol{f_S}$ and $\boldsymbol{{\cal I}_C^< (| \spacevec{k}|)}$ under field redefinitions}

We consider the non-linear field redefinition given in Ref.~\cite{Furnstahl:2000we}, 
namely 
\be
N  \to N + \eta N (N^\dagger N)\,,
\ee
where $\eta$ is an arbitrary parameter of mass dimension $-3$. 
This transformation induces a  derivative $N\!N$ interaction from the kinetic term as well as a two-nucleon current operator,  
schematically
\begin{align}
{\cal L} & \to   {\cal L}  \ +\  \eta  \, (N^\dagger N)  \,  N^\dagger \left(i \partial_0 + \frac{\spacevec{\partial}^2}{2 m_N} \right) N \ + \ 
 \eta \,   N^\dagger  \left(i \partial_0 + \frac{\spacevec{\partial}^2}{2 m_N} \right) \left[  N  \, (N^\dagger N) \right]\,,\notag
 \\
N^\dagger   J_\mu \tau^+ N  & \to  N^\dagger   J_\mu \tau^+  N   + 2 \eta  \ N^\dagger   J_\mu \tau^+ N \,  (N^\dagger N)\,.
\label{eq:etaL}
\end{align}
The resulting $\eta$-dependent four-nucleon vertex is~\cite{Furnstahl:2000we} 
\be
\langle  N_{1^\prime} N_{2^\prime} | i {\cal L} |  N_1 N_2 \rangle = i \eta \left[   
\left(\omega_{1} - \frac{\spacevec{p}_{1}^2}{2 m_N} \right) + 
\left(\omega_{2} - \frac{\spacevec{p}_{2}^2}{2 m_N} \right) + 
\left(\omega_{1^\prime} - \frac{\spacevec{p}_{1^\prime}^2}{2 m_N} \right) + 
\left(\omega_{2^\prime} - \frac{\spacevec{p}_{2^\prime}^2}{2 m_N} \right) 
 \right]\,.
\ee
This modifies the HOS form factors~\eqref{eq:HOSTffspionless} as follows
\be
 f_S  (\spacevec{q}, \spacevec{p}  )  
 \simeq    1 + \frac{C_2 + 2 (\eta/m_N)}{2 C_0}  \big( \spacevec{q}^2 - \spacevec{p}^2 \big)\,.
\ee

\begin{figure}[t]
\centering
\includegraphics[width=0.6\textwidth]{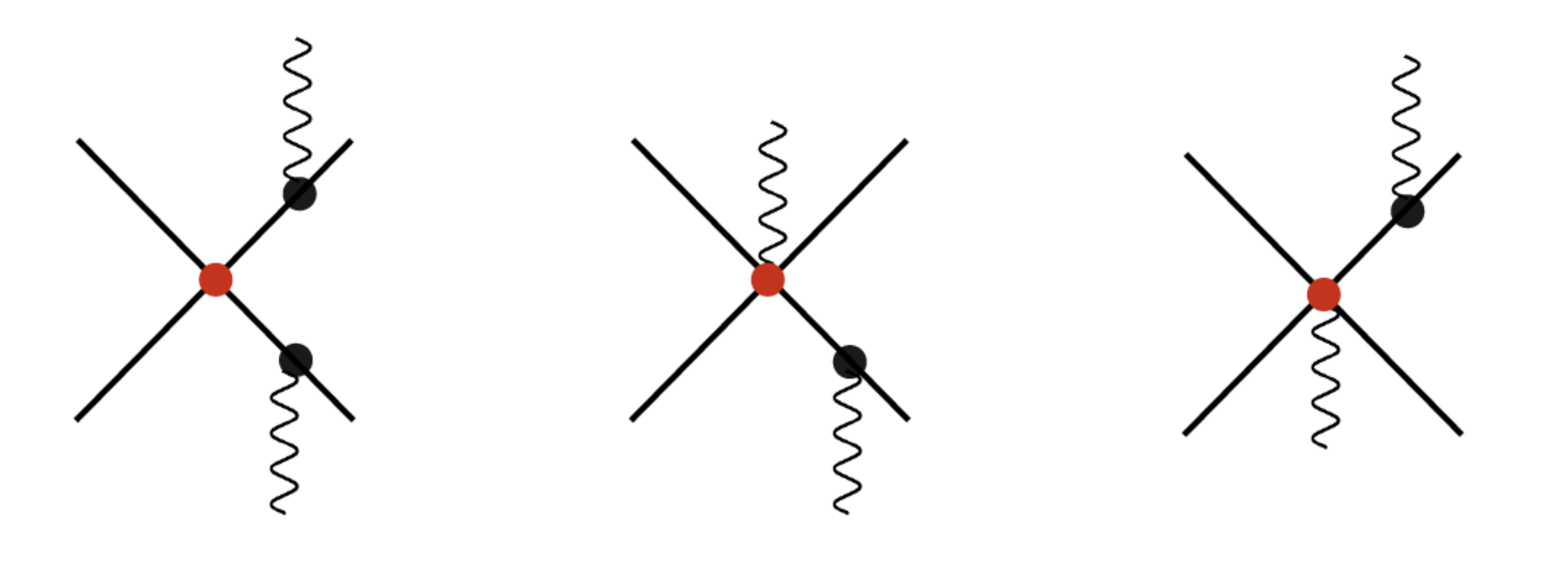}
\vspace{-0.2cm}
\caption{
Group of diagrams contributing to $n n W^+(k) \to pp W^-(k)$ in which the effects of 
field redefinitions cancels out.  
The red circles denote vertices proportional to $\eta$, as given in Eq.~\eqref{eq:etaL}.
The left-most legs denote on-shell protons. The right-most legs denote off-shell neutrons. 
The wavy lines denote weak current insertions.
}
\label{Fig:eta}
\end{figure}

These $\eta$-dependent HOS form factors combine with new contributions to the current matrix element in such a way to leave no 
$\eta$-dependent term in the amplitude $n n W^+(k) \to pp W^-(k)$ itself.  It is simple to verify that cancellations happen 
through groups of diagrams represented in Fig.~\ref{Fig:eta}. 
Therefore, the integrands in the amplitudes ${\cal M}^<_{B + \bar B}$ and ${\cal M}^<_C$ in  Eq.~\eqref{eq:ABCfull} 
(and hence ${\cal I}_C^< (| \spacevec{k}|)$ via Eqs.~\eqref{eq:MCm} and~\eqref{eq:ACmsing}) 
are individually invariant under the field redefinition considered here, i.e., they 
do not depend on  $\eta$. 
We have used this simple field redefinition of Ref.~\cite{Furnstahl:2000we} to illustrate a more generally valid point.

\subsubsection[Impact of $f_S$ on ${\cal I}_C^< (| \spacevec{k}|)$ ]{Impact of $\boldsymbol{f_S}$ on $\boldsymbol{{\cal I}_C^< (| \spacevec{k}|)}$ }

Inserting the form factor~\eqref{eq:HOSTffspionless} into Eq.~\eqref{eq:ACmsing} one obtains (using $|\spacevec{p}^\prime| = | \spacevec{p}|$)
\be
{\cal I}_C^< (| \spacevec{k}|) 
=  \frac{1}{8 |{\bf k}|} \,  \theta(|{\bf k}| - 2 |{\bf p}|) + \frac{i}{8 \pi |{\bf k}|} \, \log \left|  \frac{1+ 2 \frac{|{\bf p}|}{{|{\bf k}|}} }{ 1-  2 \frac{|{\bf p}|}{{|{\bf k}|}}}   \right|
+ \frac{1}{4 \pi} \, \frac{C_2}{C_0}   \Big(  \mu + i |\spacevec{p}|  
\Big) 
 + \Order(C_2^2)\,.
\ee
For the real part of ${\cal I}_C^< (| \spacevec{k}|)$, relevant 
for the matching, this implies 
\be 
{\rm Re} \, {\cal I}_C^< (| \spacevec{k}|)   
 =  \frac{1}{8 |{\bf k}|}  \left[ 1+ \frac{ |\spacevec{k}|}{\pi} \left( \frac{2 \mu C_2}{C_0} \right)  \right]
\approx  \frac{1}{8 |{\bf k}|}  \left[ 1 -  \frac{ |\spacevec{k}| r_0}{\pi}   \right]\,.
\label{eq:ratiopionless}
\ee
In the last step we have used  the $\slashpi$EFT expressions of $C_{0,2}(\mu)$ and taken  $\mu \sim M_\pi \gg 1/a$. 
The size of the effective range  $r_0= 2.73~{\rm fm} = 1/(72\MeV)$ implies sizable deviations from the leading $1/(8 |\spacevec{k}|)$ behavior 
already at $|\spacevec{k}| \leq M_\pi$.

While the  analysis above is promising, there are two unsatisfactory aspects:
(i) Even though it does not enter the matching, the imaginary part of  ${\cal I}_C^< (| \spacevec{k}|)$ depends on the scale $\mu$ 
through the ratio $C_2/C_0$.
(ii) The real part of  ${\cal I}_C^< (| \spacevec{k}|)$ is only approximately independent of $\mu$, 
as we have neglected  $1/(\mu a) \ll1$ in the expression 
\be
\frac{\mu C_2}{C_0} =  - \frac{r_0}{2} \frac{1}{ 1 - \frac{1}{\mu a}}\,.
\ee
Both issues can be addressed by studying how  $f_S$ impacts all components of the 
generalized Compton amplitude, i.e., the integrand of  ${\cal M}^<_{B + \bar B}$ and ${\cal M}^<_C$ in  Eq.~\eqref{eq:ABCfull}, 
and not just the integrand of  ${\cal M}^<_C$ as done above.
The final result is very simple: the contributions combine to give a real-valued and scale-independent  shift to 
 ${\cal I}_C^< (| \spacevec{k}|)$ equal to   $\delta {\cal I}_C^< (| \spacevec{k}|) = - r_0/(8 \pi)$, 
 which  agrees with the result in the second line of 
 Eq.~\eqref{eq:ratiopionless}. 

The argument proceeds as follows. 
Omitting a common factor proportional to the weak current insertions, namely $-(1 + 3 g_A^2)/\spacevec{k}^2$,  the integrand
corresponding to ${\cal M}^<_{B + \bar B} + {\cal M}^<_C$ in  Eq.~\eqref{eq:ABCfull} 
has the following structure:
\begin{align}
& m_N \Bigg(  T (E^\prime)  \, \frac{f_S(\spacevec{p}^\prime, \spacevec{k} + \spacevec{p})}{\spacevec{p}^{\prime 2} - (\spacevec{k} + \spacevec{p})^2 + i \epsilon} 
+   \frac{f_S( \spacevec{k} + \spacevec{p}^\prime, \spacevec{p})}{\spacevec{p}^{2} - (\spacevec{k} + \spacevec{p}^\prime)^2 + i \epsilon} 
\,  T (E)
\nonumber \\
& + 
T (E^\prime)  \,
m_N\, \int \frac{d^{d-1} q }{(2 \pi)^{d-1}}  
\, \frac{f_S(\spacevec{p}^\prime, \spacevec{k} + \spacevec{q})}{\spacevec{p}^{\prime 2} - (\spacevec{k} + \spacevec{q})^2 + i \epsilon} 
\ \frac{f_S(\spacevec{q}, \spacevec{p})}{\spacevec{p}^{2} -  \spacevec{q}^2 + i \epsilon} 
\  T(E) \Bigg)\,,
\label{eq:ABC1}
\end{align}
where $T(E)$ denotes the LO  on-shell $N\!N$ $T$ matrix in the $^1 S_0$ channel. 
Substituting the  form factors~\eqref{eq:HOSTffspionless},
one obtains 
\be
- m_N \Bigg(  \frac{C_2}{2 C_0}  \Big( T (E^\prime) + T(E) \Big)  +
\frac{C_2}{ 2 C_0^2}  \Big[
T(E^\prime) \  C_0 I_0 (E) T(E) \ + \ T(E^\prime) C_0 I_0(E^\prime) \ T(E) 
\Big] \Bigg)\,,
\label{eq:C20canc}
\ee
where $I_0(E)$ is given in Eq.~\eqref{eq:I0v2} and  $T(E) = C_0/(1 - C_0 I_0(E))$. 
Using the relation 
\be
T(E) - C_0 =  C_0 I_0(E) T(E)\,, 
\label{eq:ABC2}
\ee
one sees that the terms proportional to $C_2/C_0$ 
in Eq.~\eqref{eq:C20canc} cancel and one is left 
with  the scale-independent quantity $ T(E^\prime) \ (-C_2/C_0^2)  \ T(E)$. 
Upon matching the overall factors of $m_N$, this leads to 
 \be
\delta {\cal I}_C^< (| \spacevec{k}|) = - \frac{C_2}{m_N C_0^2} =  - \frac{r_0}{8 \pi}\,. 
 \ee

To conclude, we discuss several lessons that can be abstracted  from the $\slashpi$EFT analysis: first, 
the large fractional correction in  ${\cal I}_C^< (| \spacevec{k}|)$ (and hence $a_< (|\spacevec{k}|)$) 
at relatively low $|\spacevec{k}| \sim M_\pi$ (in $\slashpi$EFT) is due  to the large effective range in the  $^1 S_0$ channel.  
Second, in $\slashpi$EFT, the inclusion of HOS form factor effects from 
 ${\cal M}^<_{B + \bar B}$ was needed to cancel  a subleading $\Order( 1/(\mu a))$ scale dependence in the integrand 
corresponding to ${\cal M}^<_C$, proportional to  $ {\cal I}_C^< (| \spacevec{k}|)$.
In practice one gets the leading corrections by just including the HOS form factors 
in the  ${\cal M}^<_C$ integrand.   
This is consistent with the previous result that   ${\cal M}^<_{B + \bar B}$ 
drops out of the LO matching formula and the expectation that  the HOS form factor makes 
the  integrand in ${\cal M}^<_{B + \bar B}$ even more convergent. 
Accordingly, this suggests that the inclusion of the HOS form factors in the integrand of ${\cal M}^<_C$ 
 captures the bulk of the physical suppression of the $n n  W^+(k) \to p p W^-(k)$ amplitude
at small $|\spacevec{k}|$,   
while the inclusion of HOS form factors in  ${\cal M}^<_{B + \bar B}$ 
would lead to effects that are below the current  level of accuracy.

\subsection{Half-off-shell  effects  in chiral EFT at NLO}
In this appendix we extend the analysis to chiral EFT, thus keeping pion-exchange effects. 
To take into account off-shell effects in $\hat T_S$, we  start from 
Eq.~\eqref{eq:TS}, 
\be
\hat T_S^\chi = (1+\hat V_\pi \hat G^{(\pi)}_+(E))\hat V_S(1-\hat G^{(\pi)}_+(E) \hat V_S)^{-1}(1+ \hat G^{(\pi)}_+(E) \hat V_\pi)\,,
\ee
where $\langle \spacevec q| \hat V_S| \spacevec p\rangle = C+ C_2\frac{\spacevec q^2+\spacevec p^2}{2}$ is the NLO short-range potential, so that we can write 
\be
\hat V_S = (1+m_N E C_2/C)\,  \hat V(C)-\frac{m_NC_2}{2C}\left\{ \hat V(C),\, \left(\hat G^{(0)}_+(E)\right)^{-1}\right\}\,.
\ee
Here $\hat V(C)$ is the LO part of the short-distance potential and the last term gives rise to off-shell effects proportional to $ C_2$. Expanding  $\hat T_S^\chi$ to first order in $C_2$ we have
 \begin{align}
\langle  \spacevec q| \hat T_S^\chi | \spacevec k\rangle 
&= \chi^+_{\spacevec q}(0)\chi^+_{\spacevec k}(0)\left[K_E^{(0)} + K_E^{(1)}\right] -\frac{C_2}{C}K_E^{(0)} \left[\chi^+_{\spacevec q}(0) \frac{\spacevec p^2-\spacevec k^2}{2}+\chi^+_{\spacevec k}(0) \frac{\spacevec p^2-\spacevec q^2}{2}\right]\,,\label{eq:TSoffshell}\\
K_E^{(0)}&= \frac{C}{1-C G^{(\pi)}_E(0,0)}\,,\qquad K_E^{(1)}= \left(K_E^{(0)}\right)^2\frac{C_2}{C^2}\left[\spacevec p^2-m_N
\Big(\frac{\mu}{2}\Big)^{4-d} \int \frac{d^{d-1} q}{(2\pi)^{d-1}}
 V_\pi(q) \right]\, ,\notag
 \end{align}
 with $\spacevec p^2 = m_NE$. The scale-dependent term in $K_E^{(1)}$ can be absorbed into an NLO correction to $C\to C+C^{(1)}$, after which $K_E^{(0,1)}$ become scale independent and the NLO on-shell  matrix element of $\hat T_S^\chi$ takes the form $T_S^\chi ( \spacevec p,\spacevec p) = \left(\chi^+_{\spacevec p}(0)\right)^2\left[K_E^{(0)} + K_E^{(1)}\right]$.

Inserting the above off-shell matrix elements 
$\langle  \spacevec q| \hat T_S^\chi | \spacevec k\rangle$  into 
Eq.~\eqref{eq:ABCfull}
we can write the  low-momentum component of  the $\mathcal M_C$ amplitude as 
\begin{align}
\mathcal M_C^< & = \int \frac{d^3 \spacevec{k}}{(2 \pi)^3} \ 
\langle \spacevec p' | \left( \hat T^\chi_S  (E^\prime) \hat G_+^{(0)} (E^\prime) \right)   
\,  \hat{O}_\chi^{LL} (\spacevec{k})  \,
  \left(  \hat G_+^{(0)} (E) \, \hat T^\chi_S  (E) \right)  |\spacevec p \rangle \nn\\
  &= \int  \frac{d^3 \spacevec{k}}{(2 \pi)^3} f(\spacevec k) \chi^+_{\spacevec p}(0)\chi^+_{\spacevec p'}(0)\nn\\
&\times \Bigg\{ \left[K_{E'}^{(0)} + K_{E'}^{(1)}\right] \left[K_E^{(0)} + K_E^{(1)}\right]\int\frac{d^3 \spacevec{q}}{(2 \pi)^3}  \frac{ \chi^+_{\spacevec q}(0)}{E'-\spacevec q^2/m_N+i\epsilon} \frac{ \chi^+_{\spacevec q+ \spacevec k}(0)}{E-(\spacevec  q+\spacevec  k)^2/m_N+i\epsilon}\nn\\
&\qquad- \frac{m_N C_2}{2C}K_E^{(0)}K_{E'}^{(0)}\left( G^{(\pi)}_E(0,0)+G^{(\pi)}_{E'}(0,0)\right)
\Bigg\}\,,
\label{eq:MCNLO}
\end{align}
where $\langle \spacevec q| \hat{O}_\chi^{LL} (\spacevec{k}) | \spacevec q'\rangle = (2\pi)^3 \dt^3(\spacevec q -\spacevec q'+\spacevec k) f(\spacevec k)$, with $f(\spacevec k) = -\frac{g^2_V(\spacevec k)+3 h_{GT}(\spacevec k)}{\spacevec k^2}$. The first term in curly brackets contributes to the most singular part of $\mathcal M_C$ (the parts that diverge when integrating over $\spacevec k$). In particular, the terms with $\chi_{\spacevec q, \spacevec q +\spacevec k}^+(0)\to 1$ give rise to the  part of $\mathcal I_C$ in Eq.~\eqref{eq:ACmsing}, with $f_S  \to1$. In other words, these pieces are responsible for the  contributions involving the on-shell $T_S$ matrix. Instead, the contributions from the terms proportional to $1-\chi_{\spacevec q, \spacevec q +\spacevec k}^+(0)$ correspond to diagrams involving pion exchanges within the bubble, which are convergent.

The last line in Eq.~\eqref{eq:MCNLO} arises from the off-shell piece of $\hat T_S$, the second term in Eq.~\eqref{eq:TSoffshell}, and also contributes to the most divergent parts of $\mathcal M_C$. As in pionless EFT, these terms are $\mu$ dependent by themselves and have to be combined with pieces from the $\mathcal M_B$ diagrams in order to obtain a $\mu$-independent answer.
This part of the amplitude can be written as
\begin{align}
\mathcal M^<_{B+\bar B}& = 
 \int \frac{d^3 \spacevec{k}}{(2 \pi)^3} \Bigg[
 \langle \spacevec  p' |  \left(  \hat T^\chi_S (E^\prime)  \, \hat G_+^{(0)} (E^\prime) \right)  \, \hat O_\chi^{LL} (\spacevec{k}) \, \left( 
   I + \hat{G}_+^{(0)}  (E)  \,   \hat T^\chi_\pi (E)  \right)  | \spacevec p \rangle\notag
 \\  
&\qquad+   \langle \spacevec p' |    \left(   I + \hat T^\chi_\pi (E^\prime)  \hat{G}_+^{(0)}  (E) \right)    
 \, \hat O_\chi^{LL} (\spacevec{k}) \, \left(  \hat G_+^{(0)} (E)     \hat T^\chi_S (E)  \right)
   | \spacevec p \rangle   \Bigg]\nn\\
   &= \int \frac{d^3 \spacevec{k}}{(2 \pi)^3}  f(\spacevec k)\Bigg\{\frac{d^3 \spacevec{q}}{(2 \pi)^3} \Bigg[\frac{\langle \chi_{\spacevec p'}|\spacevec q\rangle \chi^+_{\spacevec q+ \spacevec k}(0)}{E-(\spacevec q+ \spacevec k)^2/m_N+i\epsilon} \chi^+_{\spacevec p}(0)\left(K_E^{(0)}+K_E^{(1)}\right)\nn\\
   &\qquad+\left(K_{E'}^{(0)}+K_{E'}^{(1)}\right) \chi^+_{\spacevec p'}(0)\frac{ \chi^+_{\spacevec q}(0)\langle \spacevec q+\spacevec k| \chi_{\spacevec p}\rangle}{E'-\spacevec q^2/m_N+i\epsilon}\Bigg]\nn\\
   &- \frac{m_NC_2}{2C} \chi^+_{\spacevec p'}(0) \chi^+_{\spacevec p}(0)\left(K_{E}^{(0)}+K_{E'}^{(0)}\right)\Bigg\}\,,
\end{align}
where the terms within square brackets are finite, while the last line comes from the off-shell part of  $\hat T_S^\chi$ , which is again $\mu$ dependent.
Combining the terms that have explicit $C_2$ dependence with those in $\mathcal M_C$, we obtain
\begin{align}
\mathcal M^<_{B+\bar B}+\mathcal M^<_C\Big|_{C_2}& = -\frac{m_NC_2}{2C} \chi^+_{\spacevec p'}(0) \chi^+_{\spacevec p}(0)\int \frac{d^3 \spacevec{k}}{(2 \pi)^3} f(\spacevec k)\nn\\
&\qquad\times \left[K_{E}^{(0)}+K_{E'}^{(0)}+K_E^{(0)}K_{E'}^{(0)}\left( G^{(\pi)}_E(0,0)+G^{(\pi)}_{E'}(0,0)\right)\right] \nn\\
&=- \frac{m_NC_2}{C^2}  \chi^+_{\spacevec p'}(0) \chi^+_{\spacevec p}(0)K_{E}^{(0)}K_{E'}^{(0)} \int \frac{d^3 \spacevec{k}}{(2 \pi)^3} f(\spacevec k)\,,
\label{eq:MCshift}
\end{align}
which is now $\mu$ independent and has the factors of  $\chi^+_{\spacevec p}(0) K_E$ we expect for the $\mathcal M_C$-type diagrams. Using 
\be
\mathcal M_C^< = \chi^+_{\spacevec p'}(0) K_{E'}^{(0)} \left[m_N^2 \int \frac{d^3 \spacevec{k}}{(2 \pi)^3}    f(\spacevec k)\mathcal I_C\right]   \chi^+_{\spacevec p}(0)K_{E}^{(0)}\,,
\ee
the terms in Eq.~\eqref{eq:MCshift} provide a shift to $\mathcal I^<_C$ equal to  $\dt \mathcal I^<_C =- C_2/(m_N C^2)$, which is the same as in $\slashpi$EFT, 
up to the fact that $C_2/C^2$ cannot simply be expressed in terms of the effective range parameter $r_0$.

\subsection[Half-off-shell  effects  in $N\!N$ potential models]{Half-off-shell  effects  in $\boldsymbol{N\!N}$ potential models}

Finally, in this appendix we describe the calculation of the HOS form factor and 
the ratio $r ( |\spacevec{k}|)$ in potential models. 
As a benchmark resonance model for the $^1S_0$ channel interaction we  will use the
three-Yukawa potential, known as  the Reid potential~\cite{Reid:1968sq}, 
which has also been studied in the context of analyzing the convergence of the $N\!N$ EFT~\cite{Kaplan:1999qa}.
This is a simplified version of the general resonance saturation model for $N\!N$ interactions~\cite{Epelbaum:2001fm}. 
In coordinate space the potential takes the form 
\be
V = V_\pi + V_S\,, \qquad 
V_\pi (r) = - \alpha_\pi \frac{e^{- M_\pi r}}{r}\,, 
\qquad
V_S(r) =  - \alpha_\sigma \frac{e^{- M_\sigma r}}{r}   + \alpha_\rho \frac{e^{- M_\rho r}}{r}\,, 
\label{eq:3Y}
\ee 
with the short-range term modeled through the attractive 
$\sigma$-meson  and repulsive $\rho$-meson contributions. 
Once the mass parameters are fixed ($M_\sigma = 4 M_\pi$, $M_\rho = 7 M_\pi$ in Ref.~\cite{Reid:1968sq} 
and $M_\sigma = 500\MeV$, $M_\rho = 770\MeV$ in Ref.~\cite{Kaplan:1999qa}), the couplings $\alpha_{\sigma,\rho}$ 
are tuned to reproduce the $^1S_0$ phase shift or the scattering length and effective range~\cite{Kaplan:1999qa}, 
producing an overall good representation of the data. 

In this class of models it is relatively simple to compute the HOS  $T$-matrix elements defined in Eq.~\eqref{eq:host1}  
and the associated form factors  in Eq.~\eqref{eq:fsbarnopi}: one needs to solve the LS equation 
or alternatively the Schr\"odinger equation  using $V_S$ in Eq.~\eqref{eq:3Y}.  We have performed the calculation in  both 
methods, LS equation in momentum space and Schr\"odinger equation in coordinate space. 
Useful cross-checks in the calculation are provided by  the property 
that the HOS $T$ matrix has the same
phase-shift dependence as the on-shell $T$ matrix $T( \spacevec p, \spacevec p)$~\cite{Srivastava:1975eg}
\begin{equation}
T (\spacevec q, \spacevec p)  = |T (\spacevec q, \spacevec p) | \ e^{\delta ( | \spacevec p |)}\,. 
\end{equation}

We have also performed the calculation of  $T (\spacevec q, \spacevec p)$ and  $f_S (\spacevec q, \spacevec p)$ 
with the AV18 potential~\cite{Wiringa:1994wb}, after subtracting the one-pion exchange. 
This is a representative of the class of high-quality  potentials that reproduce the $N\!N$ scattering data for $|\spacevec{p}|$  out to several hundred MeV. 

\begin{figure}[t]
\begin{center}
\includegraphics[width=0.45\linewidth]{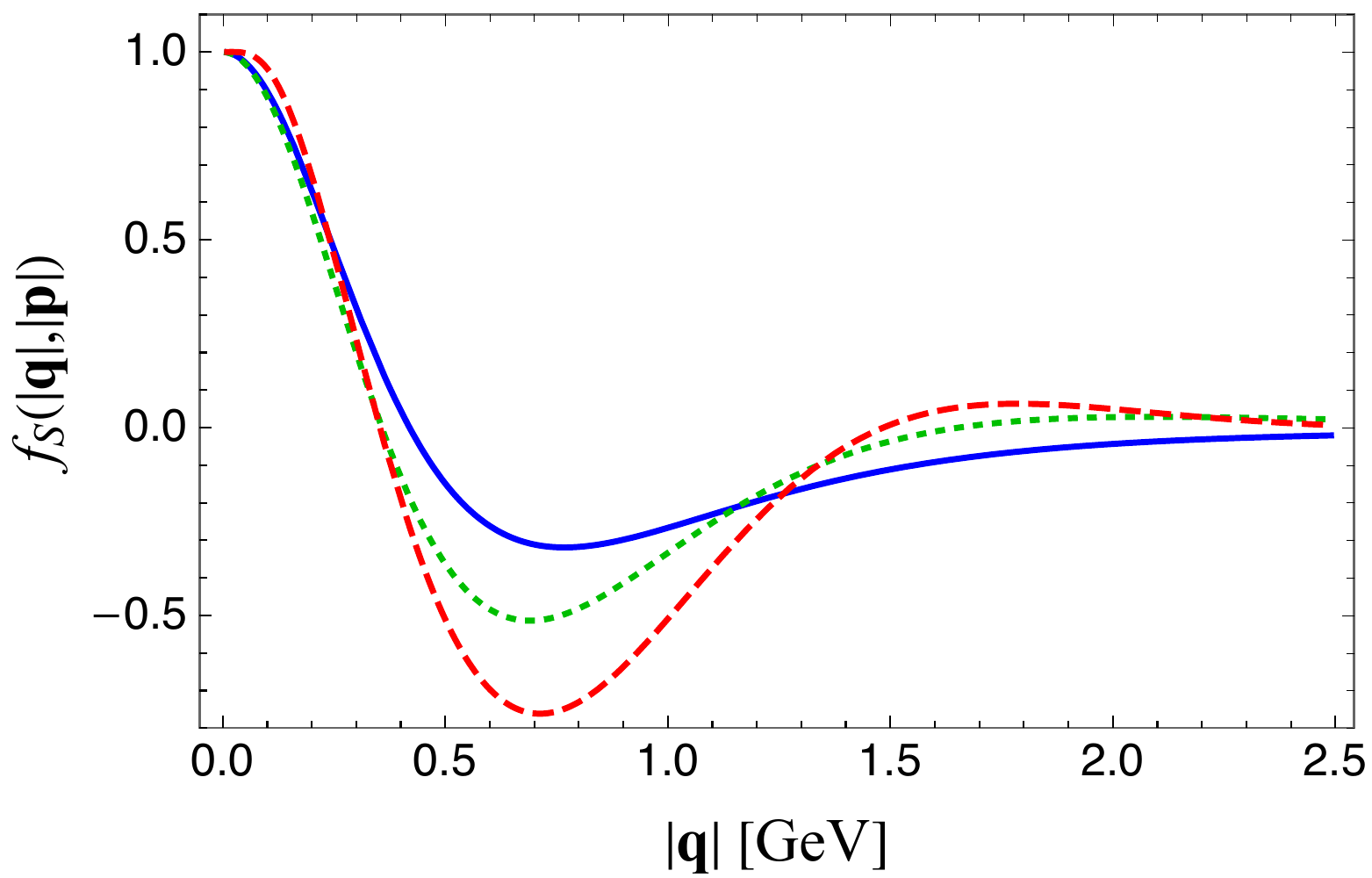}
\hspace{0.03\linewidth}
\includegraphics[width=0.45\linewidth]{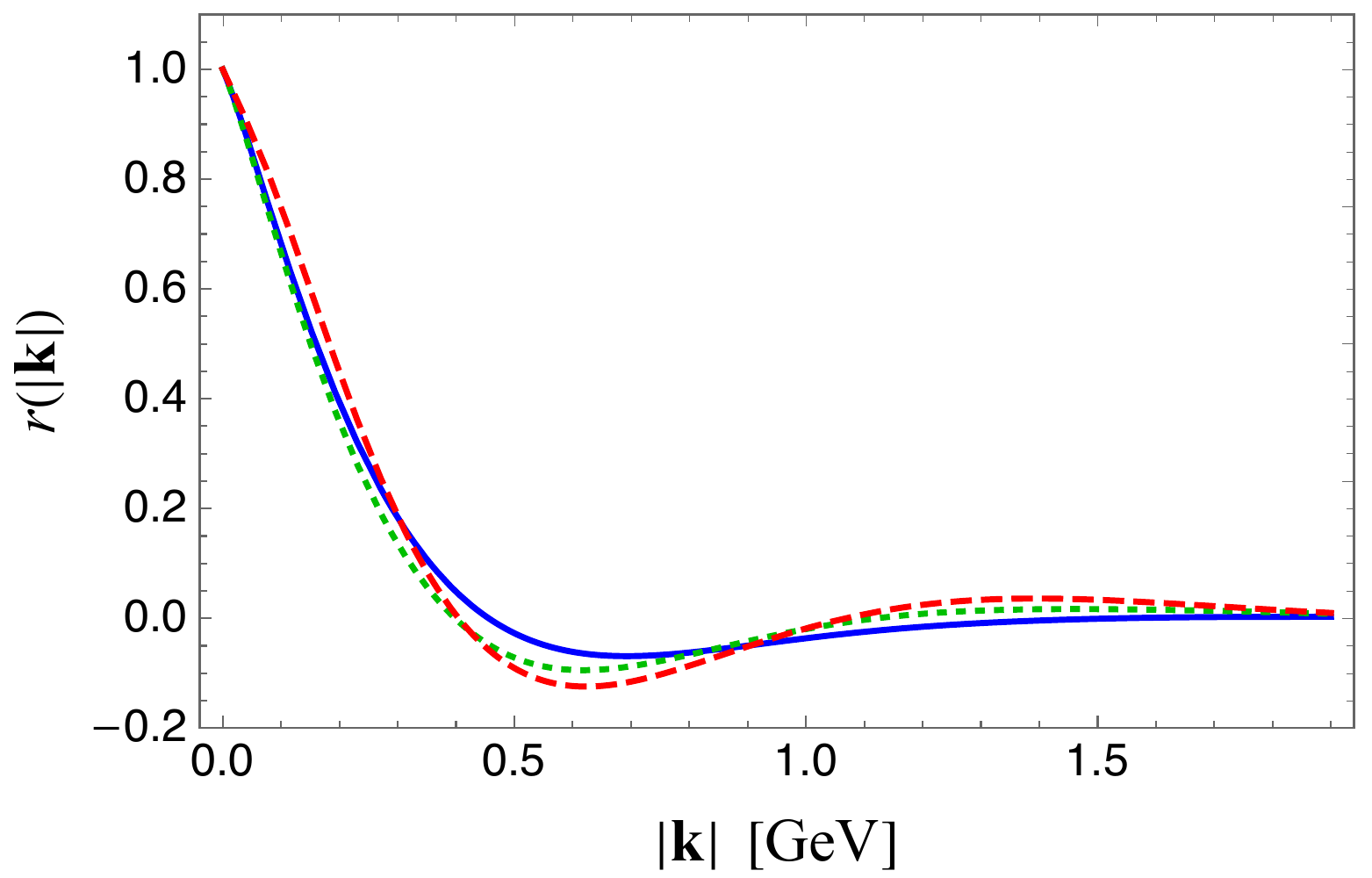}
\caption{ 
Left panel: HOS form factor $f_S (\spacevec q, \spacevec p)$ at $|\spacevec{p}|= 1\MeV$ for Kaplan--Steele (blue), 
Reid (green, dotted), and AV18 (red, dashed) potential. 
Right panel:   $r (|\spacevec{k}|)$ computed at $|\spacevec{p}|=|\spacevec{p}^\prime|= 1\MeV$ for Kaplan--Steele (blue), 
Reid (green, dotted), and AV18 (red, dashed) potential. 
\label{fig:host}
}
\end{center}
\end{figure}

Representative results from this analysis are summarized in Fig.~\ref{fig:host}.  
In the left panel we show  $f_S (\spacevec q, \spacevec p)$ versus  $|\spacevec{q}|$ 
at $|\spacevec{p}|= 1\MeV$  for the three potentials considered here, Kaplan--Steele, Reid, and AV18.
In the right panel we report the ratio $r (|\spacevec{k}|)$, see Eq.~\eqref{eq:r0},
relevant for the calculation of ${\cal A}_C^{<, \rm sing}$.  
The difference in   $r (|\spacevec{k}|)$  provides some indication of the level of model dependence in our 
matching calculation. While locally the differences can be large, 
the fall-off of the  ${\cal A}_C^{<, \rm sing}$ 
integrand multiplying  $r(|\spacevec{k}|)$ 
dilutes the impact of any feature in  $r(|\spacevec{k}|)$  
above $500 \MeV$, 
so that ultimately the impact on the extracted LEC is relatively minor. 
We expect this conclusion to hold in general, beyond the limited set of potential  models explored here.

\section{Left--right correlator}
\label{sec:LROPE}

In the main text we defined the ``left--left'' correlator, which naturally appears when dealing with the weak currents. 
In the case of electromagnetic currents and for matching purposes, we also need the 
``left--right'' correlator.  We define
\begin{align}
\hat{\Pi}_{\mu \nu}^{LL} (k,0) &=  \int d^4 r  \, e^{i k \cdot r} \     T  \Big(  \bar{u}_L \gamma_\mu d_L(r/2)   \ \bar{u}_L \gamma_\mu d_L(- r/2)  \Big)\,,\notag 
\\
\hat{\Pi}_{\mu \nu}^{LR} (k,0)  &=  \int d^4 r  \, e^{i k \cdot r} \     T  \Big(  \bar{u}_L \gamma_\mu d_L( r/2)   \ \bar{u}_R \gamma_\mu d_R(- r/2)  \Big)\,, 
\label{eq:correlatorLR}
\end{align}
and 
\begin{align}
\hat{O}^{LL}_{\alpha \beta} (0) &= \hat{O}_{\alpha \beta} (0)    =  \bar{u}_L (0) \gamma_\alpha  T^a d_L (0)\ \bar{u}_L (0)\gamma_\beta  T^a d_L(0)\,,\notag 
\\
\hat{O}^{LR}_{\alpha \beta} (0) &=   \bar{u}_L (0) \gamma_\alpha  T^a d_L (0)\ \bar{u}_R (0)\gamma_\beta  T^a d_R(0)\,.
\end{align}
With these definitions, the time-ordered products at large $k^2$ can be written as, see Eq.~\eqref{eq:opeab}, 
\begin{align}
\hat{\Pi}_{\mu \nu}^{LL} (k,0) 
&= -  \frac{4 i g_s^2}{(k^2 + i \epsilon)^3}  
\times \Big[
(k_\mu k_\nu - k^2 g_{\mu \nu} )  g^{\alpha \beta}  \, \hat{O}^{LL}_{\alpha \beta}   (0)
\nonumber \\
&+  g_{\mu \nu} k^\alpha k^\beta \hat{O}^{LL}_{\alpha \beta} (0) + k^2 \hat{O}^{LL}_{\nu \mu}  (0)
-k_\nu k^\alpha \, \hat{O}^{LL}_{\alpha \mu} (0)-  k_\mu k^\alpha \, \hat{O}^{LL}_{\nu \alpha}  (0)
\Big]\,,\notag
\\
\hat{\Pi}_{\mu \nu}^{LR} (k,0) 
&=  \frac{4 i g_s^2}{(k^2 + i \epsilon)^3}  
\times \Big[
(k_\mu k_\nu - k^2 g_{\mu \nu} )  g^{\alpha \beta}  \, \hat{O}^{LR}_{\alpha \beta}   (0)
\nonumber \\
&+  g_{\mu \nu} k^\alpha k^\beta \hat{O}^{LR}_{\alpha \beta} (0) + k^2 \hat{O}^{LR}_{\nu \mu}  (0)
-k_\nu k^\alpha \, \hat{O}^{LR}_{\alpha \mu} (0)-  k_\mu k^\alpha \, \hat{O}^{LR}_{\nu \alpha}  (0)
\Big]\,. 
\label{eq:opeab2}
\end{align}
Similarly, we obtain:
\begin{align}
g^{\mu \nu}   \, \hat{\Pi}_{\mu \nu}^{LL} (k,0)  &=   \frac{-4 i g_s^2}{(k^2  + i \epsilon)^3}  \ \Big( - 2 k^2   (\hat{O}^{LL} )_\alpha^\alpha   
 + 2 k^\alpha k^\beta \hat{O}^{LL}_{\alpha \beta} \Big)
 \to    \frac{6 i g_s^2}{(k^2  + i \epsilon)^2}    (\hat{O}^{LL} )_\alpha^\alpha\,,\notag
\\
g^{\mu \nu}   \, \hat{\Pi}_{\mu \nu}^{LR} (k,0)  &=   \frac{4 i g_s^2}{(k^2  + i \epsilon)^3}  \ \Big( - 2 k^2   (\hat{O}^{LR} )_\alpha^\alpha   
 + 2 k^\alpha k^\beta \hat{O}^{LR}_{\alpha \beta} \Big)
  \to    \frac{- 6 i g_s^2}{(k^2  + i \epsilon)^2}    (\hat{O}^{LR} )_\alpha^\alpha\,,   
\end{align}
where the respective last step holds under symmetric integration. 
Finally, using Fierz identities we can express  $(\hat{O}^{LL} )_\alpha^\alpha$ and 
$(\hat{O}^{LR} )_\alpha^\alpha$ in terms of the four-quark scalar operator basis used in 
Ref.~\cite{Cirigliano:2018yza}, obtaining 
\begin{align}
g^{\mu \nu}   \, \hat{\Pi}_{\mu \nu}^{LL} (k,0)  &=   \frac{ i g_s^2}{(k^2  + i \epsilon)^2}  \  2 O_1 \,,\notag
\\
g^{\mu \nu}   \, \hat{\Pi}_{\mu \nu}^{LR} (k,0)  &=   \frac{ i g_s^2}{(k^2  + i \epsilon)^2}  \  2 \bar{O}_{LR}\,, 
\qquad  \bar O_{LR} = \frac{1}{2}   \left(   O_4 - 3 O_5  \right) \,.
\end{align}
These results lead to:
\be
\hat O^{LL}_> (\spacevec{k}) = \frac{3 g_s^2}{4}  \frac{1}{|\spacevec{k}|^5} \ O_1\,,\qquad 
\hat O^{LR}_> (\spacevec{k}) = \frac{3 g_s^2}{4}  \frac{1}{|\spacevec{k}|^5} \
\frac{1}{2} \left(  O_4 - 3 O_5  \right)\,.
\ee

As discussed in Ref.~\cite{Cirigliano:2018yza}, the four-quark local operators $O_{1,4.5}$ 
can be treated at  low energy in chiral EFT.   
The discussion of $O_1$ is given in Section~\ref{sect:OPELL}  and we focus here on $O_{4,5}$. 
At LO only non-derivative  four-nucleon and pion--pion vertices appear, 
\be
O_{4,5}  \ \to  \  g_{4,5}^{NN} \,  \bar p n \bar p  n
\ + \ \frac{1}{2}  g_{4,5}^{\pi \pi} \,  F_\pi^2 \,  \pi^-  \pi^-  \, + \ldots
\ee
The $\pi\pi$  couplings induce a pion-range transition operator, while the $N\!N$ 
couplings give a short-range transition operator.  
The leading chiral expression for the $nn \to pp$ matrix element of $\bar{O}_{LR}$ is given by 
\be
 \langle f_- | \,  \bar{O}_{LR}  \, | i_+ \rangle 
=  \hat{\cal A }_A  + 
\chi^+_{\spacevec p^\prime}(\spacevec 0) \, K_{E^\prime} \, \hat{\cal A}_B         
+ \   \hat{\cal A}_{\bar B} \, K_E \, \chi^+_{\spacevec p}(\spacevec 0) 
+
\chi^+_{\spacevec p^\prime}(\spacevec 0) \, K_{E^\prime}  \, \left(  \hat{\cal A}_C  + \frac{2 g_{LR}^{NN}}{C^2} \right) 
\,  K_E \, \chi^+_{\spacevec p}(\spacevec 0)\,,
\label{eq:chiEFT2LR}
\ee
where  $\hat {\cal A}_{A,B,C}$ have the same formal expression as 
${\cal A}_{A,B,C}$~\cite{Cirigliano:2019vdj} with the replacement~\cite{Cirigliano:2018yza}  
\be
V^{^1S_0}_{\nu \, {\rm L}} (\spacevec{q}) \ \to \ 
V^{^1S_0}_{\bar{O}_{LR}} (\spacevec{q}) \ =
\frac{1}{2} \,  g_A^2 \, g_{LR}^{\pi \pi} \    \frac{ \spacevec{q}^2}{(\spacevec{q}^2 + M_\pi^2)^2}\,,
\ee
where~\cite{Cirigliano:2018yza}  
\be
g_{LR}^{NN} 
= \frac{1}{2} \left( g_4^{NN} - 3 g_5^{NN}  \right) \sim \Order( (4 \pi)^2)\,,
\qquad 
g_{LR}^{\pi \pi} = \frac{1}{2} \left( g_4^{\pi \pi } - 3 g_5^{\pi \pi} \right) \sim \Order( (4 \pi F_\pi)^2)\,.
\ee
The scaling of the pion couplings follows from naive dimensional analysis and is confirmed by a lattice QCD calculation~\cite{Nicholson:2018mwc}. 
On the other hand, the scaling of the $N\!N$ couplings follows from the RGE in 
chiral EFT~\cite{Cirigliano:2018yza}, which we report here for completeness in terms of the 
rescaled couplings $\tilde{g}_{4,5}^{NN}$ defined by $g_{4,5}^{NN}  = (m_N C/(4 \pi))^2 \, \tilde{g}_{4,5}^{NN}$: 
\be
\frac{d \, \tilde{g}_{4,5}^{NN} }{d \log \nu_\chi} = \frac{g_A^2}{4} \, g_{4,5}^{\pi \pi}\,.
\ee
The scaling for the nucleon couplings follows from the above RGE and the scaling  $m_N C/(4 \pi)  \sim  1/Q$.
As a consequence of the above discussion, the coupling $g_{LR}^{ NN}$ depends on both the short-distance QCD renormalization 
scale $\mu$ and on the chiral renormalization scale $\nu_\chi$. 

Using the above relations we arrive at 
\be
{\cal M}_{LR}^>  = \mathcal A^>_A  + 
\chi^+_{\spacevec p^\prime}(\spacevec 0) \, K_{E^\prime} \, \mathcal A^>_B         
+{\mathcal A}^>_{\bar B} \, K_E \, \chi^+_{\spacevec p}(\spacevec 0) 
+
\chi^+_{\spacevec p^\prime}(\spacevec 0) \, K_{E^\prime}  \, \mathcal A^>_C  \,  K_E \, \chi^+_{\spacevec p}(\spacevec 0)\,, 
\label{eq:MpLR}
\ee
with
\begin{align}
{\cal A}_{A,B, \bar B}^> &=   \frac{3 \alpha_s}{2 \pi} \, \int_{\Lambda}^{\infty} \ d |\spacevec{k}| \ \frac{1}{|\spacevec{k}|^3}  \
\hat{\cal A}_{A, B, \bar B}\,,\notag
\\
{\cal A}_C^> &=   \frac{3 \alpha_s}{2 \pi} \, \int_{\Lambda}^{\infty} \ d |\spacevec{k}| \ \frac{1}{|\spacevec{k}|^3}  \
\left( \frac{2 g_{LR}^{NN}}{C^2}   + \hat{\cal A}_C \right)\,.
\label{eq:ACpLR}
\end{align}
For matching purposes only  ${\cal A}_C^>$ matters. 
In the numerical estimates we will retain only the term proportional to $g_{LR}^{NN}$ 
and the singular part of  $\hat{\cal A}_C $, proportional to $g_{LR}^{\pi \pi}$. 
Further, for convenience in the matching analysis we express  the couplings $g_{LR}^{NN}$ and $g_{LR}^{\pi \pi}$ 
in terms of dimensionless quantities as follows
\begin{align}
g_{LR}^{NN} &=  \left( \frac{m_N}{4 \pi} C \right)^2 \, (4 \pi F_\pi)^2 \, \bar{g}_{LR}^{NN}\,, & \bar{g}_{LR}^{NN} &\sim \Order(1)\,,\notag
\\
g_{LR}^{\pi \pi} &=    (4 \pi F_\pi)^2 \, \bar{g}_{LR}^{\pi \pi}\,, & \bar{g}_{LR}^{\pi \pi}  &\sim \Order(1)\,.
\label{eq:gLRNN}
\end{align}
In numerical estimates we will use 
$\bar{g}_{LR}^{\pi \pi} =8.23 $ 
(in the $\overline{\rm MS}$ scheme at $\mu = 2\GeV$, 
based on  the lattice QCD results of Ref.~\cite{Nicholson:2018mwc})
and assume the range $\bar g_{LR}^{NN} \in [-10,+10]$.

\section{Estimating  inelastic effects}
\label{sect:inel}

As a representative of inelastic effects, we 
estimate the contribution from the  diagram in 
Fig.~\ref{Fig:inel}.\footnote{This corresponds to diagram $(G)$ in Fig.~3 of Ref.~\cite{Cirigliano:2017tvr}.}
The contribution to the generalized Compton amplitude in 
Eq.~\eqref{eq:M}  is already a two-loop diagram, which with appropriate momentum routing can be written as 
\begin{align}
\delta {\cal T}(k,p_{\rm ext}) 
&=iC^2 \  \frac{(1 + 3 g_A^2) m_N^2}{16 F_\pi^2}\notag \\
&\times  \int \frac{d^3 \spacevec{q}}{(2\pi)^3} \int \frac{d^3 \spacevec{q}^\prime}{(2\pi)^3}
\, \frac{1}{ \spacevec{p}^2 - \spacevec{q}^2 + i \epsilon}
\, \frac{1}{ \spacevec{p}^{\prime 2} - \spacevec{q}^{\prime 2} + i \epsilon}
\frac{1}{- (k^0)^2 + (\spacevec{q} - \spacevec{q}^\prime + \spacevec{k})^2 + M_\pi^2}\,.
\end{align}
Upon changing variables to $\spacevec{q}$ and $\spacevec{l} =  \spacevec{q} - \spacevec{q}^\prime$ one has 
\be
\delta {\cal T}(k,p_{\rm ext})  = iC^2 \  \frac{(1 + 3 g_A^2) m_N^2}{16 F_\pi^2} \ \int \frac{d^3 \spacevec{l}}{(2\pi)^3}   
\ \frac{1}{- (k^0)^2 + (\spacevec{l}  + \spacevec{k})^2 + M_\pi^2 - i \epsilon} \  {\cal I}_C  (\spacevec{l}^2, \spacevec{p}^2, 
\spacevec{p}^{\prime 2} ) \,.
\ee
Recalling that ${\cal I}_C \to 1/(8 |\spacevec{l}|)$ for $\spacevec{p} = \spacevec{p}^\prime \to 0$, one obtains
\be
\delta {\cal T}(k,p_{\rm ext}) 
 =  iC^2 \  \frac{(1 + 3 g_A^2) m_N^2}{128 F_\pi^2} \ \int \frac{d^3 \spacevec{l}}{ (2\pi)^3}     
\ \frac{1}{- (k^0)^2 + (\spacevec{l}  + \spacevec{k})^2 + M_\pi^2 - i \epsilon} \  \frac{1}{|\spacevec{l}|}\,.
\ee
The integral is logarithmically divergent and we  regulate the divergence with a cutoff scale  $\nu \sim \Lambda_\chi$. 
We will  mimic the effect of the associated (and unknown) counter term by varying the scale $\nu$ in order to obtain a rough estimate of the corresponding inelastic effect. 

The inelastic  contribution to ${\cal A}_C^<$  is given by 
\be
\delta {\cal A}_C^<   = 
2\,  \int \frac{d^4k}{(2 \pi)^4} 
\frac{  \delta  {\cal T} (k, p_{\rm ext}) 
 }{ k^2  + i \epsilon}\,. 
\ee
We reduce this integral by first   integrating over $k^0$ via the residues at the pion and neutrino poles, and second 
 performing the angular integral in $d^3 \spacevec{l}$, obtaining
\be
\delta {\cal A}_C^<  
= \frac{m_N^2}{(4 \pi)^2} \frac{1}{ (4 \pi F_\pi)^2}  
\frac{1}{4}   \ 
\int \, d |\spacevec{k}| \,  (1 + 3 g_A^2)     \  \int_0^\nu   d | \spacevec{l} |  
\ \log \left[ 
\frac{ |\spacevec{k}| + \sqrt{ ( |\spacevec{k}| + |  \spacevec{l} |   )^2 + M_\pi^2 } 
}{|\spacevec{k}| + \sqrt{ ( |\spacevec{k}| - |\spacevec{l}| )^2 + M_\pi^2 }}
\right]\,.
\ee
The integral in $d |\spacevec{l}|$  can be done analytically, leading to 
\begin{align}
\delta {\cal A}_C^<  &= \frac{m_N^2}{(4 \pi)^2} 
  \ 
\int \, d |\spacevec{k}| \    \delta a_< ( | \spacevec{k} |)\,,\notag
\\
\delta a_< ( | \spacevec{k} |) &=
\frac{1   + 3 g_A^2 }{ 2 (4 \pi F_\pi)^2} \
 \left[    | \spacevec{k} |   \left( 1 + \log \frac{\nu \, M_\pi}{(|\spacevec{k}| + \sqrt{\spacevec{k}^2 + M_\pi^2}  )^2}  \right)  + M_\pi F\bigg(\frac{|\spacevec{k}|}{M_\pi}\bigg)  \right]\,,\notag
 \\
 F(x) &= \sqrt{1 - x^2} \, \left[ 
 \arctan \left( \frac{x^2}{\sqrt{1 - x^4}} \right) 
 -  \arctan \left( \frac{x}{\sqrt{1 -x^2}} \right) 
 \right]\,.
 \label{eq:ainel}
\end{align}
In the $M_\pi \to 0$ limit the integrand becomes 
\be
\delta a_< ( | \spacevec{k} |) = 
\frac{1   + 3 g_A^2 }{ 2 (4 \pi F_\pi)^2} \, 
 | \spacevec{k} |   \left( 1 + \log \frac{\nu}{4 |\spacevec{k}|}   \right)\,.
\ee

The logarithmic divergence signals that there must be, at this order, a counter term for the generalized Compton amplitude   
(the divergence arose at that level, before integrating over $d^3 \spacevec{k}$). 
Naive dimensional analysis suggests that this counter term should be of the same order of the coefficient of $\log \nu$. 
To obtain a numerical estimate we vary the scale within
$\nu \sim (0.5\text{--}1.0)\GeV$.
The  new contribution to the integrand is clearly suppressed by two chiral orders compared to the leading terms, as long as $|\spacevec{k}| \ll \Lambda_\chi$, 
as one is comparing  the LO $1/|\spacevec{k}|$  to  $|\spacevec{k}|/(4 \pi F_\pi)^2$.
However, this leads to a badly divergent integral in $|\spacevec{k}|$. 
In order to estimate the contribution to the integral, and therefore to the LEC, 
\be  
\delta \tilde{C}_1 = \frac{1}{2} \ \delta  \bar{\cal A}_C^<\,,  
\ee
we explore  two options: first, one may use
the chiral EFT form of the integrand and cut it off at $\Lambda_\chi \sim 1\GeV$, or, second, one may 
replace $1+ 3 g_A^2 \to g_V^2 (\spacevec{k}^2)   + 3 g_A^2 (\spacevec{k}^2)$ in Eq.~\eqref{eq:ainel} 
and integrate up to $\Lambda$.  
Numerically, the second option leads to $|\delta \tilde{\cal C}_1 | = 0.15$, while the first option leads to $|\delta \tilde{\cal C}_1| = 0.35$ 
(the largest absolute value is obtained for $\nu = 0.5\GeV$). 
Based on these observations, we will take  into account the effects of inelastic channels  by adding a $\pm 0.5$ uncertainty to our estimate of $\tilde {\cal C}_1  (\mu_\chi)$.

\section{Evaluation of $\boldsymbol{Z^<}$}
\label{app:Zm}

Using ${\cal T}_<^{\pi \pi} (k,0)$  from Eqs.~\eqref{eq:tpipim}, \eqref{eq:Tpipim}, and  \eqref{eq:pionFF} 
we obtain 
\be
Z^< = \frac{3 i}{2 F^2} \, \int \frac{d^4k}{(2 \pi)^4} \ \frac{1}{k^2 + i \epsilon} \left( \frac{M_V^2}{M_V^2 - k^2} \right)^2\,.
\label{eq:Zint}
\ee
We  evaluate this integral in two ways.  First, performing the usual Wick rotation and introducing the variable 
$k_E^2 = (k^0)^2 + \spacevec{k}^2$ to reduce the integral 
to\footnote{One could also perform the integrals in analogy to the derivation of the Cottingham formula 
and would obtain the same result in this case, because after the change of variables the integrand does not depend on $k^0$ and 
$\int_{-k_E}^{+k_E}  dk^0 \sqrt{k_E^2 - (k^0)^2} = (\pi k_E^2)/2$. }
\be
Z^< = \frac{3}{2} \frac{1}{(4 \pi F)^2} \, \int_0^\infty  d k_E^2   \left( \frac{M_V^2}{M_V^2 + k_E^2} \right)^2  = 
 \frac{3}{2} \frac{M_V^2}{(4 \pi F)^2} \, \int_0^\infty  d x   \left( \frac{1}{1 + x} \right)^2  = 
  \frac{3}{2} \frac{M_V^2}{(4 \pi F)^2}\,.
  \label{eq:Zmstandard}
\ee
Since the integral is convergent, we have taken the integration limit in $k_E^2$ to $+ \infty$. 
Note that the overall normalization of the integral above agrees with the literature on the pion mass difference, see, e.g., Ref.~\cite{Bardeen:1988zw}.

Next, we perform the integral~\eqref{eq:Zint}  by first carrying out the $k^0$ integration via the residues' theorem, which allows us to identify 
the integrand in the variable $|\spacevec{k}|$ in Eq.~\eqref{eq:Zmatch}.
The integrand has single poles at $k^0 = \pm (|\spacevec{k}| - i \epsilon)$ and double poles at $k^0 = \pm (\omega_V  - i \epsilon)$, where 
$\omega_V = \sqrt{\spacevec{k}^2 + M_V^2}$. 
The integration gives:
\begin{align}
Z^<  &= \frac{3}{4 F^2}  \int \frac{d^3 \spacevec{k}}{(2 \pi)^3} \ \frac{1}{|\spacevec{k}|}
 \left( 
1 - \frac{|\spacevec{k}|}{\omega_V} \left( 1 + \frac{M_V^2}{2 \omega_V^2}\right) 
\right)
\nonumber \\
&=  \frac{3}{2} \frac{1}{(4 \pi F)^2}\ 
4 \int_0^\Lambda d |\spacevec{k}|   |\spacevec{k}|
 \left( 
1 - \frac{|\spacevec{k}|}{\omega_V} \left( 1 + \frac{M_V^2}{2 \omega_V^2}\right) 
\right)\,,
\label{Z<}
\end{align}
which is the result used in Eq.~\eqref{eq:Zmatch}.
As a check, we perform the integration up to $\Lambda \to \infty$, obtaining
\be
Z^< (\Lambda \to \infty) 
=  \frac{3}{2} \frac{M_V^2}{(4 \pi F)^2}\ 
4 \int_0^\infty d x x 
 \left(  1 - \frac{x}{2}  \frac{3 + 2 x^2}{(1 + x^2)^{3/2}}   \right) 
= \frac{3}{2} \frac{M_V^2}{(4 \pi F)^2}\,. 
\ee
In the last step we used the fact that the integral in the $x$ variable evaluates to $1/4$.
This result is consistent with Eq.~\eqref{eq:Zmstandard}, obtained with the more commonly used integration variable $k_E^2$. The advantage of the standard approach is that the result directly applies to an arbitrary pion vector form factor, while collecting the residues in the $k^0$ integration assumes the form dictated by vector meson dominance. For our application this approximation is completely sufficient, but could be improved by using a dispersive representation of $F_\pi^V(k^2)$~\cite{Colangelo:2018mtw,Colangelo:2020lcg} and treating the Cauchy kernel along the same lines.

\section{Details on  CIB in $\boldsymbol{N\!N}$ scattering near threshold}
\label{sect:AppCIB}

\subsection[Low-energy $N\!N$ scattering in the presence of Coulomb interactions]{Low-energy $\boldsymbol{N\!N}$ scattering in the presence of Coulomb interactions}

We begin by  recalling  some results  on the Coulomb-modified effective range expansion for $pp$ scattering~\cite{Bethe:1949yr,Jackson:1950zz,Kong:1999sf}. 
We specialize  to the $S$-wave and indicate the pure Coulomb phase shift by $\sigma$ and  
the full phase shift by $\sigma + \nu$. Further, we denote
 the total scattering amplitude by $T$,  the purely Coulomb component by $T_C$, and  the Coulomb-modified strong amplitude by 
$T_{SC}$   
\begin{align}
T  &= T_C + T_{SC} = \frac{4 \pi}{m_p} \frac{e^{2 i (\sigma (k) + \nu (k))} -1}{2 i k}\,,\notag
\\
T_C & =    \frac{4 \pi}{m_p} \frac{e^{2 i  \sigma (k) } -1}{2 i k}\,,\notag 
\\
T_{SC} &=  \frac{4 \pi}{m_p}     e^{i 2 \sigma (k)}    \frac{ \left(    e^{2 i  \nu (k) } -1\right) }{2 i k}  = 
 \frac{4 \pi}{m_p}     e^{i 2 \sigma (k)}  \frac{1 }{ k  \left(  {\rm cot} (\nu (k) ) -   i \right)}\,.
\end{align}
The phase shift $\nu (k)$ obeys a modified effective range expansion~\cite{Bethe:1949yr,Jackson:1950zz,Kong:1999sf}
\be
\frac{1 }{ k  \left(  {\rm cot} (\nu (k) ) -   i \right)}  =  \frac{C_0^2 (\eta)}{ - \frac{1}{a_C}  -  r_0 \frac{k^2}{2}  - \alpha m_p h (\eta) + \ldots }\,,
\ee
where 
\be
\eta = \frac{\alpha m_p}{2 k}\,,  \qquad C_0^2 (\eta) =  \frac{2 \pi \eta}{e^{2 \pi \eta} - 1}\,, \qquad 
h (\eta) = {\rm Re}  \left( \frac{\Gamma^\prime (i \eta)}{\Gamma (i \eta)} \right) - \log \eta\,. 
\ee
The standard effective range expansion is recovered formally by setting $\alpha \to 0$, which also implies $C_0^2 (\eta) \to 1$. 
For $\alpha \neq 0$,  the function $h(\eta)$  goes  smoothly to zero  for $k \to 0$, so that  $a_C$ can be interpreted as a scattering length. 
However, the Sommerfeld factor $C_0^2(\eta)$---the square of the Coulomb wave function at the origin---goes to zero for $k \to 0$ 
due to Coulomb repulsion  and this prevents the usual  threshold analysis of the amplitude. 
It will prove useful to work with  the momentum-dependent quantity
\be
\tilde{a}_{pp} (k) =   \frac{C_0^2 (\eta)}{  \frac{1}{a_C}  + \alpha m_p h (\eta)  }\,, 
\ee
which is a proxy for the very low-energy Coulomb-subtracted  $pp$ scattering amplitude. 

\begin{figure}[t]
\centering
\includegraphics[width=0.5\textwidth]{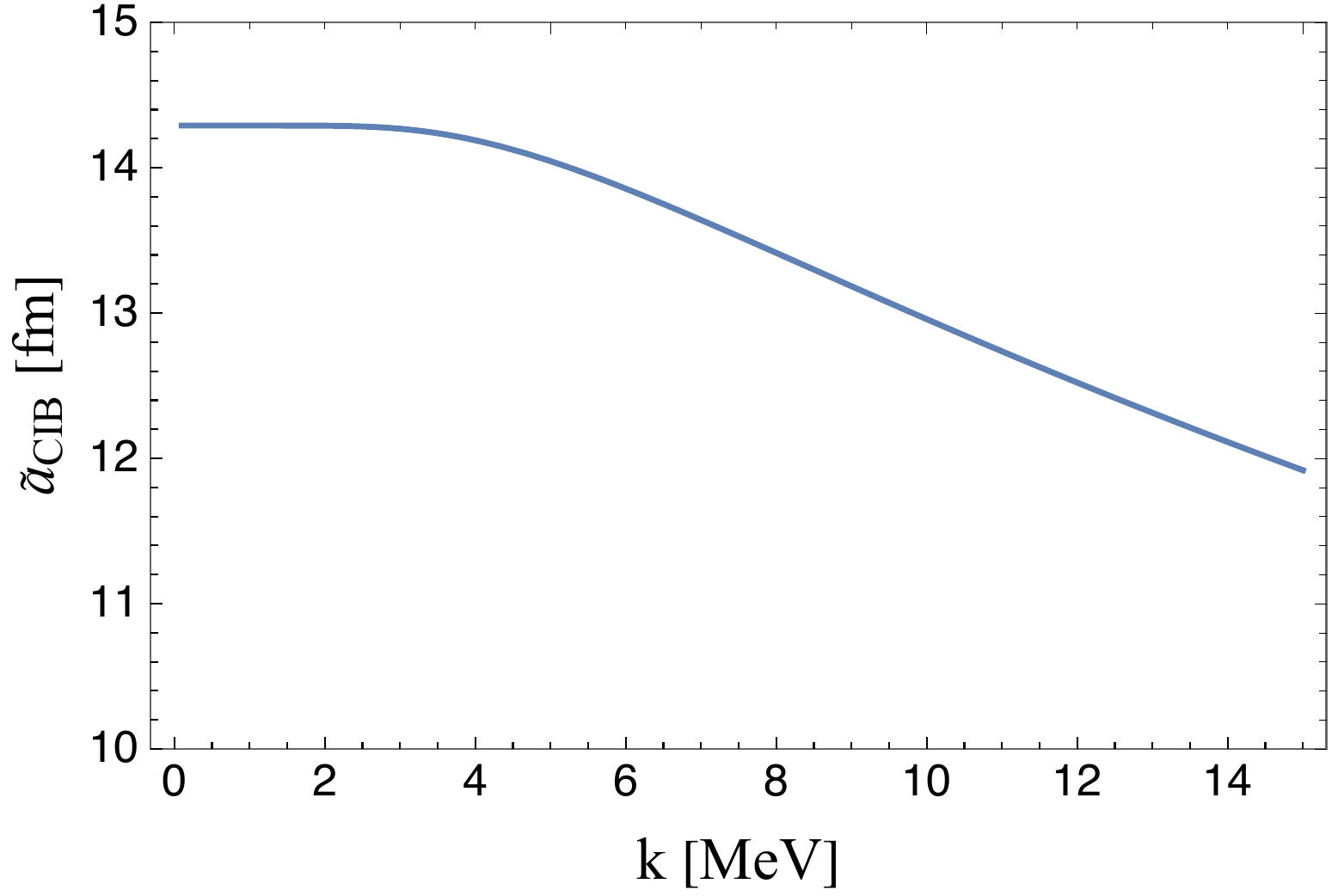}
\caption{
$\tilde{a}_{\rm CIB}  (k)$  defined in Eq.~\eqref{eq:aCIB1}  versus   $k$, using experimental input from Eq.~\eqref{eq:CIBscatt}. 
}
\label{Fig:aCIBofk}
\end{figure}

For the CIB analysis we will consider the quantities 
\be
\tilde{a}_{\rm CIB}  (k) =  \frac{ a_{nn} + \tilde{a}_{pp} (k)}{2} - a_{np}
\label{eq:aCIB1}
\ee
as well as 
\be
{a}_{\rm CIB}  =  \frac{ a_{nn} + {a}^C_{pp}}{2} - a_{np}\,,
\label{eq:aCIB2}
\ee
and we will use the  following values for the  scattering lengths determined from 
$N\!N$ data~\cite{Klarsfeld:1984es,Machleidt:2000ge,Chen:2008zzj,Bergervoet:1988zz,Reinert:2017usi}:
\begin{equation}
a_{np} = -23.74 (2)  \, \textrm{fm}\,, 
\qquad 
a_{nn} = -18.9 (4) \, \textrm{fm}\,,
\qquad 
a_{pp}^C = -7.817 (4)  \, \textrm{fm}\,.
\label{eq:CIBscatt}
\end{equation}
In Fig.~\ref{Fig:aCIBofk} we show $\tilde{a}_{\rm CIB} (k)$ using the input from Eq.~\eqref{eq:CIBscatt}. 

\subsection[Low-energy $N\!N$ amplitudes in chiral EFT with isospin breaking]{Low-energy $\boldsymbol{N\!N}$ amplitudes in chiral EFT with isospin breaking}

Using the techniques of Ref.~\cite{Kaplan:1996xu}, 
we  extract the phase shift $\nu (k)$ 
from the asymptotic behavior of the solution to the Schr\"odinger equation with inclusion 
of Coulomb potential and  the LO chiral potential. 
The $^1S_0$ chiral potential includes the LO channel-dependent  contact interaction 
(denoted by $C_{nn,pp,np}$ in the following), as well as the channel-dependent 
Yukawa potential with   the electromagnetically induced pion mass splitting. 
By including these effects in the Schr\"odinger equation we effectively iterate the 
isospin-breaking corrections to all orders.  This is mandatory for the Coulomb potential, 
while  for the other terms a perturbative treatment should be numerically close to our analysis, 
see Ref.~\cite{Cirigliano:2019vdj} for additional details. 

The phase shift $\nu(k)$ is related to the complex coefficients $y_{pp} (k)$ and $z_{pp} (k)$ 
controlling the large-$r$ behavior of the regular and irregular solutions of 
the Schr\"odinger equation with Coulomb and Yukawa potentials  
\begin{align}
\psi_\text{reg} (r)   &\to  y_{pp}  (k) \, \frac{  \ e^{i    (k r - \eta \log (2 k r) + \sigma (k))} }{ k r} + {\rm c.c.}\,,\notag
\\
\psi_\text{irr} (r) &\to  z_{pp}  (k) \, \frac{  \ e^{i    (k r - \eta \log (2 k r) + \sigma (k))} }{ k r} + {\rm c.c.}
\end{align} 
By matching the low-energy form of the amplitude  obtained in EFT to the (modified) effective range form, 
one obtains 
\begin{align}
a_{np}&= a_{np}^{\pi}  + G (C_{np}, y_{np}, z_{np})\,,\notag
\\
a_{nn}&= a_{nn}^{\pi}  + G (C_{nn}, y_{nn}, z_{nn})\,,\notag
\\
\tilde{a}_{pp} (k) &= a_{pp}^{\pi}  + G (C_{pp}, y_{pp}, z_{pp})\,,\notag 
\\
G (C_{ij} ,y_{ij},z_{ij}) &= \frac{m_{ij}}{16 \pi y_{ij}^{*2} }  \ \frac{1}{-\frac{1}{C_{ij} (\mu)} + \Delta_{ij} \big(\frac{\mu}{\lambda}\big)+ \frac{z^* (\lambda)}{y^*}}\,.
\label{eq:aij}
\end{align} 
In the above relations $a_{ij}^{\pi}$ is the Yukawa-induced scattering length (obtained in terms of the phase of $y_{ij} (k)$ at small $k$). 
$m_{ij}$ is the appropriate channel-dependent  mass for the nucleon system 
($m_{nn}= m_n$, $m_{pp} = m_p$, $m_{np } = 2 m_n m_p/ (m_n + m_p)$). 
The quantities $z_{ij}$  depend on the regulator $\lambda$ introduced when imposing boundary conditions for the irregular 
solution of the Schr\"odinger equation~\cite{Kaplan:1996xu}.
The results quoted below are obtained with $1/\lambda = 0.001\fm$ and we have checked stability  
in the range  $1/\lambda = 0.001 \to 0.05\fm$.  
The dependence on $\lambda$ is canceled by the matching factors 
$\Delta_{ij} (\mu/\lambda)$, to connect to the $\overline{\rm MS}$ scheme, in which
the contact couplings $C_{ij} (\mu)$ are defined. 
The explicit form of these scheme-changing factors is 
\begin{align}
\Delta_{nn} &= - \alpha_\pi  \frac{m_{nn}^2}{8 \pi} \left( \log \frac{\mu^2}{\lambda^2} + 2 \gamma_E - 1 \right)\,,
\qquad \alpha_\pi = \frac{g_A^2 M_{\pi^0}^2}{16 \pi F_\pi^2}\,,\notag
\\
\Delta_{np} &= - \alpha_\pi \left( 1 + 2  \frac{M_{\pi^\pm}^2 - M_{\pi^0}^2}{M_{\pi^0}^2} \right)  \frac{m_{np}^2}{8 \pi} \left( \log \frac{\mu^2}{\lambda^2} + 2 \gamma_E - 1 \right)\,,\notag
\\
\Delta_{pp} &= - (\alpha_\pi  - \alpha) \,  \frac{m_{pp}^2}{8 \pi} \left( \log \frac{\mu^2}{\lambda^2} + 2 \gamma_E - 1 \right)\,.
\end{align}

In the presence of isospin breaking, the short-range $^1S_0$ $N\!N$ couplings can be decomposed as follows 
\begin{align}
C_{np} &=  C + \frac{e^2}{3} \left( {\cal C}_1 + {\cal C}_2 \right)\,,\notag 
\\
C_{nn} &= C - \frac{e^2}{6} \left( {\cal C}_1 + {\cal C}_2 \right)  + \frac{1}{2}  {\cal C}_{\rm CSB}\,,\notag
\\
C_{pp} &= C - \frac{e^2}{6} \left( {\cal C}_1 + {\cal C}_2 \right)  - \frac{1}{2}  {\cal C}_{\rm CSB}\,,
\end{align}
where the CIB combination  ${\cal C}_1+{\cal C}_2$  
arises form $I=2$  interactions, while the  CSB term ${\cal C}_{\rm CSB}$ stems 
from $I=1$ interactions.  The latter term can originate from strong isospin-breaking $\propto (m_u - m_d)$ or electromagnetic interactions. 
For the CIB coupling, renormalization arguments (cancellation of divergences) enforces the LO scaling  $\sim e^2/Q^2$~\cite{Epelbaum:1999zn,Cirigliano:2018hja,Cirigliano:2019vdj}. 
For the CSB coupling, the Coulomb potential also induces a  UV divergence that requires it to scale as  $\sim e^2/Q^2$.

\subsection{Fitting the couplings}
Equation~\eqref{eq:aij} can be  used  to extract the three contact couplings $C_{np,nn,pp}$ (or equivalently $C$, ${\cal C}_1+{\cal C}_2$, 
${\cal C}_{\rm CSB}$).   The output is very stable  when using momenta near threshold, in the range $k \in [1\MeV, 10\MeV]$. 
For definiteness, we quote results for $k=1\MeV$. 
Using the reference scale $\mu = M_\pi \equiv (M_{\pi^0} + 2M_{\pi^\pm})/3$, 
this analysis leads to 
\begin{align}
C_{ij} (\mu) &= - \frac{1}{(\Lambda_{ij} (\mu))^2}\,, & \Lambda_{nn,pp,np} (M_\pi) &= (95.4,88.7, 98.9)\MeV\,,\notag
\\
C (\mu) &= - \frac{1}{(\Lambda_C (\mu))^2}\,, & \Lambda_C (M_\pi) &= 94.4\MeV\,,\notag 
\\
\frac{e^2}{2} ({\cal C}_1 + {\cal C}_2) (\mu)  &= \frac{1}{(\Lambda_{\rm CIB} (\mu))^2}\,, 
&   \Lambda_{\rm CIB} (M_\pi) &= 247\MeV\,, 
\qquad (\tilde{\cal C}_1 + \tilde{\cal C}_2) (M_\pi) = 5.1\,,\notag 
\\
{\cal C}_{\rm CSB} (\mu) &= \frac{1}{(\Lambda_{\rm CSB} (\mu))^2}\,, 
& \Lambda_{\rm CSB} (M_\pi) &= 254\MeV\,.
\end{align}
The result indicates that both CIB and CSB effects provide corrections to the  isosymmetric coupling $C$ at the (10--15)\% level.

\subsection{Predicting  the CIB combination of scattering lengths}

Alternatively, we can use Eq.~\eqref{eq:aij} 
to validate our calculation of 
$({\cal C}_1 + {\cal C}_2) (\mu)$ as follows:
\begin{enumerate}
\item   We use  one  linear combination of  Eq.~\eqref{eq:aij} 
to extract  the isospin-conserving coupling $C$.

\begin{figure}[t]
\begin{center}
\includegraphics[width=0.45\linewidth]{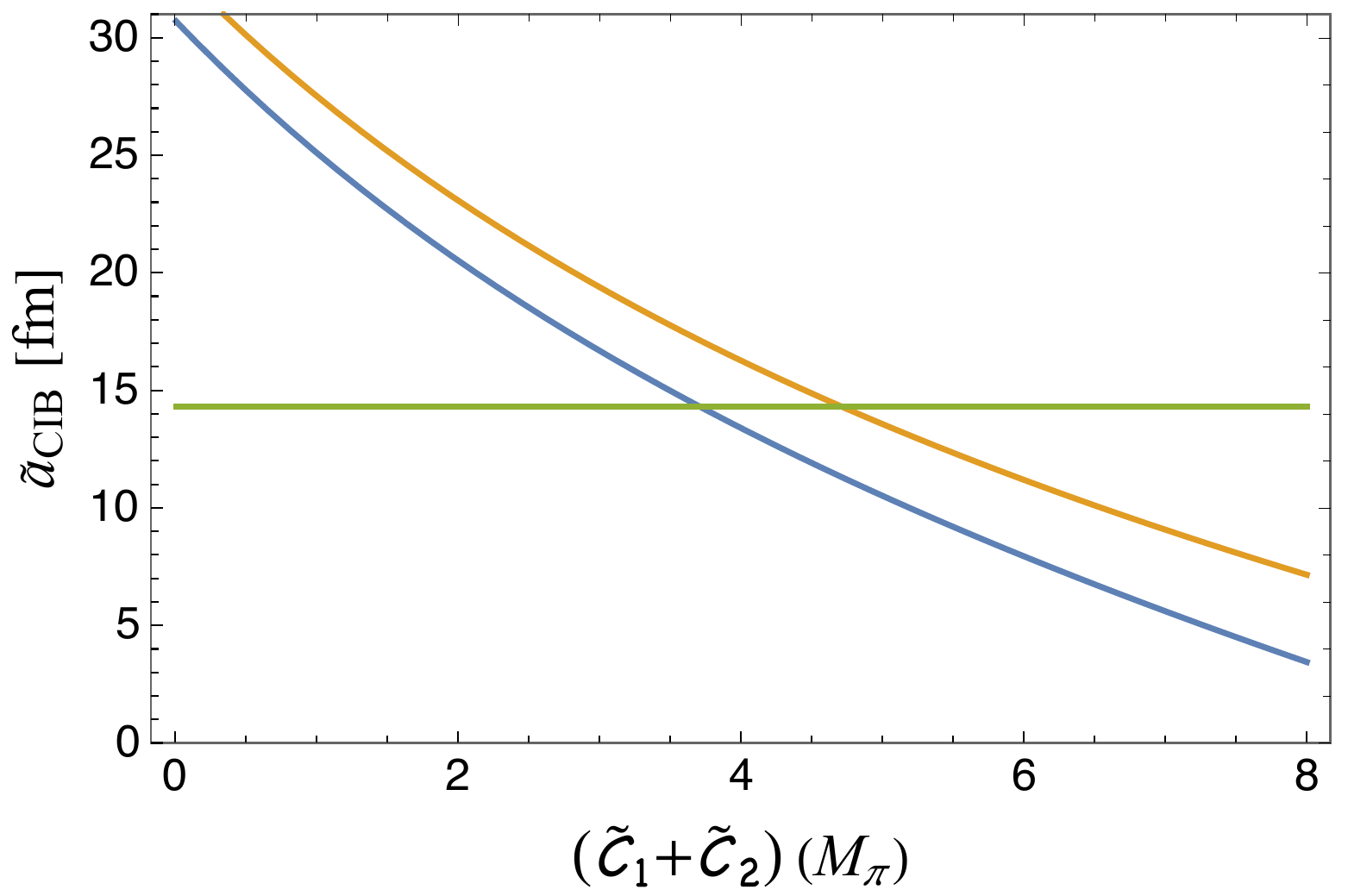}
\hspace{0.03\linewidth}
\includegraphics[width=0.45\linewidth]{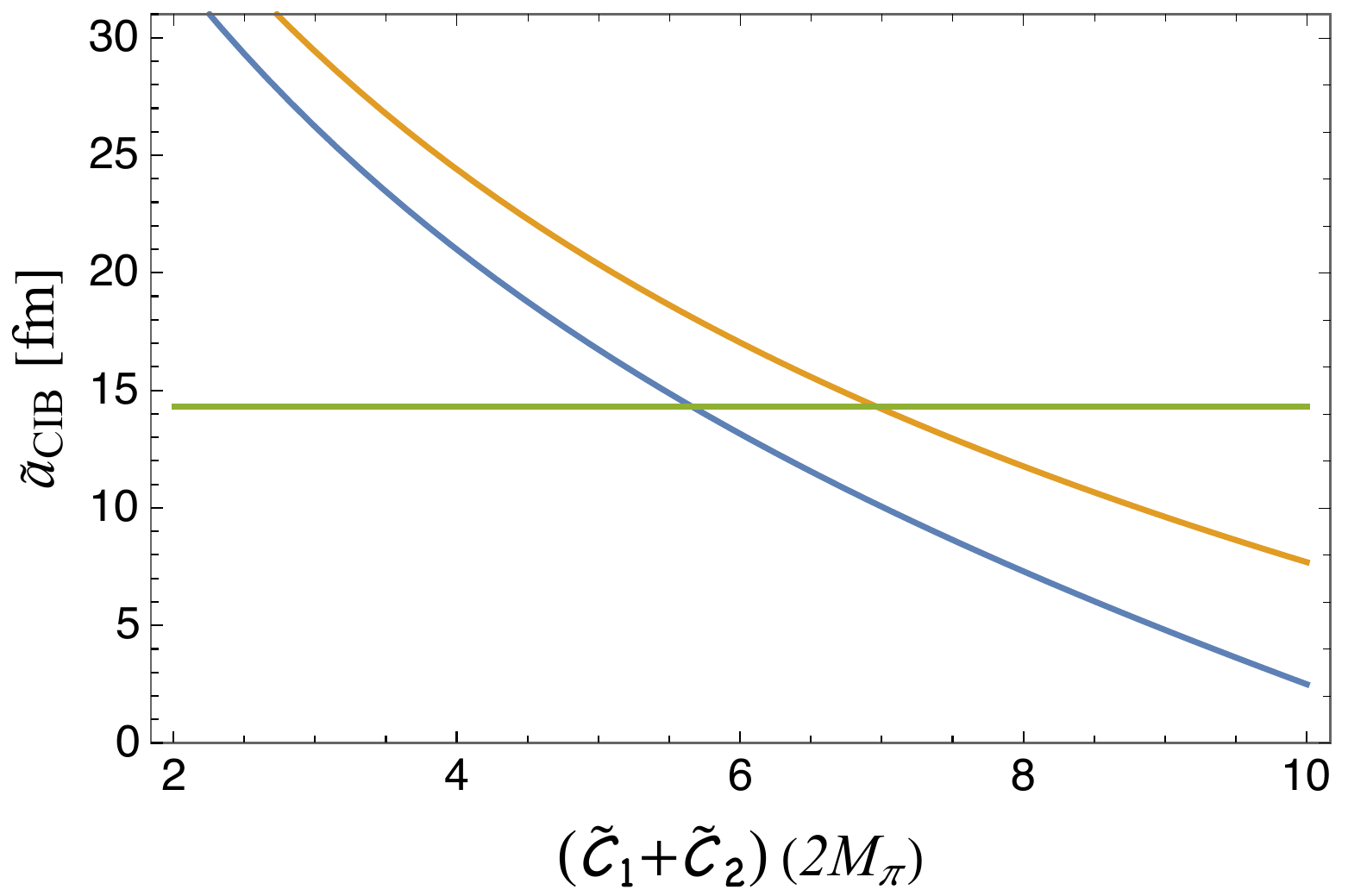}
\caption{  $\tilde{a}^{\rm th (1)}_{\rm CIB} $  (lower curve) and 
$\tilde{a}^{\rm th (2)}_{\rm CIB}$  (upper curve), 
both defined in Eq.~\eqref{eq:th12},  versus the dimensionless coupling $\tilde{\cal C}_1 + \tilde{\cal C}_2$. 
The curves  are evaluated at  $k_0 = 1\MeV$ for  $\mu = M_\pi$ (left panel)  and  $\mu = 2 M_\pi$ (right panel). 
The horizontal line represents the experimental value  $a_{\rm CIB} (k=1\MeV) = 14.25\fm$. 
For reference, our theoretical ranges  for the CIB LEC are $ (\tilde{\cal C}_1 + \tilde{\cal C}_2) (M_\pi) = 2.9(1.2)$ 
and $ (\tilde{\cal C}_1 + \tilde{\cal C}_2) (2 M_\pi) = 5.4(1.2)$. 
\label{fig:aCIB}
}
\end{center}
\end{figure}

\item We construct the CIB combinations $\tilde{a}_{\rm CIB} (k)$ and $a_{\rm CIB}$. 
The theoretical expressions  
constructed from the right-hand side of Eq.~\eqref{eq:aij} 
depend  in general on  $C$,  $({\cal C}_1 + {\cal C}_2)$, and ${\cal C}_{\rm CSB}$. 
We will fix  $C$  to the fit value (see previous step) and consider 
the dependence of the theoretical expressions $\tilde{a}^{\rm th}_{\rm CIB} (k)$ and $a_{\rm CIB}^{\rm th}$  on 
 the $I=2$ coupling $({\cal C}_1 + {\cal C}_2) (\mu)$.    
 Using our range for $({\cal C}_1 + {\cal C}_2) (\mu)$ we can predict   $\tilde{a}_{\rm CIB} (k)$  and $a_{\rm CIB}$, 
and compare to the experimental value.

\item Working to first order in isospin-breaking quantities, 
the observable $\tilde{a}^{\rm th}_{\rm CIB} (k)$  
should be insensitive to the $I=1$ coupling ${\cal C}_{\rm CSB}$. 
However, in Eq.~\eqref{eq:aij}  we are effectively  iterating  the insertions of electromagnetic  isospin-breaking sources to all orders (including  ${\cal C}_{\rm CSB}$),   since we are iterating the full couplings $C_{ij}$.  
As indicated by  ${\cal C}_{\rm CSB} / C \sim 15\%$, 
second-order effects due to two insertions of ${\cal C}_{\rm CSB}$ 
might  not  be negligible. 
To quantify the impact of this on our CIB analysis, 
in Fig.~\ref{fig:aCIB}  we show  for $k_0 = 1\MeV$,  $\mu = M_\pi$ (left panel), and $\mu = 2 M_\pi$ (right panel) 
the two functions 
\begin{align}
\tilde{a}^{\rm th (1)}_{\rm CIB}   (\tilde{\cal C}_1 + \tilde{\cal C}_2)   &=  \tilde{a}^{\rm th}_{\rm CIB} (k=k_0)  \Big\vert_{C= C_{\rm fit},\,  {\cal C}_{\rm CSB} =0}\,,\notag
\\
\tilde{a}^{\rm th (2)}_{\rm CIB}   (\tilde{\cal C}_1 + \tilde{\cal C}_2)  &= \tilde{a}^{\rm th}_{\rm CIB} (k=k_0)  \Big\vert_{C= C_{\rm fit},\,  {\cal C}_{\rm CSB} ={\cal C}_{\rm CSB, \, fit}}\,, 
\label{eq:th12}
\end{align}
where in  both diagrams the horizontal line represents the experimental value of $\tilde a_{\rm CIB} (k=1\MeV) = 14.25\fm$.

\item  A few comments are in order: 
(i) The plots clearly show the need for a LO CIB coupling. 
Not including $  (\tilde{\cal C}_1 + \tilde{\cal C}_2)  $ would lead to a scale-dependent prediction for $\tilde{a}_{\rm CIB} (k)$. 
Moreover, at $\mu = M_\pi$ the theory would predict $\tilde{a}_{\rm CIB} (k=1\MeV) \sim 30\fm$, more than a factor of two 
larger than the experimental value.  
(ii) The two choices of treating ${\cal C}_{\rm CSB}$  lead to a shift $\Delta \tilde{a}_{\rm CIB} \sim 2.5\fm$ at $\mu=M_\pi$ (growing to $\sim 4\fm$
at $\mu = 2 M_\pi$), subdominant but not negligible.  
(iii) One could use the intercept of  the curves in Fig.~\ref{fig:aCIB} with the horizontal experimental constraint to perform an 
alternative extraction of  $(\tilde{\cal C}_1 + \tilde{\cal C}_2)$.  
Due to the missing CSB contribution, 
using $\tilde{a}^{\rm th (1)}_{\rm CIB} $  implies a value of  $(\tilde{\cal C}_1 + \tilde{\cal C}_2) (M_\pi)  \sim 3.7$, to be contrasted with the $5.1$ from the previous analysis. 
(iv) Finally, we note that using  $\tilde{a}^{\rm th (2)}_{\rm CIB}$  produces results that are more stable under variation of $\mu$ (see below).

\item Our theoretical ranges  for the CIB LEC are $ (\tilde{\cal C}_1 + \tilde{\cal C}_2) (M_\pi) = 2.9(1.2)$ 
and $ (\tilde{\cal C}_1 + \tilde{\cal C}_2) (2 M_\pi) = 5.4(1.2)$. These values  can be used in conjunction with 
$a^{\rm th (1,2)}_{\rm CIB}   (\tilde{\cal C}_1 + \tilde{\cal C}_2)$ to predict $a_{\rm CIB}$.  
Using $\tilde{a}^{\rm th (1)}_{\rm CIB} $ we find 
\begin{align} 
\tilde{a}_{\rm CIB} (k=1\MeV) &= 17^{+5}_{-4}\fm      &\mu &= M_\pi\,,\notag
\\
\tilde{a}_{\rm CIB} (k=1\MeV) &= 15.3^{+5}_{-4}\fm     &\mu &= 2M_\pi\,,
\end{align}
while the second variant $\tilde{a}^{\rm th (2)}_{\rm CIB} $ gives 
\begin{align}
\tilde{a}_{\rm CIB} (k=1\MeV) &= 19.6^{+4.5}_{-4}\fm  &  \mu &= M_\pi\,,\notag 
\\
\tilde{a}_{\rm CIB} (k=1\MeV) &= 19.0^{+4.5}_{-4}\fm   & \mu &= 2 M_\pi\,.
\end{align}
These values compare well  with $\tilde{a}^{\rm exp}_{\rm CIB} (k=1\MeV) =  14.25\fm$.
Using $k=10\MeV$, the experimental value decreases to $13.0\fm$ and the prediction is also reduced by slightly over $1\fm$. 

\item 
We can repeat the same analysis using the momentum-independent combination of scattering lengths 
\be
a_{\rm CIB}  =  \frac{ a_{nn} + a_{pp}^C}{2} - a_{np} \approx 10.4\fm\,, 
\label{eq:aCIB3}
\ee
by constructing theoretical quantities in complete analogy to Eq.~\eqref{eq:th12}.
The results are illustrated in Fig.~\ref{fig:aCIB2}. 
Using $a^{\rm th (1)}_{\rm CIB} $ we find 
\begin{align} 
a_{\rm CIB}   &= 13.5^{+5}_{-4}\fm   & \mu &= M_\pi\,,\notag 
\\
a_{\rm CIB}&= 12^{+5}_{-4}\fm  & \mu &= 2M_\pi\,,
\end{align}
and for $a^{\rm th (2)}_{\rm CIB} $
\begin{align}
a_{\rm CIB} &= 16.0^{+4.5}_{-4}\fm &  \mu &= M_\pi\,,\notag 
\\
a_{\rm CIB} &= 15.4^{+4.5}_{-4}\fm &   \mu &= 2M_\pi\,.
\end{align}

\end{enumerate}

\begin{figure}[t]
\begin{center}
\includegraphics[width=0.45\linewidth]{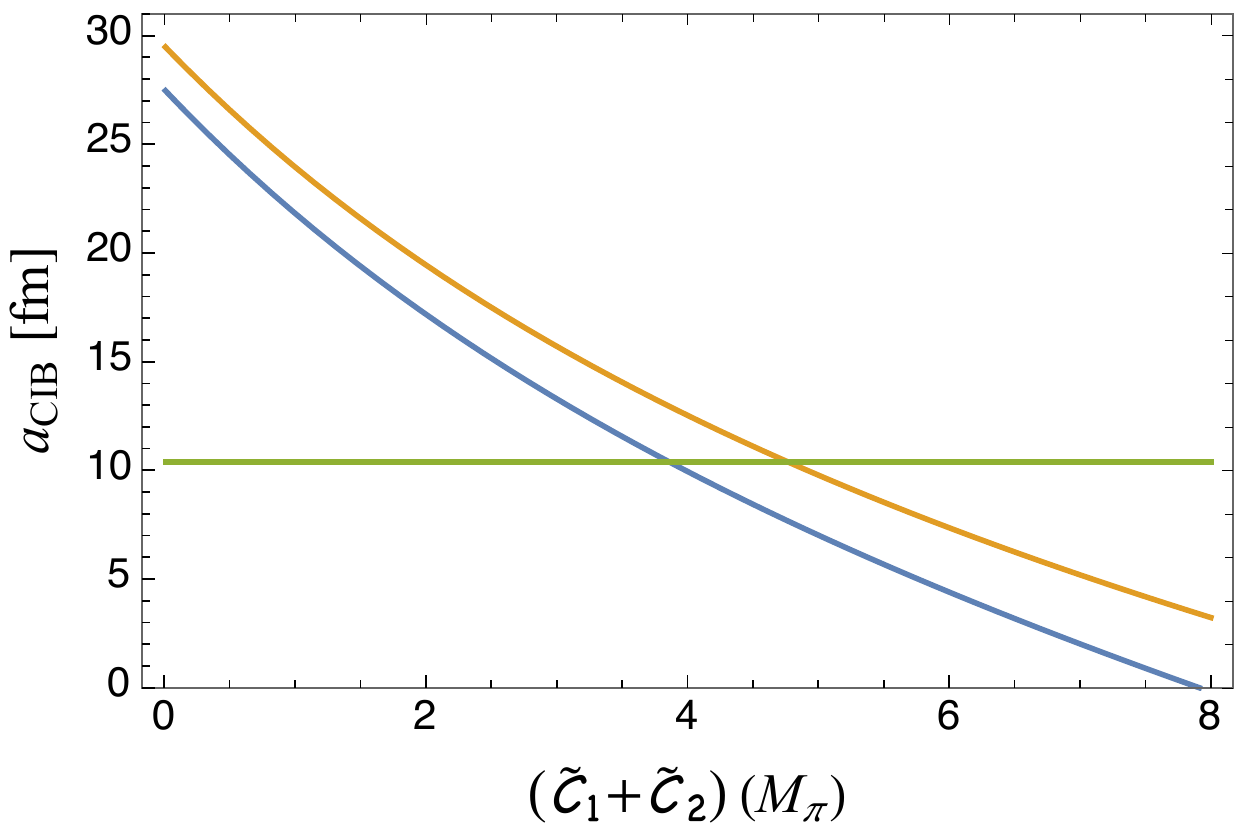}
\hspace{0.03\linewidth}
\includegraphics[width=0.45\linewidth]{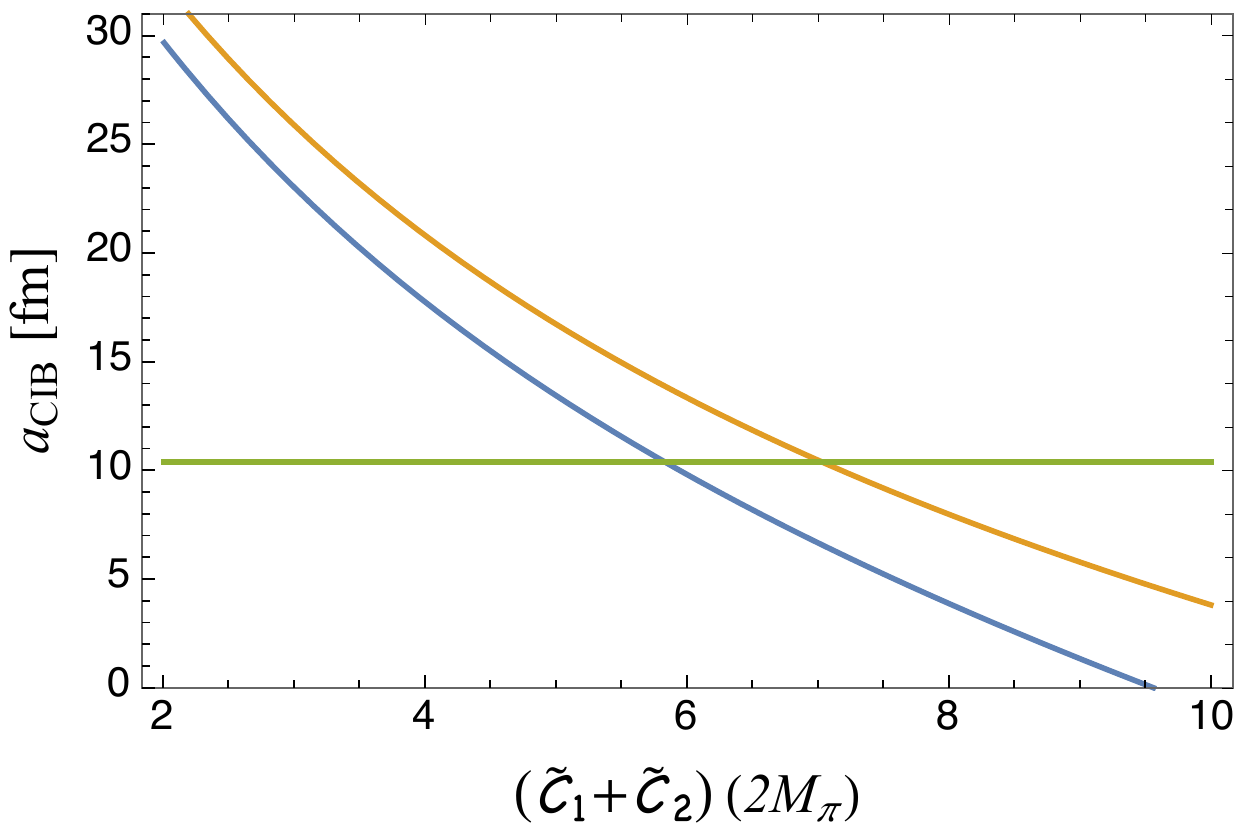}
\caption{  $a^{\rm th (1)}_{\rm CIB} $  (lower curve) and 
$a^{\rm th (2)}_{\rm CIB} $  (upper curve) 
versus the dimensionless coupling $\tilde{\cal C}_1 + \tilde{\cal C}_2$. 
The curves are evaluated at  $\mu = M_\pi$ (left panel)  and  $\mu = 2 M_\pi$ (right panel).
The horizontal line represents the experimental value  $a_{\rm CIB}  = 10.35\fm$. 
For reference, our theoretical ranges  for the CIB LEC are $ (\tilde{\cal C}_1 + \tilde{\cal C}_2) (M_\pi) = 2.9 (1.2)$ 
and $ (\tilde{\cal C}_1 + \tilde{\cal C}_2) (2 M_\pi) = 5.4(1.2)$. 
\label{fig:aCIB2}
}
\end{center}
\end{figure} 

In summary, we have derived a prediction of the CIB scattering length that overshoots the experimental value by about $35\%$ and has 
a comparable uncertainty,  thus comparing  well with data.   
This is a significant  phenomenological success of our theoretical approach and supports the validity of our 
uncertainty estimates.

\bibliographystyle{h-physrev3} 
\bibliography{bibliography}
\end{document}